\documentclass[longauth]{aa} 
\usepackage{xcolor}
\usepackage{support-caption}
\usepackage{subcaption}
\usepackage{graphicx}
\usepackage{threeparttable}
\usepackage{float}
\usepackage{booktabs}
\usepackage{longtable}
\usepackage{lscape}

\newcommand{\ha}{\ifmmode {\rm H}\alpha \else H$\alpha$\fi}
\newcommand{\hb}{\ifmmode {\rm H}\beta \else H$\beta$\fi}
\newcommand{\lya}{\ifmmode {\rm Ly}\alpha \else Ly$\alpha$\fi}
\newcommand{\pg}{\ifmmode {\rm P}\gamma \else Pa$\gamma$\fi}
\newcommand{\lyb}{\ifmmode {\rm Ly}\beta \else Ly$\beta$\fi}
\newcommand{\lyg}{\ifmmode {\rm Ly}\gamma \else Ly$\gamma$\fi}

\newcommand{\flyc}{\ifmmode \mathrm{f}_\mathrm{esc}\mathrm{(LyC)} \else $\mathrm{f}_\mathrm{esc}\mathrm{(LyC)}$\fi}

\def\kmsmpc{km s$^{-1}$ Mpc$^{-1}$}

\def\ergs{\ifmmode \mathrm{erg\hspace{1mm}s}^{-1} \else erg s$^{-1}$\fi}

\def\micron{\ifmmode \mu\mathrm{m} \else $\mu$m\fi}
\def\msun{\ifmmode \mathrm{M}_{\odot} \else M$_{\odot}$\fi}
\def\msunyr{\ifmmode \mathrm{M}_{\odot} \hspace{1mm}{\rm yr}^{-1} \else $\mathrm{M}_{\odot}$ yr$^{-1}$\fi}
\def\zsun{\ifmmode Z_{\odot} \else Z$_{\odot}$\fi}
\def\lsun{\ifmmode L_{\odot} \else L$_{\odot}$\fi}
\def\mstar{\ifmmode \mathrm{M}_{\star} \else M$_{\star}$\fi}

\newcommand{\jwst}{JWST}

\newcommand{\NIRSpec}{NIRSpec}
\newcommand{\NIRCam}{NIRCam}

\usepackage{txfonts}
\newcommand{\orcid}[1]{\href{https://orcid.org/#1}{\includegraphics[width=10pt]{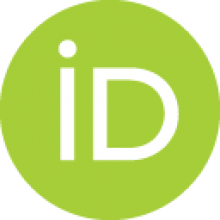}}}
\usepackage{hyperref}
\hypersetup{colorlinks, 
 linkcolor = blue,
 urlcolor = blue,
 citecolor = blue,
 anchorcolor = blue
}
\begin{document}

\title{Ly$\alpha$ visibility from z = 4.5 to 11 in the UDS field: Evidence for a high neutral hydrogen fraction and small ionized bubbles at z $\sim$ 7}

\titlerunning{\lya\ visibility evolution in the UDS field at z = 4.5 - 11}
\authorrunning{L. Napolitano et al.}

 \subtitle{}
     \author{L. Napolitano \orcid{0000-0002-8951-4408}
 \inst{1,2}
 \and L. Pentericci \orcid{0000-0001-8940-6768}
 \inst{1}
 \and M. Dickinson \orcid{0000-0001-5414-5131}
 \inst{3}
 \and P. Arrabal Haro \orcid{0000-0002-7959-8783}
 \inst{4}
 \and A. J. Taylor \orcid{0000-0003-1282-7454}
 \inst{5}
 \and A. Calabrò \orcid{0000-0003-2536-1614}
 \inst{1}
 \and \\ A. Bhagwat \orcid{0000-0003-0275-5506}
 \inst{15}
 \and P. Santini \orcid{0000-0002-9334-8705}
 \inst{1}
 \and F. Arevalo-Gonzalez \orcid{0009-0005-5968-8553}
 \inst{1}
 \and R. Begley \orcid{0000-0003-0629-8074}
 \inst{6}
 \and M. Castellano \orcid{0000-0001-9875-8263}
 \inst{1}
 \and B. Ciardi \orcid{0000-0002-5037-310X}
 \inst{15}
 \and \\ C. T. Donnan \orcid{0000-0002-7622-0208}
 \inst{3}
 \and D. Dottorini \orcid{0009-0004-5511-3309}
 \inst{1}
 \and J. S. Dunlop \orcid{0000-0002-1404-5950}
 \inst{6}
 \and S. L. Finkelstein \orcid{0000-0001-8519-1130}
 \inst{5}
 \and A. Fontana \orcid{0000-0003-3820-2823}
 \inst{1}
 \and M. Giavalisco \orcid{0000-0002-7831-8751}
 \inst{7}
 \and \\ M. Hirschmann \orcid{0000-0002-3301-3321}
 \inst{8}
 \and I. Jung \orcid{0000-0003-1187-4240}
 \inst{9}
 \and A. M. Koekemoer \orcid{0000-0002-6610-2048} 
 \inst{9}
 \and V. Kokorev \orcid{0000-0002-5588-9156}
 \inst{5}
 \and M. Llerena \orcid{0000-0003-1354-4296}
 \inst{1}
 \and R. A. Lucas \orcid{0000-0003-1581-7825}
 \inst{9}
 \and \\ S. Mascia \orcid{0000-0002-9572-7813}
 \inst{10}
 \and E. Merlin \orcid{0000-0001-6870-8900}
 \inst{1}
 \and P. G. P\'erez-Gonz\'alez \orcid{0000-0003-4528-5639}
 \inst{11}
 \and T. M. Stanton \orcid{0000-0002-0827-9769}
 \inst{6}
 \and R. Tripodi \orcid{0000-0002-9909-3491}
 \inst{1,12,13}
 \and X. Wang \orcid{0000-0002-9373-3865}
 \inst{16,17,18}
 \and  B. J. Weiner \orcid{0000-0001-6065-7483}
 \inst{14}
 }
 \institute{\textit{INAF – Osservatorio Astronomico di Roma, via Frascati 33, 00078, Monteporzio Catone, Italy}\\ 
 \email{lorenzo.napolitano@inaf.it}
 \and 
 \textit{Dipartimento di Fisica, Università di Roma Sapienza, Città Universitaria di Roma - Sapienza, Piazzale Aldo Moro, 2, 00185, Roma, Italy} 
 \and 
 \textit{NSF's National Optical-Infrared Astronomy Research Laboratory, 950 N. Cherry Ave., Tucson, AZ 85719, USA} 
 \and
 \textit{Astrophysics Science Division, NASA Goddard Space Flight Center, 8800 Greenbelt Rd, Greenbelt, MD 20771, USA} 
 \and
 \textit{Department of Astronomy, The University of Texas at Austin, Austin, TX, USA} 
 \and
 \textit{Institute for Astronomy, University of Edinburgh, Royal Observatory, Edinburgh EH9 3HJ, UK} 
 \and
 \textit{University of Massachusetts Amherst, 710 North Pleasant Street, Amherst, MA 01003-9305, USA} 
 \and
 \textit{Institute of Physics, Laboratory of Galaxy Evolution, Ecole Polytechnique Federale de Lausanne (EPFL), Observatoire de Sauverny, 1290 Versoix, Switzerland} 
 \and
 \textit{Space Telescope Science Institute, 3700 San Martin Drive, Baltimore, MD 21218, USA} 
 \and
 \textit{Institute of Science and Technology Austria (ISTA), Am Campus 1, A-3400 Klosterneuburg, Austria} 
 \and
 \textit{Centro de Astrobiolog\'{\i}a (CAB), CSIC-INTA, Ctra. de Ajalvir km 4, Torrej\'on de Ardoz, E-28850, Madrid, Spain} 
 \and 
 \textit{University of Ljubljana FMF, Jadranska 19, 1000 Ljubljana, Slovenia} 
 \and 
 \textit{IFPU - Institute for Fundamental Physics of the Universe, via Beirut 2, I-34151 Trieste, Italy} 
 \and
 \textit{MMT/Steward Observatory, University of Arizona, 933 N. Cherry St., Tucson, AZ 85721, USA} 
 \and 
 \textit{Max Planck Institut für Astrophysik, Karl Schwarzschild Straße 1, D-85741 Garching, Germany} 
 \and 
 \textit{School of Astronomy and Space Science, University of Chinese Academy of Sciences (UCAS), Beijing 100049, China} 
 \and 
 \textit{National Astronomical Observatories, Chinese Academy of Sciences, Beijing 100101, China} 
 \and
 \textit{Institute for Frontiers in Astronomy and Astrophysics, Beijing Normal University, Beijing 102206, China} 
}

\date{Received 19 August 2025 / Accepted 21 February 2026}

\abstract{The resonant scattering nature of Ly$\alpha$ photons interacting with neutral hydrogen makes Ly$\alpha$ emitters (LAEs) robust tracers of the intergalactic neutral hydrogen fraction, and thus sensitive probes of cosmic reionization. We present an extensive study of the Ly$\alpha$ evolution from galaxies at 4.5 $\leq$ z $\leq$ 11 in the UDS field, observed as part of the CAPERS survey, and complemented with spectra from the DAWN JWST Archive. The combined sample includes 651 spectroscopically confirmed Ly$\alpha$-break galaxies, among which we find 73 S/N>3 LAEs in JWST-NIRSpec PRISM spectra. We trace the redshift evolution of the LAE fraction with EW$_0$ >25\AA\ (X$_{\mathrm{Ly\alpha}}$) between z = 5 and z = 9, extending such an analysis to the UDS field for the first time. At z = 5 and 6, the UDS results agree with the average JWST X$_{\mathrm{Ly\alpha}}$ values from multiple fields. However, JWST measurements are consistently lower than ground-based results. To investigate this, we compare JWST observations to a population of star-forming galaxies at z$\sim$6 observed with VLT-FORS2. We find that a Ly$\alpha$ slit-loss of 35 $\pm$ 10\% in JWST spectra accounts for the offset, as the resonant Ly$\alpha$ emission is more spatially extended than the stellar continuum. From z = 6 to 7, the UDS field shows a significant drop in Ly$\alpha$ visibility, from which we infer a neutral hydrogen fraction of X$_{\mathrm{HI}}$ = 0.7--0.9. Finally, we identify two robust ionized bubbles at z = 7.29 and 7.77, with radii of $R_{\mathrm{ion}}$ = 0.6 and 0.5 physical Mpc and photometric overdensities of N/$\langle$N$\rangle$ = 3 and 4, based on candidate counts down to the photometric completeness limit. Compared to the large ionized region at z$\sim$7 in the EGS field, these results indicate significant field-to-field variation, supporting a patchy, inhomogeneous reionization process.}


 \maketitle

\section{Introduction} \label{sec:intro}
The main goals of observational studies of the epoch of reionization (EoR) are twofold. First, they aim to identify the earliest sources responsible for reionizing their surroundings and to constrain their ionizing nature, intrinsic luminosity, and spatial clustering. A second goal is to trace the time evolution of the volume-averaged neutral hydrogen fraction (X$_{\mathrm{HI}}$) in the intergalactic medium (IGM) across multiple independent lines of sight. 
One of the most effective probes of the evolving IGM is the Lyman-$\alpha$ (\lya, $\lambda_{\mathrm{rest}}$ = 1215.67~\AA) hydrogen emission line from star-forming galaxies \citep{Miralda1998, Malhotra2006, Dijkstra2014}. In the post-reionization Universe, the fraction of \lya-break galaxies showing \lya\ in emission (X$_{\mathrm{Ly\alpha}}$) is observed to increase with redshift, reflecting both younger stellar populations and lower dust content that let more \lya\ escape from the interstellar medium \citep[ISM, e.g.,][]{Kornei2010, Dijkstra2017, Napolitano2023}. However, due to the resonant nature of \lya, even modest amounts of neutral hydrogen in the IGM can significantly attenuate the line, making \lya\ emitters (LAEs) sensitive probes of reionization \citep{Dijkstra2014, Ouchi_2020}.
Ground-based spectroscopic surveys conducted on \lya-break selected sources found a significant drop in the fraction of LAEs at z > 6--7 \citep[][]{Fontana2010, Stark2010, Pentericci2011, Ono2012, Pentericci2014, Schenker2014, Tilvi2014, DeBarros2017, Mason2018, Pentericci_2018b, Bolan2022}, indicating a partially neutral IGM. Despite the progressive decline in the visibility of \lya, several LAEs have been spectroscopically identified at z > 7, suggesting the presence of ionized bubbles in overdensities, regions where \lya\ photons can escape IGM absorption due to cosmological redshifting within locally ionized zones \citep[e.g.,][]{Finkelstein2013, Castellano2016, Hu2017, Jung2020, Tilvi2020, Endsley2022, Jung2022, Larson2022, Leonova2022}. These bubbles, typically spanning $R_{\mathrm{ion}} \sim$ 0.1–3 physical Mpc \citep[pMpc, e.g.,][]{Mason2020}, are believed to originate from local overdensities of bright galaxies and are considered the seeds of spatially inhomogeneous reionization \citep[e.g.,][]{Keating2020, bosman22, Becker2024}.

Ground-based efforts to detect \lya\ at high redshift were restricted to following-up on the brightest sources \citep[e.g.,][]{Ono2012, Oesch2015, Roberts-Borsani2016} and, in case of a non-detection, atmospheric telluric lines and limited spectral coverage hindered the actual redshift determination of photometrically selected candidates. The advent of the James Webb Space Telescope \citep[\jwst,][]{Gardner2006, Gardner2023} has fundamentally changed this picture. Its Near InfraRed Spectrograph \citep[NIRSpec,][]{Jakobsen2022} provides continuous wavelength coverage in the 0.6--5.3 $\mu$m wavelength range, enabling simultaneous detection of both rest-frame UV and optical features, removing the observational bias associated with the detection of \lya. This has led to the spectroscopic confirmation of a substantial population of bright galaxies \citep[][]{Arrabal_Haro2023Nature, Bunker2023B, Curtis-Lake2023, DEugenio2023, Fujimoto2023, Wang2023, Castellano2024, Carniani2024, Hainline2024, Harikane2024, Hsiao2024, Donnan2025, Witstok2025Nature, Kokorev2025b, Naidu2025B, Napolitano2025a, Pollock2025, Tang2025} and active galactic nuclei \citep[AGNs,][]{Maiolino2024, Bogdan2024, Kovacs2024, Napolitano2025b, Taylor2025} already in place within 500 million years of cosmic time.\\
While the extreme frontier of z > 15 photometric candidates remains unconfirmed \citep[e.g.,][]{Castellano2025, Kokorev2025a, PerezGonzalez2025, Whitler2025}, we now have large samples of spectroscopically confirmed galaxies at z > 4, where the \lya-break falls within the wavelength range of \jwst-\NIRSpec. These observations enable us to trace the progression of cosmic reionization from its end phase \citep[z $\sim$ 5.2--5.7,][]{becker15b, bosman22, Spina2024} to its midpoint \citep[z $\sim$ 7.7,][]{Planck2020} and toward the dawn of galaxy formation. Furthermore, these datasets allow for direct comparisons with the post-reionization galaxy population at z = 4--5 using the same instrumentation and methodology.\\
\jwst\ has already enabled the systematic characterization of LAEs during the EoR over a wide luminosity range, from M$_{\mathrm{UV}}$ < -20 down to M$_{\mathrm{UV}} \sim$ -17 \citep{Tang2023, Chen2024, Jones2024, Jung2024, Nakane2024, Napolitano2024, Saxena2024, Tang2024B, Chen2025, Jones2025, Kageura2025, Morishita2025, Witstok2025Nature, Witstok2025}. The availability of data from multiple fields has improved constraints on cosmic variance and revealed significant field-to-field variations, reinforcing the picture of a patchy, inhomogeneous reionization \citep[e.g.,][]{Napolitano2024, Tang2024B, Runnholm2025}. \\
In this paper, we analyze for the first time all publicly available \jwst\ spectra of z > 4.5 \lya-break detected galaxies in the Ultra-deep Survey (UDS) field \citep{Lawrence2007} to constrain the effect of the reionization process on \lya\ visibility, the neutral hydrogen fraction, and the spatial distribution of ionized regions. We compare UDS results with those obtained in other \jwst\ fields and ground-based surveys. \\
The paper is organized as follows. We discuss the parent sample construction in Sect.~\ref{sec:Data_and_sample_selection} and the methodology used in Sect.~\ref{sec:Method}. We present the analysis of the observed redshift evolution of \lya\ fraction $X_{\lya}$ in Sect.~\ref{sec:XlyA}. The mismatch between the \jwst-based and ground-based \lya\ visibilities and potential \jwst\ \lya\ slit losses are discussed in Sect.~\ref{sec:slitloss}. We present constraints on the UDS neutral hydrogen fraction X$_{\mathrm{HI}}$ and ionized bubbles in Sect.~\ref{sec:XHI} and Sect.~\ref{sec:bubble}, respectively. We summarize our conclusions in Sect.~\ref{sec:Conclusion}.\\
In the following, we adopt the $\Lambda$CDM concordance cosmological model ($H_0 = 70$ \kmsmpc, $\Omega_M = 0.3$, and $\Omega_{\Lambda} = 0.7$). We report all magnitudes in the AB system \citep{Oke1983} and equivalent widths (EWs) to rest-frame values.

\section{Data} \label{sec:Data_and_sample_selection}
To investigate \lya\ visibility during the Epoch of Reionization, we compiled the largest possible sample of galaxies with \jwst\ spectroscopic observations in the UDS field. We focused on sources with secure redshifts (z > 4.5, see Sect.~\ref{sec:zspec}) for which \lya\ falls within the blue-sensitive range of the \NIRSpec-PRISM configuration. Additionally, since we analyzed spectroscopic overdensities near LAEs, we included galaxies observed with higher-resolution \NIRSpec\ configurations, using their spectra only to obtain reliable redshift confirmations.
In total we considered 651 unique galaxies at 4.5 $\leq$ z $\leq$ 11: 135 are from the CAPERS survey and 516 from other programs whose data are included in the public DAWN JWST Archive\footnote{\url{https://dawn-cph.github.io/dja/index.html}}\citep[DJA,][]{Heintz2024dja, deGraaff2025}. Among the total sample, 531 spectra were obtained from the \NIRSpec-PRISM (R $\sim$ 30 - 300) configuration, and 120 with medium-resolution configurations (R $\sim$ 1000), including G140M/F100LP, G235M/F170LP, and G395M/F290LP.

\subsection{UDS-CAPERS data}\label{sec:CAPERS}
We utilized publicly released \jwst-\NIRSpec-PRISM data from the CANDELS-Area Prism Epoch of Reionization Survey (GO-6368, CAPERS; PI Mark Dickinson), an ongoing Cycle 3 Treasury Program. CAPERS is designed to observe a total of 21 pointings, uniformly distributed across the Cosmic Evolution Survey (COSMOS), the Ultra-deep Survey (UDS), and the Extended Groth Strip (EGS) fields, using the PRISM/CLEAR configuration. In this work, we focus on the four pointings already observed in the UDS field. These UDS observations were conducted between UT 2024 December 31 and 2025 January 10, and originally included three MSA configurations for each of the seven scheduled pointings, for a total of 21 planned configurations. 
Due to technical issues and communication restrictions with JWST, only eight of the planned MSA observations were completed. The remaining UDS observations are scheduled for completion by January 2026. A detailed description of CAPERS target selection and observational strategy will be presented in a forthcoming survey paper. Here, we summarize the key aspects relevant to our analysis. 

Targets were selected based on their detection in the \NIRCam\ F277W, F356W, and F444W bands. 
By design, the CAPERS UDS pointings have non-overlapping footprints on the sky. Exposure times vary with target priority: within a given pointing, sources were assigned to one, two, or all three MSA configurations, each providing 5690 s of integration over the same sky footprint. Among the 135 z > 4.5 CAPERS galaxies, 24, 56, and 55 have total exposure times of 17070 s, 11380 s, and 5690 s, respectively.\\
We processed the data using the STScI Calibration Pipeline\footnote{\url{https://jwst-pipeline.readthedocs.io/en/latest/index.html}} version 1.16.1 \citep{Bushouse2025} with CRDS version \texttt{1312.pmap}, following the general procedures outlined in \cite{ArrabalHaro2023}, with two key modifications: we employed the \texttt{clean\_flicker\_noise} module\footnote{\url{https://jwst-pipeline.readthedocs.io/en/stable/jwst/clean\_flicker\_noise/main.html\#correction-algorithm}} in the \texttt{calwebb\_detector1} stage to mitigate the effect of $1/f$ noise; and we applied a modified flat-field during \texttt{calwebb\_spec2} stage. For slits with multiple sources (primary and known or serendipitous secondary), we combined the 2D spectra using an asymmetric nodding pattern in the \texttt{calwebb\_spec2} stage as described in \cite{Napolitano2025a} to mitigate contamination. We extracted the 1D spectra using optimal extraction described in \cite{Horne1986}.\\
To check for potential residual issues in the absolute flux calibration caused by slit losses or other inaccuracies in flux calibration files, we compared the extracted spectra with broadband photometry from \citet{Merlin2024} by integrating the spectra across the \NIRCam\ filter bandpasses of F090W, F115W, F150W, F200W, F277W, F356W, and F444W. We only considered the resulting synthetic photometry with a signal-to-noise ratio (S/N) of > 5. For each spectrum, we computed a correction factor per filter by comparing synthetic and observed fluxes, then applied a weighted average correction. This correction was independent of wavelength and was applied uniformly to fluxes and associated uncertainties. We note that the photometric correction does not affect EW and UV $\beta$ slope measurements.

\subsection{Other UDS data from DJA archive} \label{sec:otherJWST}
To complement the CAPERS data, we collected all publicly available \jwst\ spectroscopic observations in the UDS field from version 3 of DJA. We limited our selection to the most robust identifications (flagged as grade = 3 in DJA) at z > 4.5, to match the redshift range of interest.\\
This selection includes: 459 spectra from GO-4233 \citep[RUBIES; PI deGraaff,][]{deGraaff2025}, of which 361 and 98 are in the \NIRSpec-PRISM and \NIRSpec-G395M/F290LP configurations, respectively; 31 PRISM spectra from GTO-1215 (PI Luetzgendorf); 4 PRISM spectra from GO-2565 \citep[PI Glazebrook and Nanayakkara,][]{Kawinwanichakij2025}; and 22 medium resolution (G140M/F100LP, G235M/F170LP, and G395M/F290LP) spectra from GO-3543 \citep[EXCELS; PI Carnall,][]{Carnall2025}. 
For consistency, we applied the same photometric correction method to all PRISM spectra as described for CAPERS.
We note that the combination of RUBIES and CAPERS data accounts for 91\% of the final sample. We verified that both RUBIES and CAPERS rely on detections in the same set of \NIRCam\ bands (F277W, F356W, and F444W), ensuring broadly consistent target selection across the two datasets. 

\section{Methods} \label{sec:Method}

\subsection{Spectroscopic redshift} \label{sec:zspec}
\begin{figure*}[!ht]
\begin{minipage}{0.5\textwidth}
\centering
\includegraphics[width=\linewidth]{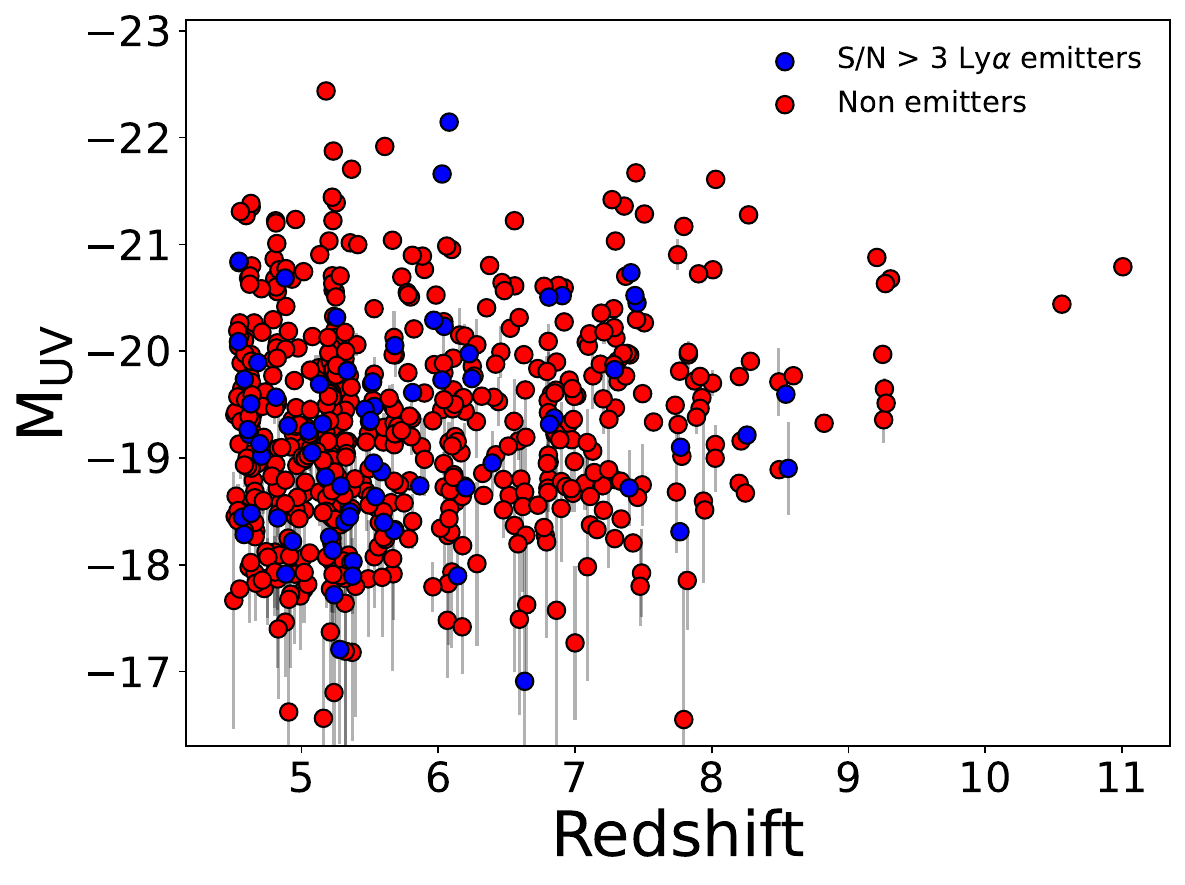}
\end{minipage}
\begin{minipage}{0.5\textwidth}
\centering
\includegraphics[width=\linewidth]{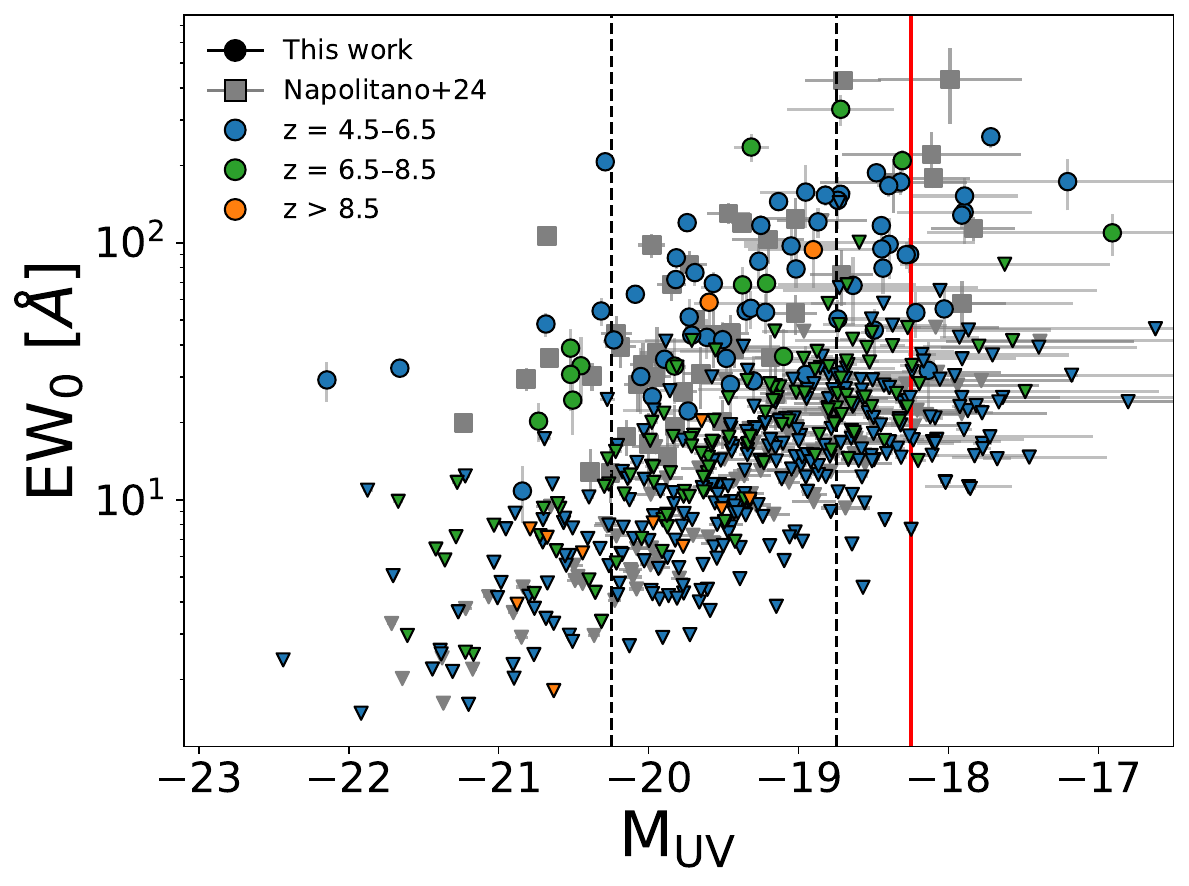}

\end{minipage}
\caption{Left: Absolute UV magnitude as a function of redshift in the UDS sample. The 73 galaxies with robust S/N > 3 \lya\ detection are shown in blue, while the rest of the population is shown in red. Right: Distribution of \lya\ EW$_0$ as a function of M$_{\mathrm{UV}}$. Galaxies in the UDS with S/N > 3 are shown as circles, while upper limits EW$_{\mathrm{0,lim}}$ are shown as triangles. They are color-coded by spectroscopic redshift. For comparison, we report CEERS-EGS emitters (gray squares) and upper limits (gray triangles) from \cite{Napolitano2024}. The dashed black lines at M$_{\mathrm{UV}}$ = –20.25 and –18.75 define the bright and faint regimes. The UDS sample is complete down to M$_{\mathrm{UV}}$ = -18.25, indicated by the solid red line.}
\label{fig:UVcompare}
\end{figure*}

Spectroscopic redshifts in the CAPERS dataset were determined using a combination of automated tools, interactive inspection, and visual vetting. Each redshift was assigned a quality flag to indicate its reliability, with flags 3 and 4 designating the most secure solutions, supported by multiple emission line detections. Full details of the procedure will be provided in a forthcoming CAPERS survey paper. For galaxies retrieved from the DJA archive, we used the secure spectroscopic redshifts (flagged as grade = 3) provided by \texttt{msaexp} \citep{Brammer2022_msaexp} as a reference. Based on secure CAPERS and DJA redshift solutions, we preselected sources with z > 4.5. We then visually inspected the 1D and 2D spectra of all selected objects to validate the spectroscopic redshift solution based on rest-frame optical, UV emission lines, and the \lya-break. When emission lines with S/N > 3 are detected, we computed redshifts using their observed centroids \citep[e.g.,][]{Castellano2024, Napolitano2025a}.
The expected observed emission line profile was broadened by the instrumental resolution, with a standard deviation given by $\sigma_R (\lambda_{\mathrm{obs}}) [\AA] = \lambda_{\mathrm{obs}} / 2.355 R(\lambda_{\mathrm{obs}})$. The resolution was provided by the \jwst\ documentation\footnote{\url{https://jwst-docs.stsci.edu/jwst-near-infrared-spectrograph/nirspec-instrumentation/nirspec-dispersers-and-filters\#gsc.tab=0}} with the assumption of a source that illuminates the slit uniformly. The final spectroscopic redshift was computed as a weighted average of all significant line-based estimates. For reference, we used the following key rest-frame optical lines when available: H$\alpha$, [OIII]$\lambda$5007, [OIII]$\lambda$4959, H$\beta$, and [OII]$\lambda \lambda$3727,29 in 510, 572, 271, 533, and 199 cases, respectively. Spectroscopic redshifts are reported in the left panel of Fig.~\ref{fig:UVcompare} and Table \ref{tab:summary_data}. All sources observed with \NIRSpec-PRISM show a robust detection of the \lya-break. The spectroscopic redshifts we find always agree with the CAPERS and DJA redshift solutions used as preselection.

\subsection{AGN identification} \label{sec:AGN}
To identify AGNs within the sample for further selection (see Sect.~\ref{sec:EoR}), we first visually inspected all spectra for the presence of broad Balmer emission lines. We used H$\alpha$ as the reference line, while at z > 7 where H$\alpha$ falls outside the spectral coverage, we instead considered H$\beta$. We then performed a quantitative fit to the emission profile, testing two models on top of a linear continuum: a single narrow Gaussian with the standard deviation fixed by the instrumental resolution $\sigma_R (\lambda_{\mathrm{obs}})$, corresponding to a full width at half maximum (FWHM) of $\sim$ 1000--2000 km/s; and a combination of a narrow and a broad Gaussian component. Uniform priors were adopted for the amplitudes, and the broad component was allowed to have a FWHM up to 10,000 km/s. The model parameters were sampled using the \textsc{emcee} Markov chain Monte Carlo (MCMC) sampler \citep{Foreman_Mackey2013}, with 30 walkers and 10,000 steps. The best-fit parameters and uncertainties were derived from the posterior median and the 68-th percentile intervals, respectively. Following \cite{Juodzbalis2025}, to evaluate the necessity of the broad component, we used the Bayesian information criterion (BIC), which incorporates both the goodness of fit ($\chi^2$) and the number of free parameters. To accept the model with the broad emission line, we required a $\Delta$BIC > 0. Additionally, we required the best-fit broad FWHM to exceed the instrumental resolution by more than 2$\sigma$ to ensure robustness. This two-step criterion minimizes false positives and guarantees that the broad feature is both statistically and physically significant. With this approach, we identified 24 sources as broad-line AGNs (BLAGNs), 20 of which are from the UDS-RUBIES sample. Among the RUBIES sources, eight were previously classified as BLAGN by \cite{Taylor2024} and nineteen by \cite{Hviding2025}. \\
Additionally, we cross-matched the UDS dataset with the X-ray-detected sources from X-UDS \citep{Kocevski2018}, which provides $\sim$ 600 ks of Chandra coverage across the field. However, we note that the current X-ray data do not allow us to identify heavily obscured or X-ray weak AGNs \citep{Madau2024B}, which may still be present in the UDS sample.
In particular, narrow-line AGNs (NLAGNs) may represent up to $\sim$ 20\% of spectroscopic galaxy samples, as shown in the JADES and CEERS surveys by \cite{Scholtz2023} and \cite{Mazzolari2024b}, respectively. However, identifying NLAGNs requires multiple robust line detections in both the UV and optical rest-frame. These emission lines are necessary to constrain key physical properties (e.g., metallicity, ionization parameter, abundance ratios) and place sources in diagnostic diagrams designed to distinguish AGNs from star-forming galaxies \citep[e.g.,][]{Hirschmann2019, Hirschmann2023, Mazzolari2024a}, and to fairly compare with photoionization models \citep[e.g.,][]{Feltre2016, Gutkin2016, Nakajima2022}. At high redshift, even for sources with extreme line emission and high S/N, such as GHZ2 \citep{Castellano2024}, current low-resolution PRISM data have yielded inconclusive classifications. In the UDS sample, most galaxies are missing the key line transitions needed for robust NLAGN diagnostics. Given these limitations, we do not attempt NLAGN identification in this work.\\
In the following, we consider as AGNs only the 24 sources identified through the broad-line selection. The list of confirmed AGNs is provided in Table \ref{tab:summary_data}. 

\subsection{M$_{\mathrm{UV}}$ and $\beta$ } \label{sec:Muv}
Following \cite{Napolitano2025a}, we estimated the UV slope $\beta$ for all the \NIRSpec-PRISM sources by fitting a power-law model ($f_{\lambda} \propto \lambda^\beta$) to the continuum flux in the rest-frame range 1350--2600~\AA. This wavelength range avoids the damping wing region (1200--1350~\AA), which can bias the UV slope, especially at high redshift \citep[e.g.,][]{Dottorini2025}. To avoid contamination, we masked known bright features (i.e., CIV$\lambda\lambda$1548,51 and CIII]$\lambda$1909) using line widths broadened according to the instrumental resolution (see Sect.~\ref{sec:zspec}). The fitting was performed using the \textsc{emcee} with 30 walkers and 500,000 steps. We adopted a flat prior on $\beta$ between -3.5 and 0. The best-fit value and uncertainty were derived from the posterior median and standard deviation, respectively. Each fit is visually inspected to ensure reliability. \\
If the fit was successful (86\% of cases), we computed the absolute UV magnitude (M$_{\mathrm{UV}}$) from the model flux averaged over 1450--1550~\AA\ in the rest-frame. In the remaining cases (14\% of total) in which the posterior distribution implies a S/N<3, we derived M$_{\mathrm{UV}}$ directly from the median observed flux in the same wavelength range, and the UV $\beta$ slope from the relation derived by \cite{Dottorini2025}:
\[
\beta = 
\begin{cases}
  (-0.110 \ \pm \ 0.019) \ \mathrm{M}_{\mathrm{UV}} - (4.00 \ \pm 0.38), & z < 6.5 \\
  (0.021 \ \pm \ 0.045) \ \mathrm{M}_{\mathrm{UV}} - (1.6 \ \pm 0.9), & z \geq 6.5.
\end{cases}
\]\\
For galaxies observed in medium-resolution configurations, where the continuum is typically undetected, we estimated M$_{\mathrm{UV}}$ from the photometry, considering only the relevant bands whose throughput\footnote{\url{https://jwst-docs.stsci.edu/jwst-near-infrared-camera/nircam-instrumentation/nircam-filters\#gsc.tab=0}} includes the 1500~\AA\ rest-frame information. We note the considered UDS sample is complete down to M$_{\mathrm{UV}}$ = -18.25.
All results are reported in the left panel of Fig.~\ref{fig:UVcompare} and Table \ref{tab:summary_data}. While the UV slope is unaffected by the photometric correction, we report M$_{\mathrm{UV}}$ using the flux-corrected spectra. 

\subsection{\lya\ emission line measurements} \label{sec:LyAmodel}
We briefly summarize the \lya\ fitting procedure previously adopted in \cite{Napolitano2024}. Due to the low spectral resolution of the \NIRSpec-PRISM, \lya\ emission was modeled within a $\sim$ 4--5 pixel window centered on the observed peak. As a first step, we computed the line flux via direct integration over this window. The continuum was estimated by a linear fit to the red side of \lya, from 1900~\AA\ to 3 pixels redward of the emission peak. \cite{Chen2024} and \cite{Napolitano2024} showed that this method underestimates the intrinsic \lya\ flux by 30--50\%. \\
To refine the measurement, we adopted a forward-modeling approach. We constructed a library of Gaussian emission with FWHMs uniformly sampled in the 100--1500 km/s range. For each profile, the amplitude was drawn from a uniform distribution constrained to yield an integrated flux within a factor of 0.2--5 of the value obtained via direct integration. Each Gaussian profile was combined with a step-function continuum: the red side is given by the linear fit, while the blue side is fixed to the median flux blueward of the line, accounting for IGM absorption. 
The full model was convolved with a Gaussian kernel ($\sigma_R (\lambda_{\mathrm{obs}})$) matching the instrumental resolution. We performed the fit using \textsc{emcee} with 10 walkers and 20,000 steps. The best-fit \lya\ flux and EW$_0$ were derived from the posterior median, and uncertainties from the 68-th percentile intervals.\\
In total, we identify 73 robust \lya\ emitters with S/N > 3. They are reported in the left panel of Fig.~\ref{fig:UVcompare}.
In the appendix (see Sect.~\ref{sec:app_figures}) we show the fit \lya\ line profiles for all sources together with the associated continuum fits, and their residuals (Fig.~\ref{fig:Lya_emitters}), which show no evidence of systematic deviations. For all galaxies, including non-emitters, we derived the limiting rest-frame equivalent width that could be measured (EW$_{0,\mathrm{lim}}$) following Equation 2 from \cite{Jones2024}, which incorporates the flux uncertainty at the \lya\ peak ($E(\lambda ^{\lya})$), the continuum level ($F_{\lambda}^{\mathrm{cont\ }}$), and the Gaussian kernel that accounts for the instrumental resolution:
\begin{equation*} 
    EW_{\mathrm{0,lim}} = \frac{\sqrt{2 \pi} \ E(\lambda ^{\lya}) \sigma_R (\lambda_{\mathrm{obs}})}{(1+z)F_{\lambda}^{\mathrm{cont\ }}}.
\end{equation*}
We used these limits to assess the \lya\ sample completeness, as described in Sect~\ref{sec:XlyA}. 
We show the measured rest-frame equivalent width and limits as a function of M$_{\mathrm{UV}}$ in the right panel of Fig.~\ref{fig:UVcompare}. Results are reported in Table \ref{tab:summary_data}. We note that \cite{Chen2025} discuss the \lya\ emission of RUBIES-24303, RUBIES-930869, and CAPERS-142615, finding \lya\ EW$_0$ compatible with our measurements, within 1--2$\sigma$ uncertainty.

\subsection{SED fitting} \label{sec:SED_fitting}
Stellar masses were derived from a spectral energy distribution (SED) fit of the observed photometry \citep{Merlin2024}, fixing the redshift of each source to the spectroscopic value. We fit synthetic stellar templates with the SED fitting code \textsc{zphot} \citep{fontana00}, following the method described in \cite{Santini2023}. We adopted \cite{Bruzual2003} models, assumed a \cite{Chabrier2003} IMF, and modeled delayed star formation histories (SFH($t$) $\propto (t^2/\tau) \cdot \exp(-t/\tau)$), with the e-folding timescale, $\tau$, ranging from 100 Myr to 7 Gyr. The time elapsed since the onset of star formation was constrained to be between 10 Myr and the age of the Universe at each galaxy redshift, while for the metallicity we assumed values of 0.02, 0.2, or 1 times the solar metallicity. For the dust extinction, we used the \cite{Calzetti2000} law with $E(B-V)$ ranging from 0 to 1.1. Nebular emission was included following the prescriptions of \cite{Castellano2014} and \cite{Schaerer2009}. Briefly, hydrogen lines were computed from the number of hydrogen-ionizing photons predicted by the stellar SED, assuming case B recombination, a null LyC escape fraction, an electron temperature of T$_{\mathrm{e}}$ = 10000 K, and an electron density of N$_{\mathrm{e}}$ = 100 cm$^{-3}$. Nebular E(B-V) was assumed to be equal to the stellar reddening. He and metal lines were added by scaling the H$\beta$ flux using the tabulated values as a function of metallicity from \cite{Anders2003}. 

\section{\lya\ constraints on reionization in the UDS field}\label{sec:EoR}
After three years of operations, \jwst\ has yielded a statistically significant number of \lya-break galaxy spectra at z > 4, enabling measurements of the evolving neutral hydrogen fraction across several extragalactic fields, including EGS, GOODS-S, GOODS-N, and Abell-2744 \citep[e.g.,][]{Jones2024, Nakane2024, Napolitano2024, Umeda2024, Tang2024B, Jones2025, Kageura2025, Umeda2025}. 
In this section, we extend such analyses to the UDS field for the first time, presenting a systematic measurement of the \lya\ emitter fraction and the corresponding neutral hydrogen fraction using \jwst\ spectroscopic observations. We compare the UDS results with both \jwst-based and ground-based estimates from the relevant literature.

\subsection{The evolution of the \lya\ fraction} \label{sec:XlyA}
\begin{figure}[t]
\centering
\includegraphics[width=\linewidth]{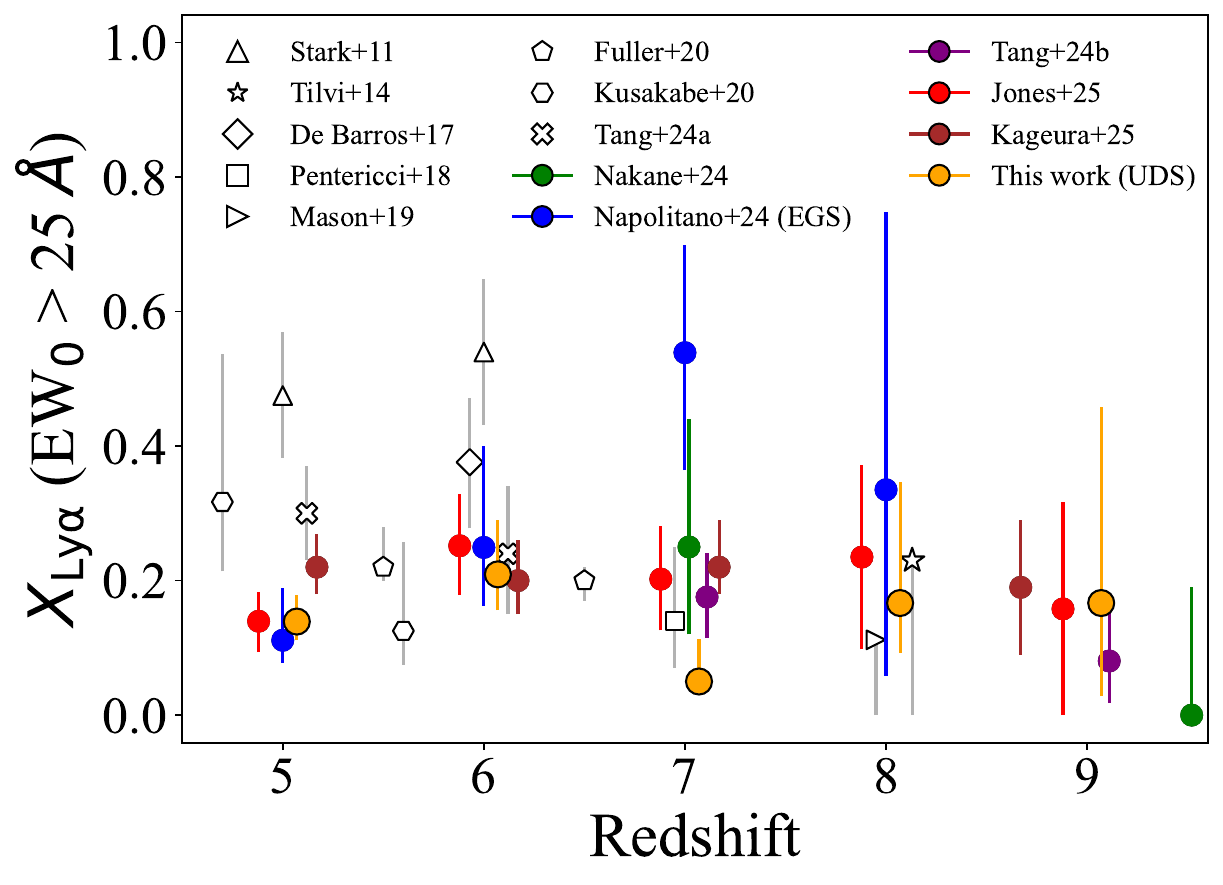}
\caption{Redshift evolution of the fraction of galaxies with observed EW$_0$ > 25~\AA. Results from this work are shown as orange circles. No correction for \lya\ slit losses was applied. Error bars were computed using binomial statistics following \cite{Gehrels1986}. Colored points represent \jwst\ estimates \citep[][]{Nakane2024, Napolitano2024, Tang2024B, Jones2025, Kageura2025}, while open symbols with black edges correspond to ground-based literature results \citep[][]{Stark2011, Tilvi2014, DeBarros2017, Pentericci_2018b, Mason2019, Fuller2020, Kusakabe2020, Tang2024}. Data points have been slightly shifted in redshift for clarity.} \label{fig:XLya}
\end{figure}

We defined the \lya\ emitter fraction, X$_{\mathrm{Ly\alpha}}$, as the ratio between the number of galaxies with a detected S/N > 3 \lya\ emission with EW$_0$ > 25~\AA\ and the total number of \lya-break galaxies with no signs of AGN activity (Sect.~\ref{sec:AGN}), restricted to the absolute UV magnitude range -20.25 < M$_{\mathrm{UV}}$ < -18.75. All galaxies within this range have a secure UV continuum detection (see Sect.~\ref{sec:Muv}). 
The same UV selection is commonly adopted in the literature \citep[e.g.,][]{Stark2011, Pentericci2011, Ono2012} to ensure a consistent and fair reference sample for comparisons. This is motivated by the well-established correlation between M$_{\mathrm{UV}}$ and \lya\ EW$_0$ \citep[e.g.,][see also Fig.~\ref{fig:UVcompare}]{Nakane2024, Napolitano2024, Jones2025}, driven by intrinsic galaxy properties that regulate \lya\ radiative transfer through the interstellar and circumgalactic media. To ensure completeness, we only included \lya-break galaxies with a limiting rest-frame equivalent width EW$_{0,\mathrm{lim}}$ < 25~\AA. Galaxies with EW$_{0,\mathrm{lim}}$ > 25~\AA\ were excluded from the statistical analysis, regardless of whether \lya\ emission is formally detected. This results in a total sample of 216 galaxies. In practice, we restricted the sample to spectra with sufficient sensitivity to confidently detect a line above the adopted EW$_0$ threshold, given the continuum level and redshift. 
This prevents upward bias from bright lines detected in shallow spectra, as well as downward bias from continuum-faint sources in which \lya\ with such a small EW$_0$ would be undetectable even if present. We find that 46\% of the \lya-break galaxies in the UDS sample meet the above criteria. 
We remark that X$_{\mathrm{Ly\alpha}}$ was calculated using the measured EW$_0$ values, without correcting for additional \lya\ slit losses.\\
We measured X$_{\mathrm{Ly\alpha}}$ across five redshift bins centered at z = 5, 6, 7, 8, and 9, each with a width of $\Delta$z = 1. Given the small-number statistics, uncertainties were estimated following \cite{Gehrels1986}. The redshift evolution of the \lya\ emitter fraction in the UDS field is presented in Table \ref{tab:Xlya} and Fig.~\ref{fig:XLya}.
\begin{table}
\caption{Observed fraction of \lya-emitting galaxies with EW$_0$ > 25~\AA\ as a function of redshift.}
    \label{tab:Xlya}
    \centering
    \begin{tabular}{ccc}
    \hline \noalign{\smallskip}
         & \multicolumn{2}{c}{X$_{\mathrm{Ly\alpha}}$ (EW$_0$ > 25 \AA)} \\
        \noalign{\smallskip}
        Redshift & UDS (This work) & Average JWST fields \\
        \hline \noalign{\smallskip}
        5 & 13.9$_{-2.7}^{+4.0}$ & 15.3 $\pm$ 2.1 \\ [3pt]
        6 & 21$_{-5}^{+8}$ & 21.8 $\pm$ 3.5 \\ [3pt]
        7 & 5.0$_{-1.9}^{+6}$ & 14.8 $\pm$ 2.7 \\ [3pt]
        8 & 17$_{-7}^{+18}$ & 21 $\pm$ 9 \\ [3pt]
        9 & 17$_{-14}^{+29}$ & 10 $\pm$ 6 \\ [3pt]
        \hline
    \end{tabular}
    \tablefoot{Results from the UDS field are not corrected for \lya\ slit losses. For comparison, we report the average estimates from all available \jwst\ fields obtained when considering also literature results \citep{Nakane2024, Napolitano2024, Tang2024B, Jones2025, Kageura2025}.}
\end{table}
We observe an increase in X$_{\mathrm{Ly\alpha}}$ between z = 5 and z = 6 and then a decrease to z = 7, as has previously been observed in many other studies. At higher redshift (i.e., the  z = 8 and z = 9 bins) we have an apparent increase although the results are particularly affected by small-number statistics, with only 12 and 6 \lya-break galaxies, respectively. 

We compared our results with other \jwst-based estimates \citep[][]{Nakane2024, Napolitano2024, Tang2024B, Jones2025, Kageura2025} as well as with ground-based surveys \citep[][]{Stark2011, Tilvi2014, DeBarros2017, Pentericci_2018b, Mason2019, Fuller2020, Kusakabe2020, Tang2024} that adopt similar selection criteria. Previous \jwst\ measurements were based on multiple-fields data, including the EGS, Abell-2744, GOODS-S and GOODS-N, with the exception of \cite{Napolitano2024}, which reported EGS results. 
To better assess the comparison, we computed the weighted average \jwst-based \lya\ emitter fraction, X$_{\mathrm{Ly\alpha}}$, using all available estimates, where the weights were defined as the inverse variances from the original papers. As the estimates from different studies rely on partially overlapping datasets, we caution that they are not statistically independent and that the weighted average should be considered as a descriptive reference of literature findings on the combined JWST fields. The results are summarized in Table~\ref{tab:Xlya}. In particular we note that at z = 5 and z = 6, average results from multiple  fields agree very well with those derived in the UDS field alone, as well as the EGS field alone.
The lack of significant field-to-field variations at these redshifts confirms that the Universe is largely ionized, with little residual neutral hydrogen \citep[e.g.,][]{bosman22, Spina2024}, and therefore the visibility of \lya\ is primarily driven by the intrinsic properties of galaxies, rather than by the IGM.

On the other hand, at z = 7, there is a significant scatter between the fractions of \lya\ obtained from different fields. In the EGS field, it is anomalously high, while in the UDS it is comparatively low, with the two measurements differing at the $\sim$2.7$\sigma$ level. The other points reported in the figure were computed on more than one field: for example, the fraction by \cite{Jones2025} is from GOODS-S and GOODS-N, while \cite{Kageura2025} also includes the EGS and Abell-2744 fields, therefore providing only median visibilities across different lines of sight. This highlights that at z = 7 (and above)  the visibility of \lya\ is also driven by local IGM conditions. For example, the high incidence of \lya\ emitters at z $\sim$ 7 in the EGS field has been previously reported \citep[e.g.,][]{Chen2024, Napolitano2024, Tang2024B} and attributed to the presence of large ionized region(s) \citep[ionized bubbles identified in][]{Tilvi2020, Jung2022, Leonova2022, Chen2024, Napolitano2024, Chen2025} that locally enhance \lya\ transmission \citep[e.g.,][]{Ouchi2010}. In contrast, the low \lya\ fraction found in the UDS field at z $\sim$ 7 would support a scenario with a largely neutral IGM condition. 
In this sense, the EGS and UDS fields would represent extreme and opposite conditions in the IGM neutral hydrogen content at z = 7. \\
We further derived the implications for the neutral hydrogen fraction on these two fields in Sect.~\ref{sec:XHI}. These findings highlight the importance of wide-field, multi-pointing surveys to accurately quantify the global visibility of \lya\ during reionization. They also support a scenario of highly spatially inhomogeneous reionization, in which the presence of ionized bubbles can significantly alter the \lya\ visibility \citep[e.g.,][]{Taylor2014}, as has also been suggested by LAE clustering analyses \citep[e.g.,][]{Ouchi2010, Ouchi2018, Umeda2025a}.

At redshift 8 and above, the scatter between various results is also very large, but we emphasize that all fractions are based on very low number statistics. \\
Although all \jwst-based estimates are in good agreement in the post-ionization Universe a tension persists at z = 5 and z = 6 between the new JWST results and previous ground-based measurements. We discuss this discrepancy in the next section.

\subsection{Comparison between \jwst\ and ground-based spectroscopic \lya-break samples at z $\sim$ 6} \label{sec:slitloss}

\begin{figure*}[h]
\begin{minipage}{0.5\textwidth}
\centering
\includegraphics[width=\linewidth]{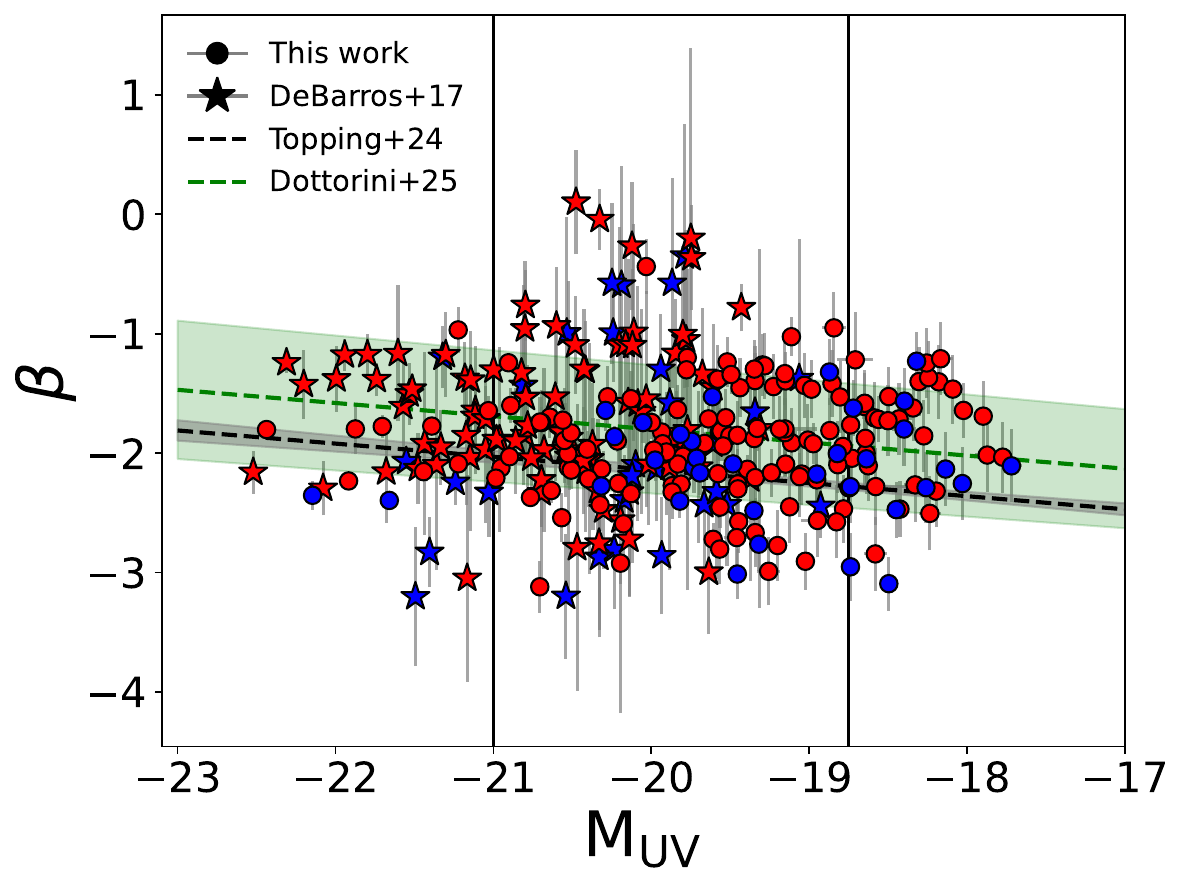}
\end{minipage}
\begin{minipage}{0.5\textwidth}
\centering
\includegraphics[width=\linewidth]{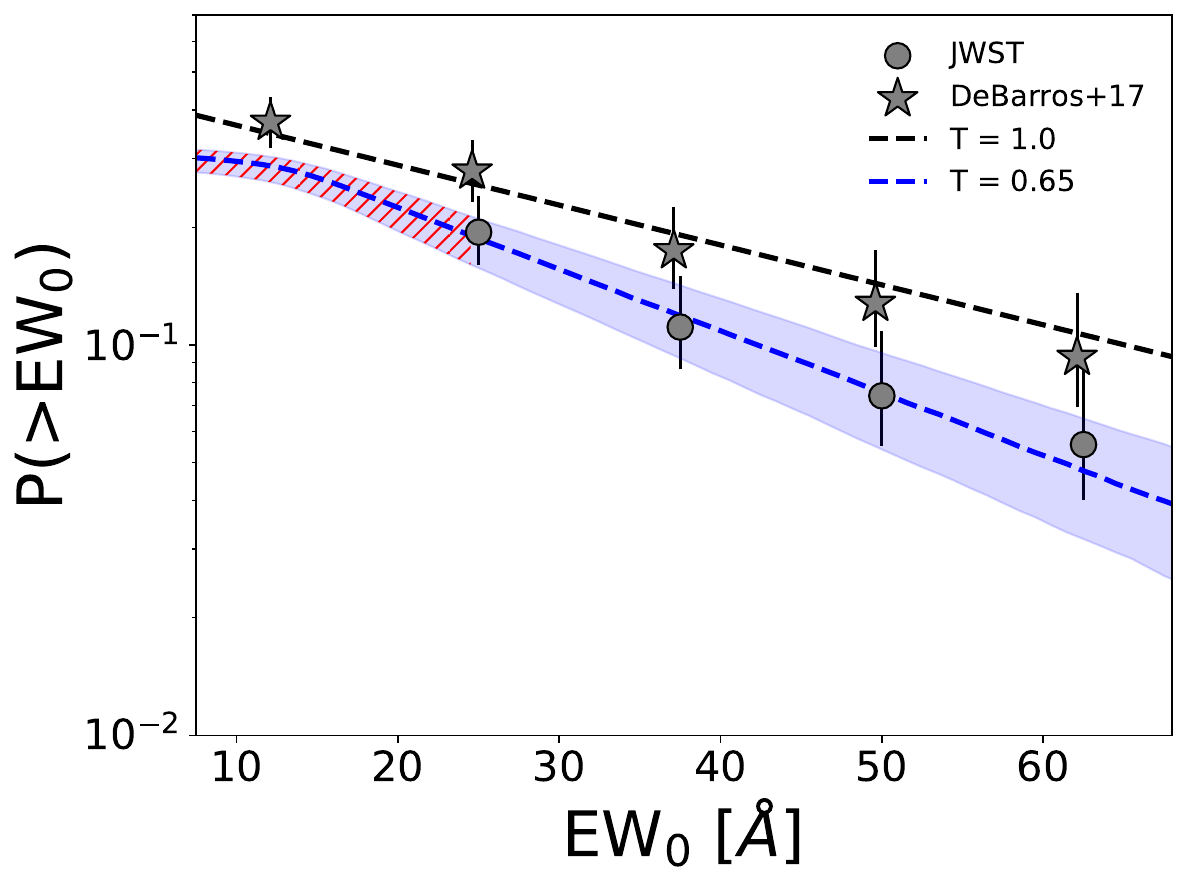}

\end{minipage}
\caption{Left: UV $\beta$ slope as a function of M$_{\mathrm{UV}}$ for galaxies in the redshift range 5 < z < 6.5 (circles), matching the selection in \cite{DeBarros2017} (stars). The continuous black lines at M$_{\mathrm{UV}}$ = -21 and –18.75 mark the range considered. Galaxies with \lya\ EW$_0$ > 25~\AA\ are shown in blue, while the rest of the population is shown in red. The best-fit relation and 1$\sigma$ uncertainty at z $\sim$ 6 from \cite{Topping2024c} and \cite{Dottorini2025} are overplotted in black and green, respectively. Right: Cumulative distribution functions of \lya\ EW$_0$ at z $\sim$ 6. The DB17 and \jwst\ \lya-break samples are shown as stars and circles, respectively. Error bars were computed following \cite{Gehrels1986}. The dashed black line shows the best-fit exponentially declining function to the DB17 data, corresponding to a case with no \lya\ slit loss (T = 1). The dashed blue line and shaded blue region represent the \jwst\ extracted CDF assuming T = 0.65 $\pm$ 0.10. The dashed red area indicates the EW$_0$ range where a direct comparison is not possible due to the \jwst\ completeness limit.}
\label{fig:CDF_slitloss}
\end{figure*}

At z $\sim$ 6 the cosmic reionization process is nearly complete, making the selected \lya-break populations a valuable reference point for the early post reionization Universe. 
In the previous section, we showed that \jwst\ observations consistently give a fraction of \lya\ emitters around 21\% at z = 6, with very little field-to-field variation as expected in the (almost completely) reionized Universe.
However, we note that ground-based results provide a substantially larger number of \lya\ emitters \citep{DeBarros2017, Schenker2012} at the same redshift and in the same M$_{\mathrm{UV}}$ range, a result further supported by ground-based narrow-band selections \citep[e.g.,][]{Arrabal_Haro2018}. 

Any statistically significant differences in the \lya\ properties between two sets of \lya-break samples at z $\sim$ 6 including data from multiple fields that limits the effect of cosmic variance are likely driven by observational biases, rather than intrinsic astrophysical effects. To search for such biases, we compared \lya\ emission from \jwst\ spectra (including \cite{Jones2024}, \cite{Napolitano2024}, and this work, i.e., data from four fields) to the population of star-forming galaxies observed through ground-based spectroscopy in \cite{DeBarros2017} (hereafter DB17) at z $\sim$ 6. DB17 provides a comprehensive catalog of galaxies at 5.1 < z < 6.6, including sky coordinates, redshift, M$_{\mathrm{UV}}$, UV $\beta$ slope, and \lya\ EW$_0$, regardless of the presence of \lya\ emission. This makes it a robust complete sample for \lya-break selected galaxies. 
To enable a fair comparison between the \lya\ statistics of the two datasets, but still keep significant numbers of sources, we restrict our analysis to galaxies in the ranges -21 < M$_{\mathrm{UV}}$ < -18.75 and 5.5 < z < 6.5. Galaxies within these ranges are well represented in both the JWST and ground-based samples. In particular, at z $\sim$ 6 the JWST sample extends significantly fainter than M$_{\mathrm{UV}}$ = -18.75 (see Fig.\ref{fig:UVcompare}, left), ensuring completeness across the adopted magnitude range. Overall, the sample includes 108 \jwst\ galaxies and 86 galaxies from DB17. We note that the DB17 sample combines data from five separate fields. \\
A possible cause of the diversity between the two samples could be due to the selection bands. The ground-based sample from DB17 mostly comes from the CANDELSz7 program, based on the H band selection by \cite{Pentericci2018} in CANDELS \citep{Grogin2011,Koekemoer2011}, with additional sources also coming from other programs selected in the Y band centered at 1.05~$\mu$m. This corresponds to selecting z $\sim$ 6 galaxies based on their rest-frame UV emission (1500--2300~\AA). In contrast, as discussed in Sect.~\ref{sec:Data_and_sample_selection}, \jwst\ targets were selected using a combination of the \NIRCam\ F277W, F356W, and F444W bands, which trace the rest-frame optical (4000--6300~\AA). As a result, a potential selection bias could be present, with the ground-based sample favoring bluer galaxies in the rest-frame UV, while the \jwst\ sample may preferentially include galaxies with higher stellar masses due to the use of rest-frame optical selection bands. Since broad correlations between \lya\ and the spectral slope and stellar masses are known to exist \citep[e.g.,][]{Schenker2014, Napolitano2023}, the different selection would then cause a different prevalence of strong \lya\ emitters. For the M$_{\mathrm{UV}}$ matched ground based and \jwst\ samples, we performed a Kolmogorov-Smirnov test on the distributions of UV $\beta$ slopes and stellar masses. In both cases, we find no statistically significant evidence that the \jwst\ and ground-based samples are drawn from different parent distributions. A direct comparison of the UV $\beta$ slope is shown in the left panel of Fig.~\ref{fig:CDF_slitloss}.\\
Another potential difference between the ground-based and \jwst\ samples is that the redshift identification of the former was based on the presence of the \lya\ line and the \lya-break, with no other spectral feature. In the absence of detectable \lya\ emission or break, the photometric redshift was assumed to be the correct one. We verified the accuracy of the redshifts reported in DB17 by cross-matching their catalog with available \jwst\ spectra from the DJA archive. 
This step is essential, particularly for low-S/N and unconfirmed sources whose redshift is based on the photometric estimate. Among the 35 DB17 galaxies with available \jwst\ spectra, we confirm the ground-based redshifts within the quoted uncertainties in 32 cases: 20 were associated with \lya\ emitters, 4 were confirmed by DB17 as \lya-break galaxies, and 8 had only a photometric redshift in DB17 that is now confirmed spectroscopically. In three further cases, in which DB17 used  the photometric redshift, the \jwst\ spectroscopic redshift differs significantly from the previously assumed value, revealing a foreground (z = 0.7) interloper and two background (z = 7.2, 8.4) interlopers. For the remaining DB17 galaxies, no \jwst\ spectra are currently available. Given the small number of interlopers, we conclude that the ground-based sample was accurate in tracing the z $\sim$ 6 population. Note that in \cite{Pentericci_2018b} the interloper fraction was assumed to be 10\%, which is very close to our findings. \\
Finally, we note that AGN identification in ground-based samples relied solely on cross-matching with available X-ray catalogs. Prior to \jwst, the peculiar high AGN fraction at high redshifts \citep[10--20\%, e.g.,][]{Scholtz2023, Mazzolari2024b, Juodzbalis2025} was not known. To our knowledge, AGN contamination was only marginally considered in previous analyses. 
In any case, to ensure consistency in the sample selection, for the only purpose of this comparison  of \lya\ emission at z $\sim$ 6, we include confirmed \jwst\ AGNs. 

Given that we do not find any evidence of bias in the sample selection, or inaccuracies in the ground-based redshift determination, the significant differences in the \lya\ properties between the \jwst\ and DB17 samples at z $\sim$ 6 must therefore be driven by the observational \lya\ slit loss \citep[e.g.,][]{Melinder2023, Nakane2024, Tang2024} that is caused by differences in the slit widths of ground-based ($\sim$ 0.7--1 arcsec) and \jwst\ \citep[0.2 arcsec,][]{Jakobsen2022} spectrographs, combined with the resonant nature of \lya\ emission. Specifically, \lya\ photons can scatter to spatial scales larger than the stellar continuum, producing emission that is offset by $\sim$ 0.1 arcsec \citep{Ning2024, Bhagwat2025} or extended over larger regions \citep[e.g.,][]{leclercq2017, kusakabe2022}.\\
To independently estimate the \jwst\ \lya\ slit loss, we compared the cumulative distribution functions (CDFs) of \lya\ EW$_0$ between the DB17 and the combined \jwst\ sample from \cite{Jones2024}, \cite{Napolitano2024}, and this work.
 
The DB17 CDF was assumed as a reference to the case with no significant slit loss, where the transmission factor (T) equals unity.
We modeled the DB17 CDF with an exponentially declining function, fit by weighting the measured uncertainties. We derived the associated normalized probability density function (PDF), which was then used to generate 1000 \lya\ EW$_0$ draws for each galaxy in the \jwst\ sample.
To incorporate the \jwst\ \lya\ detection completeness (see Fig.~\ref{fig:UVcompare}), we accounted for the EW$_0$ sensitivity limit of each galaxy. Based on the observed M$_{\mathrm{UV}}$, we assigned non-detections (EW$_0$ = 0) to all randomly drawn values falling below the threshold EW$_{0,\mathrm{lim}}^{\mathrm{fit}}$. This threshold was determined via a linear fit to the upper limits observed in the \jwst\ sample, and is typically around 10--20~\AA\ at z $\sim$ 6.\\
The resulting EW$_0$ distribution thus represents a simulated population consistent with \jwst\ selection criteria, but unaffected by \lya\ slit losses. We then applied varying transmission factors to this simulated EW$_0$ sample and compared the resulting CDFs to the observed \jwst\ distribution. We found the best match when T = 0.65 $\pm$ 0.10, implying an average \lya\ slit loss of 35 $\pm$ 10\% when comparing \jwst-\NIRSpec\ and ground-based VLT-FORS2 EW$_0$ measurements (see Fig.~\ref{fig:CDF_slitloss}). This estimate is broadly consistent with recent results at z $\sim$ 6. For instance, \cite{Nakane2024} report \lya\ slit losses of 28 $\pm$ 8\% from a forward-modeling analysis comparing \jwst\ and VLT-FORS2, assuming different scale lengths for the UV and \lya\ spatial profiles (see their Appendix B). 
Similarly, \cite{Tang2024} found $\sim$ 20\% \lya\ slit loss when comparing \jwst-\NIRSpec\ and ground-based VLT-MUSE EW$_0$ values. Finally, \cite{Bhagwat2025} used the SPICE radiation-hydrodynamical simulation \citep{Bhagwat2024} to investigate the \lya\ \jwst–FORS2 mismatch. Selecting galaxies in the range of -21 < M$_{\mathrm{UV}}$ < -18.75, with no spatial offset between the UV and \lya\ emission, they found \lya\ slit losses of 26--37\%, depending on the stellar feedback model adopted. The best agreement with our observational result was found for their \textsc{bursty-sn} model, which assumes that all supernovae explode after 10 Myr, releasing 2 $\times$ 10$^{51}$ erg in a single event. However, we note that, within the uncertainty, our observational constraint remains consistent with all their supernova feedback models.

In the following section, we apply the derived \lya\ slit loss correction to the z = 7 UDS \lya-break galaxy population observed with \jwst, assuming that the correction does not evolve with redshift. This assumption is supported by the observed lack of evolution in the spatial offset between UV and \lya\ emission from z = 4 to z = 7 \citep[e.g.,][]{Hoag2019, Lemaux2021}.

\subsection{The evolution of the neutral hydrogen fraction at z $\sim$ 7} \label{sec:XHI}
To derive the evolution of \lya\ transmission in the IGM between z = 6 and z = 7, we adopted the DB17 sample at z = 6 as a complete representation of the \lya-break population in a post-reionization Universe. At this epoch, the IGM is already highly ionized, with a volume-averaged neutral hydrogen fraction of $X_{\mathrm{HI}} \sim 5 \times 10^{-3}$, and is therefore assumed to be effectively transparent to \lya\ photons \citep[see Fig. 2 in][]{Ellis2025}. 
We selected the parent \lya-break subsample in the range of -20.25 < M$_{\mathrm{UV}}$ < -18.75 and compared the resulting EW$_0$ CDF to the \jwst\ measurements at z = 7. The UDS CDF was derived from the EW$_0$ values corrected for the observed \lya\ slit loss derived in the previous section (Sect.~\ref{sec:slitloss}) after selecting the sample in the same M$_{\mathrm{UV}}$ range, further requiring no AGN activity and EW$_{0,\mathrm{lim}}$ < 25~\AA. \\
The impact of a neutral IGM on \lya\ visibility was derived by considering the IGM transmission models of \cite{Dijkstra2011}, which combine galactic outflow effects with large-scale semi-numeric simulations of reionization. These models assume N$_{\mathrm{HI}}$ = 10$^{20}$ cm$^{-2}$, outflow velocities of 25--200 km/s, no dust absorption, and an intrinsic exponentially declining CDF at z = 6 with an e-folding scale of 50~\AA. This value is in excellent agreement with our best-fit to the DB17 CDF (50 $\pm$ 10~\AA). 

\begin{figure}[t]
\centering
\includegraphics[width=\linewidth]{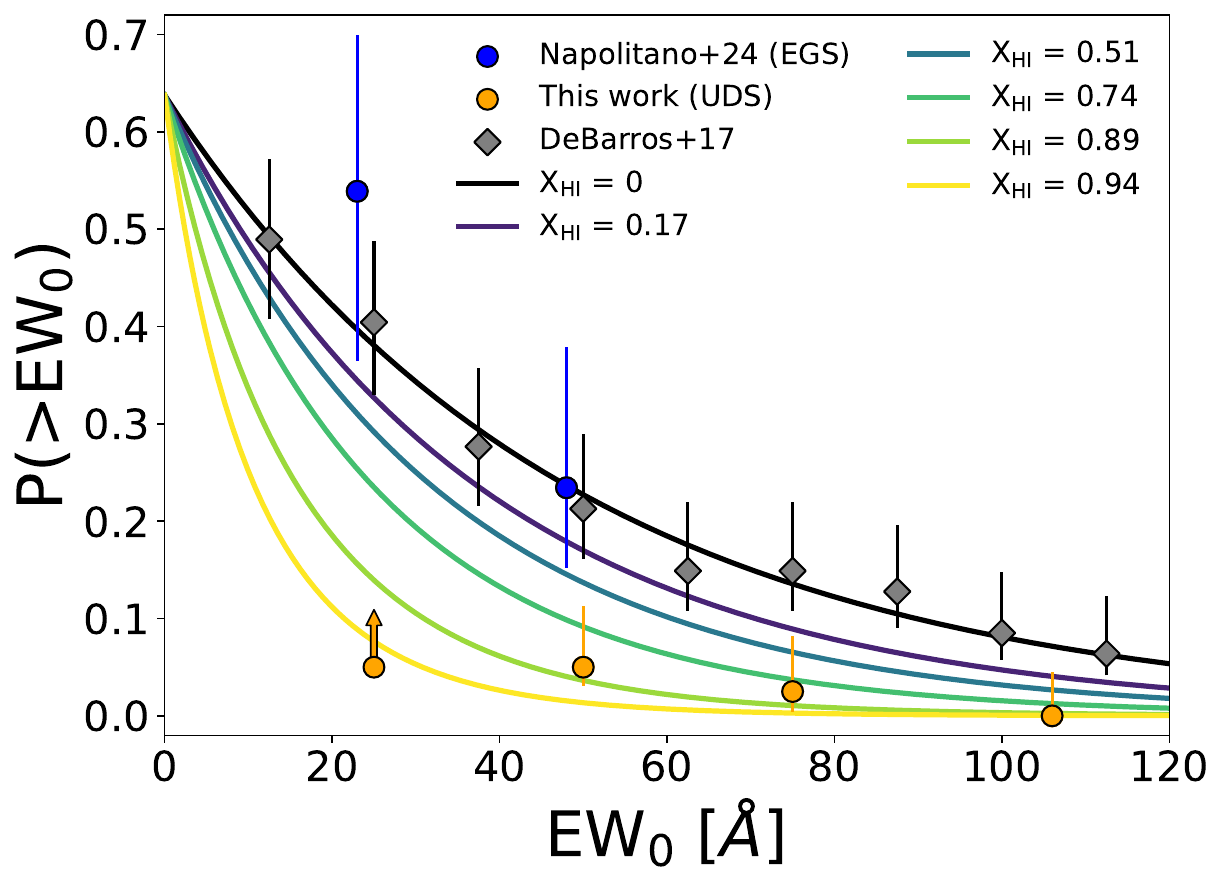}
\caption{Cumulative distribution functions of \lya\ EW$_0$ as a function of the neutral hydrogen fraction, X$_{\mathrm{HI}}$. Blue and orange circles represent the \jwst\ samples at z = 7 from the EGS \citep{Napolitano2024} and the UDS (this work) fields, respectively. The arrow represents the lower limit we could derive for EW$_0$ = 25~\AA\, due to incompleteness after \lya\ slit loss correction. The DB17 sample at z = 6 (gray diamonds) is shown as a reference for the fully ionized Universe. Error bars were computed following \cite{Gehrels1986}. Data points have been slightly shifted in redshift for an easier visualization. The solid black line shows the best-fit exponentially declining function to the DB17 data, corresponding to a null X$_{\mathrm{HI}}$ value. Colored lines represent theoretical models derived from \cite{Dijkstra2011}, which show the impact of an increasingly neutral IGM.
} \label{fig:xHI}
\end{figure}

We applied the modeled IGM transmission as a function of the neutral hydrogen fraction to the DB17 CDF and compared the resulting curves to the observed \jwst\ CDFs in the UDS and EGS fields (Fig.~\ref{fig:xHI}). The UDS sample is consistent with a significantly neutral IGM, with  X$_{\mathrm{HI}}$ $\sim$ 0.7--0.9. We note that the \lya\ slit loss correction shifts the completeness limit of the CDF from the observed $\sim$ 20~\AA\ to $\sim$ 30~\AA; therefore, we report only lower limits below this threshold. The EGS measurements are instead consistent with a fully ionized IGM. In practice, the z = 7 EGS CDF is fully consistent with the z = 6 CDF measured by DB17. 
This highlights substantial field-to-field variation in the \lya\ transmission at z = 7, supporting a  highly inhomogeneous reionization scenario and underscoring the importance of estimating X$_{\mathrm{HI}}$ across multiple independent fields. 

For comparison, recent \jwst\ studies that considered multiple fields have independently reported high neutral fractions of $\sim$ 0.5--0.8 \citep[e.g.,][]{Nakane2024, Tang2024B, Jones2025, Kageura2025, Umeda2025} in agreement with results from z $\sim$ 7 quasars \citep[e.g.,][]{Durovcikova2024}. Recently, \cite{Chen2025} discussed how the extreme \lya\ transmission in the EGS field could be explained by the presence of a single $\sim$ 12 pMpc ionized bubble. We emphasize that tighter constraints on X$_{\mathrm{HI}}$ require access to the low-EW$_0$ regime of the CDF, where theoretical models show the largest spread. Probing the necessary EW$_0$ values, on the order of 10--15 \AA\ after accounting for \lya\ slit losses, appears to be within reach with moderately deeper \jwst-\NIRSpec\ observations.\\

Finally, we note that assuming a non-evolving stellar population and ISM conditions across cosmic time is critical when comparing \lya\  visibility in the partially ionized Universe (z $\sim$ 7 and above) to the post-reionization Universe (z $\sim$ 6) to infer the IGM neutral hydrogen fraction. However, this assumption may be overly simplistic. For example, \cite{Ferrara2023} proposed the so called attenuation-free model in which early galaxies experience significantly reduced dust attenuation in their ISM once they exceed the Eddington limit. This model accounts for several JWST results, including the overabundance of bright blue sources at early epochs \citep[e.g.,][]{Fujimoto2023, Harikane2024, Napolitano2025a}. It also predicts that \lya\ emission from \lya-break galaxies at z > 9--10 could be intrinsically enhanced \citep[][see the evolution of the predicted super-Eddington galaxy fraction in their Figure 3]{Ferrara2024}.
The apparent lack of evolution in the \lya\ visibility at z = 8 and above, as is shown in Fig.~\ref{fig:XLya} (despite the large uncertainties; see Table \ref{tab:Xlya}), may suggest that specific high-redshift galaxy evolution effects, such as those described in \cite{Ferrara2024}, begin to play a critical role at these epochs. These effects will be investigated in detail using hydrodynamical simulations of reionization in a future work.

\subsection{Ionized bubbles in the UDS field} \label{sec:bubble}

\begin{figure}[t]
\centering
\includegraphics[width=\linewidth]{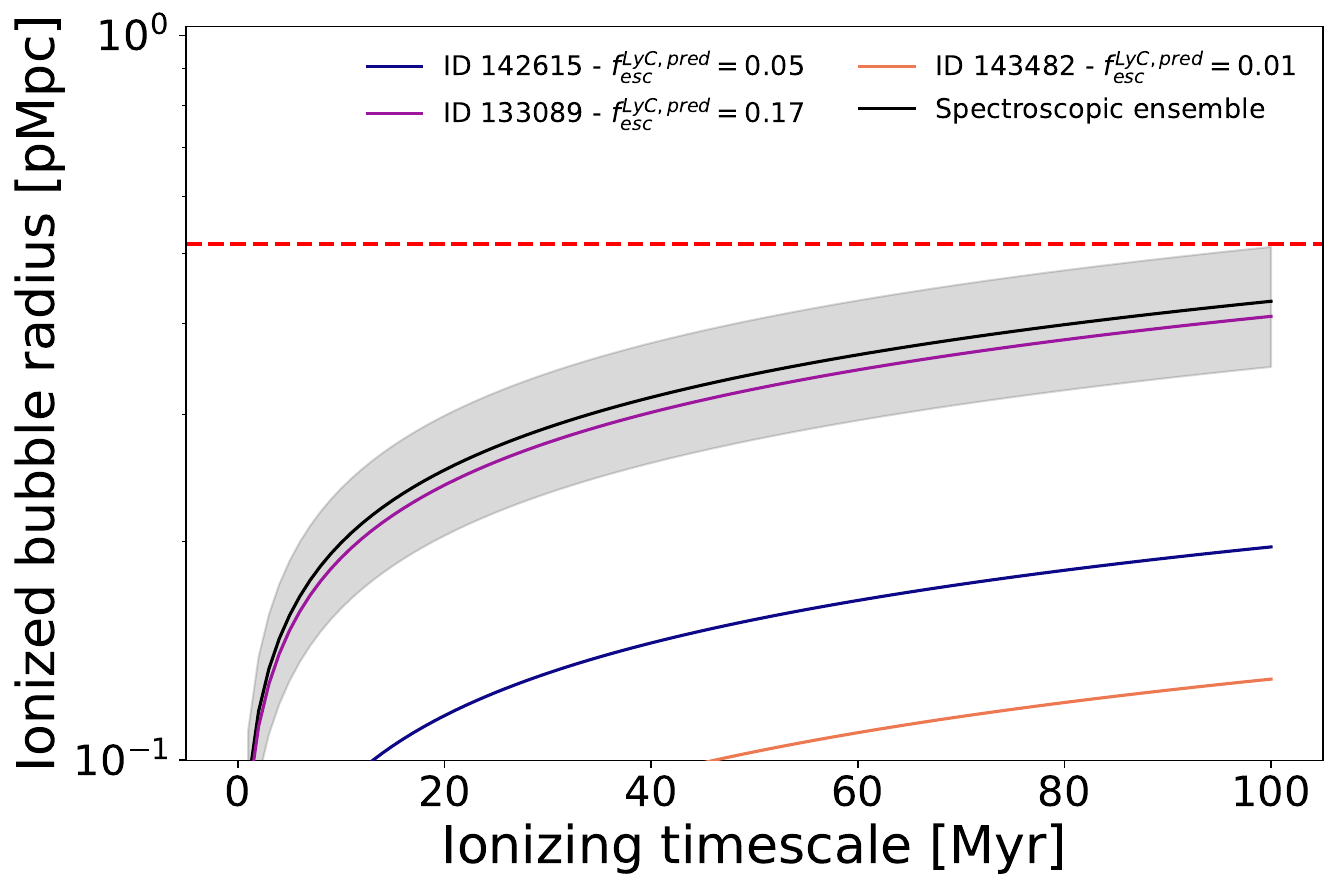}
\caption{Predicted size of the ionized bubble at z = 7.77 as a function of time since ionizing radiation is switched on. Colored solid lines show the contribution from individual sources, while the black solid line and shaded region represent the cumulative predicted radius and its associated uncertainty. The horizontal red dashed line marks the maximum physical distance between the central \lya\ emitting source and its furthest companion.} \label{fig:Rion_z7p7}
\end{figure}

In a patchy reionization scenario, \lya\ photons emitted by a galaxy surrounded by a highly neutral IGM, as described in the previous section, are expected to undergo significant absorption, unless the galaxy resides within a locally ionized region with a radius of $\sim$ 0.1--3 pMpc \citep{Mason2020}. To investigate the galaxy overdensity conditions that enable \lya\ transmission through the IGM, we searched for ionized bubbles at z > 7. In this analysis, we considered all galaxies with detected \lya\ emission reported in Table \ref{tab:summary_data}, without applying additional selection criteria based on AGN classification, M$_{\mathrm{UV}}$, or EW$_{0,\mathrm{lim}}$. We assumed that each \lya-emitting galaxy lies at the center of its own ionized bubble and implemented an iterative approach to identify the most probable set of spectroscopic companions, defining the ionized region. We began by identifying all spectroscopic galaxies within a 2 pMpc sphere \citep[e.g.,][]{Lu2024, Runnholm2025} around the central \lya-emitting galaxy, using the distance prescription from \cite{Liske2000}. The physical distance, $R_{\mathrm{phys}}$, to the most distant companion was taken as an upper limit to the bubble size. We then compared $R_{\mathrm{phys}}$ to the ionized bubble radius, $R_{\mathrm{ion}}$, produced after 100 Myrs by the combined ionizing output of the central galaxy and its companions.
At each iteration, the furthest companion from the central \lya\ emitter was excluded and the procedure was repeated, reassessing $R_{\mathrm{phys}}$ and $R_{\mathrm{ion}}$. We determined the best ionizing bubble configuration as the one that minimizes the quantity $|R_{\mathrm{ion}} - R_{\mathrm{phys}}| / \sigma$, where $\sigma$ is the uncertainty of $R_{\mathrm{ion}}$ (see below).
\begin{figure}[t]
\centering
\includegraphics[width=\linewidth]{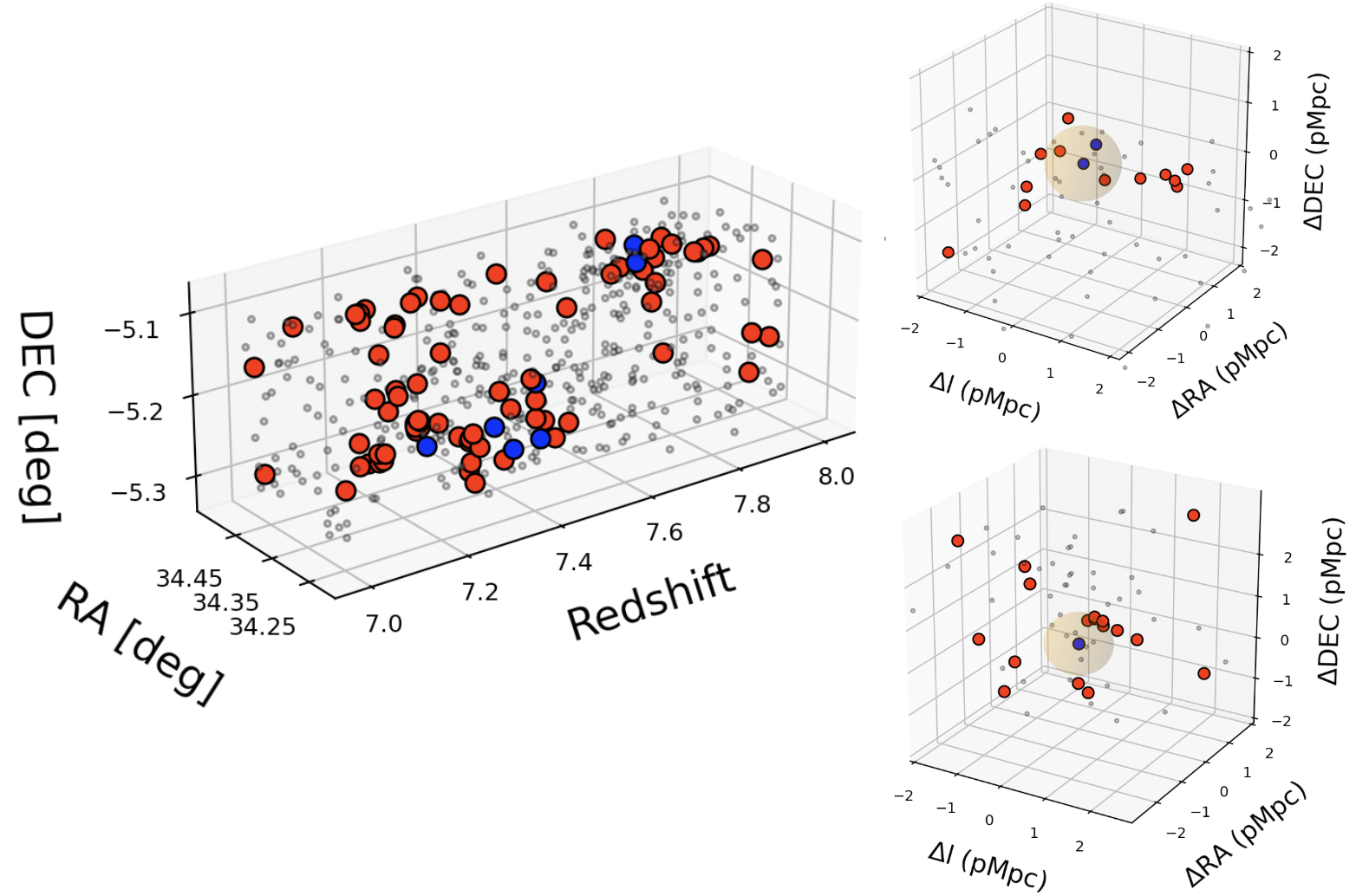}
\caption{Spatial distribution of spectroscopically confirmed galaxies in the UDS field at 7 < z < 8. Blue and red circles indicate \lya-emitting and \lya-break galaxies, respectively. Gray circles mark photometric candidates from the \cite{Merlin2024} catalog. The two robust, spectroscopically confirmed ionized bubbles in the UDS field at z = 7.29 (lower panel) and z = 7.77 (upper panel) are shown as zoom-ins. The extent of each ionized region is represented by a shaded sphere.} \label{fig:Bubbles}
\end{figure}
To estimate $R_{\mathrm{ion}}$, we followed the prescription from \cite{Shapiro-giroux1987} and \cite{Cen2000}, whereby the bubble size is determined by the total number of emitted ionizing photons, neglecting the Hubble expansion and recombination terms, whose timescales are longer than star-formation episodes at z $\sim$ 7 \citep[see][]{Witstok2024}. The ionizing emissivity of each central source and its spectroscopically confirmed companions was computed using Equation 9 of \cite{Mason2020}, based on the observed M$_{\mathrm{UV}}$ and UV $\beta$ slope:
\begin{equation*} \label{eq:Nion}
    \dot{N}_{\mathrm{ion}} = \frac{3.3 \times 10^{54}}{\alpha} 10^{-0.4 (M_{\mathrm{UV}}+20)} \left( \frac{912}{1500} \right)^{\beta +2} \text{s}^{-1},
\end{equation*}
where we adopted a spectral slope $\alpha = 1$ for the ionizing continuum, consistent with measurements of high-redshift galaxies \citep[e.g.,][]{Steidel2014, Feltre2016}. The predicted escape fraction of Lyman continuum photons ($f_{\mathrm{esc}}^{\mathrm{LyC, pred}}$) was estimated using the empirical relation from \cite{chisholm2022} and was assumed to remain constant over time. For the z > 7 UDS sub-sample, we found a median predicted escape fraction of 3\%, consistent with the median 1--3\% range reported by \cite{Papovich2025}. While different methods \citep[e.g.,][]{Mascia2023_GLASS, Jaskot2024, Giovinazzo2025} yield a range of predictions for $f_{\mathrm{esc}}^{\mathrm{LyC}}$, as is noted by \cite{Witstok2025}, the inferred ionized bubble size depends weakly on this parameter. Specifically, since $R_{\mathrm{ion}} \propto (f_{\mathrm{esc}}^{\mathrm{LyC, pred}})^{1/3}$, even an order-of-magnitude variation in the assumed escape fraction would affect the derived radius by a factor of $\sim$ 2. 

We identify two robust (|$R_{\mathrm{ion}}$-$R_{\mathrm{phys}}$| $\leq$ 1$\sigma$) ionized bubbles centered on RUBIES-22104 (z = 7.29, including six galaxies in total) and CAPERS-142615 (z = 7.77, including three galaxies in total) with corresponding $R_{\mathrm{ion}}$ = 0.6 and 0.5 pMpc at 100 Myr, respectively. These values are obviously  lower limits, as additional unconfirmed sources, as well as those below the detection limit, also contribute to the ionizing photon budget, further expanding the ionized structures. Figure~\ref{fig:Rion_z7p7} shows the bubble growth as a function of time for the z = 7.77 bubble, which contains two \lya\ emitting galaxies (CAPERS-142615 and CAPERS-133089).\\
We estimated the ionizing contribution of photometric candidates using the ASTRODEEP-JWST catalog \citep{Merlin2024}.
To assess the presence of overdensities associated with the ionized bubbles, we evaluated the number of photometric candidates within a cylindrical volume of the Universe spanning [z$_{\mathrm{bubble}}$-0.5, z$_{\mathrm{bubble}}$+0.5]. We compared this result with the expected average number of galaxies based on the UVLF from \cite{Bouwens2015}, integrated down to the observed completeness limit of M$_{\mathrm{UV}}$ = -18.5 in \cite{Merlin2024}. We found significant overdensities associated with the two bubbles at z = 7.29 and 7.77, with N/$\langle$N$\rangle$ = 3 and 4, respectively. Figure~\ref{fig:Bubbles} illustrates the positions of spectroscopically confirmed galaxies, photometric candidates, and zoom-ins of the robust ionized regions found at 7 < z < 8 in the UDS field.\\
No other robust ionized bubbles were identified with our fiducial method around the remaining high-redshift \lya\ emitting galaxies. However, if we relax the agreement criterion between $R_{\mathrm{ion}}$ and $R_{\mathrm{phys}}$ down to 3$\sigma$, we find a tentative ionized bubble of 1 pMpc at z = 7.44, composed of three \lya\ emitting galaxies (RUBIES-59990, RUBIES-930869, and RUBIES-24303) and a non-emitter (RUBIES-971810). The associated N/$\langle$N$\rangle$ is only 1.3; therefore, this structure remains unconfirmed in our analysis. \\
Compared to the large EGS ionized regions at z $\sim$ 7 \citep[2--12 pMpc,][]{Chen2025}, the relatively small ionized bubbles we identify in the UDS field are expected for highly neutral hydrogen IGM environments according to simulations from \cite{Lu2024} \citep[see also,][]{Neyer2024, Neyer2025}. This picture is consistent with the result discussed in Sect.~\ref{sec:XHI}.

We note that \cite{Chen2025} recently identified overdense structures in the UDS field that are fully consistent with the spatial distribution of galaxies reported in this work. They interpret them as two larger-scale overdensities spanning z = 7.24--7.43 and z = 7.74--7.83, extending over more than 7 and 3 pMpc in radial distance, respectively. Their findings support a consistent picture in which overdense regions are more likely to host strong \lya-emitting galaxies. 
The second part of the CAPERS-UDS spectroscopic campaign, scheduled for winter 2025--2026, will significantly improve constraints by expanding the number of confirmed galaxies. This will help distinguish between the two scenarios: isolated, small ionized bubbles or extended, interconnected large-scale overdensities. A refined empirical method of mapping ionized bubbles using \lya\ transmission has recently been suggested by \cite{Lu2025} and \cite{Nikolic2025}. It requires a spectroscopic identification of > 0.004 galaxies/cMpc$^3$, a 5$\sigma$ upper limit of $\sim$ 30~\AA\ for the \lya\ EW$_0$ of the \lya-break population, and a redshift precision of $\Delta$z < 0.015. Among these, the most challenging requirement is the depth, i.e., reaching an EW$_{0,\mathrm{lim}} of \sim$ 6~\AA\ that is far below current constraints.

\section{Summary} \label{sec:Conclusion}
We have presented the results of a systematic study of the evolution of \lya\ emission from galaxies at 4.5 $\leq$ z $\leq$ 11 in the UDS field. The sample consists of 651 galaxies: 135 are from the CAPERS survey and 516 from other spectroscopic \jwst\ programs whose data are included in the public DAWN JWST Archive. Among the total sample, 531 spectra were obtained from \NIRSpec-PRISM configuration, and 120 with medium-resolution configurations. Each source is associated with a secure spectroscopic redshift identified through multiple optical and UV line detections. 
We flagged broad-line AGN sources by fitting significant Balmer emission lines and using the Bayesian information criterion. Among the 485 galaxies with available \lya-break information, we identified 73 secure (S/N > 3) \lya\ emitting galaxies. Their spectroscopic information is reported in Table \ref{tab:summary_data}. We summarize our main results as follows:
\begin{itemize}
\item We traced the evolution of the number of galaxies with  \lya\ emission with EW$_0$ > 25~\AA\  (X$_{\mathrm{Ly\alpha}}$) across five redshift bins centered at z = 5, 6, 7, 8, and 9. We restricted this calculation to the 216 galaxies with -20.25 < M$_{\mathrm{UV}}$ < -18.75, EW$_{0,\mathrm{lim}}$ < 25~\AA, and no AGN signatures. In the UDS, between z = 5--6 (6--7), the increase (decrease) in X$_{\mathrm{Ly\alpha}}$ is significant at a level of 1$\sigma$ (2$\sigma$). At z = 5 and z = 6, the average \jwst\ results from multiple fields agree very well with those derived in the UDS field alone, as well as the EGS field alone from \cite{Napolitano2024}. This was expected since in the post-reionization Universe the visibility of the \lya\ line in galaxies is only driven by their physical properties and not by IGM conditions.
On the other hand, at z = 7, field-to-field variations become significant and the difference is exacerbated when comparing the EGS and UDS results individually.
\item A tension remains between the average \jwst\ and ground-based X$_{\mathrm{Ly\alpha}}$ results at z = 5--6. Comparing to the VLT-FORS2 ground-based sample from \cite{DeBarros2017} (DB17), we find no evidence for any bias in sample selection. Instead we suggest that an average \lya\ slit loss of $\sim$ 35 $\pm$ 10\% in \jwst\ observations can account for the discrepancy. 
This is likely due to an instrumental bias introduced when comparing \jwst\ pseudo-slit spectra (0.2 arcsec $\times$ 0.46 arcsec) with ground-based data obtained using 1 arcsec slit widths. The results are in agreement with the radiation-hydrodynamical simulation by \cite{Bhagwat2025}, when selecting galaxies with no spatial offset between the UV and \lya\ emission.
\item We adopted DB17 as a reference sample for the \lya-break population in an almost fully-ionized Universe at z = 6, and compared it to the z = 7 CDF of \lya\ EW$_0$ from \jwst. We infer a high neutral hydrogen fraction in the UDS ($X_{\mathrm{HI}}$ $\sim$ 0.7--0.9), while EGS remains compatible with a nearly completely ionized IGM ($X_{\mathrm{HI}}$ $\sim$ 0). This extreme field-to-field variation indicates a highly spatially inhomogeneous reionization process.
\item We identify two ionized bubbles in the UDS field centered on RUBIES-22104 (z = 7.29) and CAPERS-142615 (z = 7.77), with radii of 0.6 and 0.5 pMpc, respectively. These structures show significant photometric overdensities (N/$\langle$N$\rangle$ = 3 and 4) compared to UVLF expectations. Their smaller sizes, relative to the $\sim$ 2--12 pMpc bubble recently identified in the EGS at z $\sim$ 7, further support the very high X$_{\mathrm{HI}}$ in the UDS field. 
\end{itemize}

A systematic census of \lya\ emission from spectroscopically confirmed \lya-break galaxies at high redshift across \jwst\ fields is currently missing. In particular, the spatially inhomogeneous nature of cosmic reionization makes it difficult to converge on a unique reference value for the neutral hydrogen fraction at z $\sim$ 7. Additionally, the identified instrumental bias between ground-based and \jwst-based \lya\ measurements introduces further uncertainty when combining results. 
A critical next step is to compare the current \jwst\ \NIRSpec\ MSA spectroscopic data with a statistical sample of \lya-emitting galaxies observed using \jwst\ \NIRSpec-IFU. Such a comparison would enable the construction of a completely independent, space-based reference sample in the post-reionization Universe, unbiased by \lya\ slit losses, which is essential for robustly constraining X$_{\mathrm{HI}}$ at high redshifts.

Further progress in characterizing the epoch of reionization will also require deeper or higher-resolution spectra of the \lya-break population to reach a higher completeness in the space-based CDF of \lya\ EW$_0$, down to 10~\AA\ in EW$_{0,\mathrm{lim}}$. As is shown in Fig.~\ref{fig:xHI}, the intermediate \lya\ EW$_0$ range (10--40~\AA) is where model predictions for X$_{\mathrm{HI}}$ diverge the most, making it the regime with the greatest potential for observational constraints.

\begin{acknowledgements}
We thank the anonymous referee for the constructive feedback provided. LN acknowledges support from grant ``Progetti per Avvio alla Ricerca - Tipo 1, Unveiling Cosmic Dawn: Galaxy Evolution with CAPERS" (AR1241906F947685) and from PRIN 2022 MUR project 2022CB3PJ3 - First Light And Galaxy aSsembly (FLAGS) funded by the European Union – Next Generation EU. LN, LP, AB, and BC acknowledge support from the ERC synergy grant 101166930 - RECAP.\\ 
This work is based on observations made with the NASA/ESA/CSA James Webb Space Telescope, obtained at the Space Telescope Science Institute, which is operated by the Association of Universities for Research in Astronomy, Incorporated, under NASA contract NAS5-03127. These observations are associated with programs \#6368, \#4233, \#1215, \#2565, and \#3543.
Support for program number GO-6368 was provided through a grant from the STScI under NASA contract NAS5-03127. The data were obtained from the Mikulski Archive for Space Telescopes (MAST) at the Space Telescope Science Institute. These observations can be accessed via \href{http://dx.doi.org/10.17909/0q3p-sp24}{DOI}.

Some of the data products presented in this work were retrieved from the Dawn JWST Archive (DJA). DJA is an initiative of the Cosmic Dawn Center (DAWN), which is funded by the Danish National Research Foundation under grant DNRF140.
\end{acknowledgements}

\bibliographystyle{aa}
\bibliography{biblio.bib}

\begin{appendix}
\onecolumn
\section{Spectroscopic properties and \lya\ line profiles of UDS \lya\ emitting galaxies}\label{sec:app_figures}
In this appendix, we present in Table \ref{tab:summary_data} the measured properties of the 73 \lya\ emitting galaxies at z > 4.5 identified in the UDS sample. Their fit \lya\ line profiles are shown in Fig.~\ref{fig:Lya_emitters}. We report galaxies in order of descending redshift. 

\small
\begin{longtable}{lcccccc}
\caption{\label{tab:summary_data}Spectroscopic properties of the S/N$>$3 \lya\ emitting galaxies in the UDS sample.}\\
\hline\hline
ID & RA [deg] & DEC [deg] & $z_{\mathrm{spec}}$ & EW$_0$ [\AA] & M$_{\mathrm{UV}}$ [mag] & $\beta$ \\
\hline
\endfirsthead

\caption{continued.}\\
\hline\hline
ID & RA [deg] & DEC [deg] & $z_{\mathrm{spec}}$ & EW$_0$ [\AA] & M$_{\mathrm{UV}}$ [mag] & $\beta$ \\
\hline
\endhead

\hline
\endfoot

RUBIES-23219$^{\star}$ & 34.33032 & -5.28124 & 8.5596 $\pm$ 0.0025 & 94 $\pm$ 27 & -18.90 $\pm$ 0.43 & -2.0 $\pm$ 0.9 \\ 
RUBIES-927815 & 34.23072 & -5.27243 & 8.5425 $\pm$ 0.0022 & 59 $\pm$ 13 & -19.60 $\pm$ 0.04 & -2.12 $\pm$ 0.26 \\
RUBIES-53873 & 34.45964 & -5.23151 & 8.2583 $\pm$ 0.0027 & 69 $\pm$ 19 & -19.21 $\pm$ 0.06 & -1.90 $\pm$ 0.29 \\
CAPERS-133089$^{\dag}$ & 34.42481 & -5.11393 & 7.772 $\pm$ 0.005 & 36 $\pm$ 10 & -19.10 $\pm$ 0.04 & -2.53 $\pm$ 0.22 \\
CAPERS-142615$^{\dag +}$ & 34.41388 & -5.13355 & 7.7678 $\pm$ 0.0039 & 209 $\pm$ 20 & -18.31 $\pm$ 0.06 & -2.05 $\pm$ 0.28 \\
RUBIES-59990 & 34.31000 & -5.20920 & 7.4538 $\pm$ 0.0013 & 33 $\pm$ 10 & -20.45 $\pm$ 0.04 & -1.89 $\pm$ 0.23 \\
RUBIES-930869$^{+}$ & 34.28053 & -5.26837 & 7.439 $\pm$ 0.015 & 39 $\pm$ 7 & -20.52 $\pm$ 0.03 & -2.53 $\pm$ 0.20 \\
RUBIES-29954 & 34.36448 & -5.27026 & 7.4092 $\pm$ 0.0036 & 20.2 $\pm$ 3.5 & -20.73 $\pm$ 0.01 & -1.79 $\pm$ 0.14 \\
RUBIES-24303$^{+ \star}$ & 34.30025 & -5.27933 & 7.396 $\pm$ 0.005 & 330 $\pm$ 46 & -18.72 $\pm$ 0.36 & -2.0 $\pm$ 1.2 \\ 
RUBIES-22104$^{\dag}$ & 34.39805 & -5.28321 & 7.2887 $\pm$ 0.0008 & 33 $\pm$ 10 & -19.83 $\pm$ 0.03 & -2.42 $\pm$ 0.21 \\
RUBIES-3545 & 34.37429 & -5.31694 & 6.9066 $\pm$ 0.0031 & 31 $\pm$ 9 & -20.52 $\pm$ 0.03 & -2.17 $\pm$ 0.19 \\
RUBIES-40329 & 34.38866 & -5.25353 & 6.8476 $\pm$ 0.0042 & 69 $\pm$ 12 & -19.37 $\pm$ 0.03 & -3.13 $\pm$ 0.22 \\
CAPERS-147066 & 34.42724 & -5.13864 & 6.814 $\pm$ 0.007 & 236 $\pm$ 30 & -19.32 $\pm$ 0.12 & -2.17 $\pm$ 0.30 \\
RUBIES-39024 & 34.28039 & -5.24841 & 6.8089 $\pm$ 0.0016 & 24 $\pm$ 7 & -20.51 $\pm$ 0.02 & -1.82 $\pm$ 0.14 \\
CAPERS-151887$^{\star}$ & 34.49586 & -5.15224 & 6.6317 $\pm$ 0.0025 & 109 $\pm$ 20 & -16.9 $\pm$ 1.2 & -1.9 $\pm$ 1.1 \\ 
RUBIES-30525$^{\star}$ & 34.43508 & -5.26946 & 6.394 $\pm$ 0.005 & 157 $\pm$ 44 & -18.95 $\pm$ 0.30 & -1.9 $\pm$ 0.5  \\ 
RUBIES-159869 & 34.38714 & -5.12160 & 6.2462 $\pm$ 0.0047 & 120 $\pm$ 7 & -19.74 $\pm$ 0.02 & -1.90 $\pm$ 0.13 \\
GO1215-2182 & 34.46487 & -5.15544 & 6.2263 $\pm$ 0.0045 & 25 $\pm$ 8 & -19.98 $\pm$ 0.03 & -2.06 $\pm$ 0.17 \\
CAPERS-149441 & 34.26308 & -5.14745 & 6.2018 $\pm$ 0.0015 & 154 $\pm$ 14 & -18.72 $\pm$ 0.04 & -1.62 $\pm$ 0.24 \\
CAPERS-134200$^{\star}$ & 34.27536 & -5.11668 & 6.141 $\pm$ 0.012 & 132 $\pm$ 33 & -17.90 $\pm$ 0.45 & -2.0 $\pm$ 0.5 \\ 
RUBIES-25381 & 34.35766 & -5.27760 & 6.0794 $\pm$ 0.0027 & 29 $\pm$ 5 & -22.15 $\pm$ 0.02 & -2.35 $\pm$ 0.12 \\
RUBIES-174752 & 34.20581 & -5.10050 & 6.0427 $\pm$ 0.0024 & 41.8 $\pm$ 4.3 & -20.23 $\pm$ 0.02 & -1.86 $\pm$ 0.09 \\
CAPERS-24063 & 34.50378 & -5.19384 & 6.0279 $\pm$ 0.0035 & 32.5 $\pm$ 0.5 & -21.66 $\pm$ 0.01 & -2.40 $\pm$ 0.02 \\
RUBIES-33009 & 34.35461 & -5.26536 & 6.0268 $\pm$ 0.0018 & 51 $\pm$ 9 & -19.73 $\pm$ 0.03 & -2.12 $\pm$ 0.21 \\
GO1215-1259$^{\ddag}$ & 34.42231 & -5.25071 & 5.9672 $\pm$ 0.0029 & 207 $\pm$ 8 & -20.29 $\pm$ 0.03 & -1.64 $\pm$ 0.17 \\
RUBIES-27089 & 34.33960 & -5.27475 & 5.8648 $\pm$ 0.0031 & 146 $\pm$ 34 & -18.74 $\pm$ 0.07 & -2.28 $\pm$ 0.31 \\
RUBIES-47308 & 34.24220 & -5.23299 & 5.8123 $\pm$ 0.0049 & 43 $\pm$ 8 & -19.61 $\pm$ 0.02 & -1.53 $\pm$ 0.15 \\
RUBIES-143046 & 34.31046 & -5.14566 & 5.682 $\pm$ 0.006 & 30 $\pm$ 9 & -20.05 $\pm$ 0.03 & -1.74 $\pm$ 0.18 \\
RUBIES-19521$^{\ddag *}$ & 34.38367 & -5.28773 & 5.6744 $\pm$ 0.0027 & 172 $\pm$ 19 & -18.32 $\pm$ 0.04 & -1.23 $\pm$ 0.24 \\
RUBIES-170624 & 34.32651 & -5.10630 & 5.600 $\pm$ 0.008 & 99 $\pm$ 26 & -18.40 $\pm$ 0.06 & -1.56 $\pm$ 0.27 \\
RUBIES-172350$^{\ddag *}$ & 34.36895 & -5.10394 & 5.5845 $\pm$ 0.0032 & 121 $\pm$ 16 & -18.87 $\pm$ 0.06 & -1.32 $\pm$ 0.28 \\
GO2565-15294 & 34.31094 & -5.22438 & 5.5410 $\pm$ 0.0019 & 68 $\pm$ 20 & -18.64 $\pm$ 0.06 & -2.05 $\pm$ 0.29 \\
RUBIES-61052 & 34.24815 & -5.20683 & 5.5314 $\pm$ 0.0019 & 36 $\pm$ 6 & -19.48 $\pm$ 0.02 & -2.09 $\pm$ 0.15 \\
RUBIES-154518 & 34.28918 & -5.12934 & 5.52589 $\pm$ 0.00031 & 31 $\pm$ 9 & -18.95 $\pm$ 0.04 & -2.18 $\pm$ 0.23 \\
CAPERS-23419$^{\ddag}$ & 34.47111 & -5.19046 & 5.5193 $\pm$ 0.0038 & 43.7 $\pm$ 3.8 & -19.71 $\pm$ 0.01 & -2.04 $\pm$ 0.08 \\
CAPERS-127878 & 34.27856 & -5.10278 & 5.5026 $\pm$ 0.0042 & 54 $\pm$ 9 & -19.35 $\pm$ 0.04 & -2.48 $\pm$ 0.28 \\
GO1215-5674 & 34.42093 & -5.27198 & 5.4640 $\pm$ 0.0046 & 28 $\pm$ 8 & -19.46 $\pm$ 0.03 & -3.01 $\pm$ 0.21 \\
RUBIES-148683 & 34.21677 & -5.13721 & 5.377 $\pm$ 0.007 & 55 $\pm$ 13 & -18.03 $\pm$ 0.06 & -2.25 $\pm$ 0.30 \\
RUBIES-44439$^{\star}$ & 34.23105 & -5.23858 & 5.374 $\pm$ 0.010 & 152 $\pm$ 27 & -17.89 $\pm$ 0.36 & -2.04 $\pm$ 0.49 \\ 
CAPERS-83687 & 34.46527 & -5.24196 & 5.3540 $\pm$ 0.0029 & 46 $\pm$ 10 & -18.49 $\pm$ 0.03 & -3.09 $\pm$ 0.23 \\
CAPERS-142605 & 34.41013 & -5.12752 & 5.3498 $\pm$ 0.0032 & 117 $\pm$ 10 & -18.45 $\pm$ 0.03 & -2.47 $\pm$ 0.23 \\
RUBIES-33660 & 34.48128 & -5.26433 & 5.3348 $\pm$ 0.0010 & 71.9 $\pm$ 4.7 & -19.82 $\pm$ 0.02 & -2.40 $\pm$ 0.14 \\
RUBIES-10293 & 34.47240 & -5.30330 & 5.33315 $\pm$ 0.00034 & 87 $\pm$ 8 & -19.82 $\pm$ 0.03 & -1.84 $\pm$ 0.19 \\
RUBIES-157664 & 34.20792 & -5.12481 & 5.3149 $\pm$ 0.0049 & 167 $\pm$ 16 & -18.40 $\pm$ 0.08 & -1.80 $\pm$ 0.31 \\
RUBIES-4483 & 34.36079 & -5.31504 & 5.287 $\pm$ 0.007 & 51 $\pm$ 9 & -18.74 $\pm$ 0.05 & -2.95 $\pm$ 0.29 \\
RUBIES-53692$^{\star}$ & 34.45538 & -5.23181 & 5.2796 $\pm$ 0.0011 & 173 $\pm$ 39 & -17.2 $\pm$ 0.9 & -2.1 $\pm$ 0.5\\ 
RUBIES-116442 & 34.30884 & -5.18177 & 5.2559 $\pm$ 0.0017 & 54 $\pm$ 7 & -20.32 $\pm$ 0.03 & -2.27 $\pm$ 0.19 \\
CAPERS-111594 & 34.52191 & -5.17482 & 5.2373 $\pm$ 0.0029 & 258 $\pm$ 24 & -17.72 $\pm$ 0.07 & -2.11 $\pm$ 0.30 \\
RUBIES-180765 & 34.33417 & -5.09082 & 5.2276 $\pm$ 0.0026 & 32 $\pm$ 9 & -18.14 $\pm$ 0.06 & -2.13 $\pm$ 0.31 \\
RUBIES-37741 & 34.29722 & -5.25038 & 5.2047 $\pm$ 0.0020 & 90 $\pm$ 14 & -18.26 $\pm$ 0.05 & -2.29 $\pm$ 0.30 \\
RUBIES-36955 & 34.31490 & -5.25880 & 5.1756 $\pm$ 0.0018 & 153 $\pm$ 15 & -18.82 $\pm$ 0.05 & -2.00 $\pm$ 0.27 \\
GO1215-15 & 34.46877 & -5.14984 & 5.1534 $\pm$ 0.0033 & 56 $\pm$ 9 & -19.32 $\pm$ 0.03 & -2.76 $\pm$ 0.23 \\
GO1215-3982 & 34.55130 & -5.19914 & 5.1314 $\pm$ 0.0013 & 77 $\pm$ 8 & -19.69 $\pm$ 0.03 & -2.17 $\pm$ 0.19 \\
RUBIES-158198$^{\star}$ & 34.38526 & -5.12401 & 5.075 $\pm$ 0.006 & 97 $\pm$ 23 & -19.05 $\pm$ 0.31 & -1.9 $\pm$ 0.5 \\ 
RUBIES-56138 & 34.31158 & -5.21696 & 5.0502 $\pm$ 0.0018 & 117 $\pm$ 11 & -19.25 $\pm$ 0.04 & -2.82 $\pm$ 0.26 \\
CAPERS-139408 & 34.45489 & -5.12763 & 4.934 $\pm$ 0.006 & 54 $\pm$ 11 & -18.22 $\pm$ 0.04 & -2.42 $\pm$ 0.26 \\
GO1215-3115 & 34.45782 & -5.16698 & 4.9008 $\pm$ 0.0025 & 29 $\pm$ 7 & -19.30 $\pm$ 0.03 & -2.32 $\pm$ 0.18 \\
CAPERS-151496 & 34.48890 & -5.15116 & 4.8832 $\pm$ 0.0006 & 128 $\pm$ 15 & -17.91 $\pm$ 0.05 & -2.65 $\pm$ 0.30 \\
RUBIES-41445 & 34.38360 & -5.25190 & 4.8793 $\pm$ 0.0023 & 48 $\pm$ 5 & -20.69 $\pm$ 0.02 & -1.95 $\pm$ 0.13 \\
RUBIES-9407 & 34.26865 & -5.30020 & 4.8247 $\pm$ 0.0032 & 80 $\pm$ 17 & -18.44 $\pm$ 0.06 & -1.81 $\pm$ 0.31 \\
RUBIES-31552 & 34.38041 & -5.26773 & 4.8121 $\pm$ 0.0030 & 69 $\pm$ 8 & -19.57 $\pm$ 0.03 & -1.74 $\pm$ 0.19 \\
RUBIES-9779 & 34.40306 & -5.30429 & 4.706 $\pm$ 0.007 & 79 $\pm$ 12 & -19.02 $\pm$ 0.04 & -1.59 $\pm$ 0.26 \\
GO1215-4446$^{\ddag}$ & 34.27071 & -5.21767 & 4.6943 $\pm$ 0.0014 & 145 $\pm$ 9 & -19.13 $\pm$ 0.03 & -1.13 $\pm$ 0.16 \\
RUBIES-18912 & 34.33640 & -5.28889 & 4.680 $\pm$ 0.005 & 35 $\pm$ 6 & -19.89 $\pm$ 0.02 & -2.30 $\pm$ 0.14 \\
RUBIES-114160 & 34.30688 & -5.18549 & 4.6307 $\pm$ 0.0036 & 187 $\pm$ 13 & -18.48 $\pm$ 0.05 & -2.95 $\pm$ 0.29 \\
GO1215-4257 & 34.57211 & -5.20520 & 4.6283 $\pm$ 0.0017 & 42 $\pm$ 7 & -19.51 $\pm$ 0.03 & -2.06 $\pm$ 0.18 \\
RUBIES-53188 & 34.26913 & -5.22232 & 4.618 $\pm$ 0.006 & 54 $\pm$ 10 & -19.22 $\pm$ 0.03 & -2.00 $\pm$ 0.24 \\
GO1215-4982 & 34.36467 & -5.18481 & 4.6074 $\pm$ 0.0037 & 85 $\pm$ 14 & -19.27 $\pm$ 0.04 & -2.12 $\pm$ 0.28 \\
GO1215-3162 & 34.58588 & -5.18297 & 4.5813 $\pm$ 0.0040 & 22 $\pm$ 6 & -19.74 $\pm$ 0.02 & -1.20 $\pm$ 0.11 \\
RUBIES-171824 & 34.38148 & -5.10465 & 4.57941 $\pm$ 0.00027 & 90 $\pm$ 11 & -18.28 $\pm$ 0.04 & -2.48 $\pm$ 0.27 \\
RUBIES-33550 & 34.41445 & -5.26447 & 4.5670 $\pm$ 0.0047 & 95 $\pm$ 12 & -18.45 $\pm$ 0.05 & -2.98 $\pm$ 0.28 \\
GO1215-5695 & 34.46037 & -5.19995 & 4.5424 $\pm$ 0.0049 & 10.8 $\pm$ 2.7 & -20.84 $\pm$ 0.01 & -1.95 $\pm$ 0.06 \\
CAPERS-9648 & 34.46768 & -5.13095 & 4.538 $\pm$ 0.005 & 63.0 $\pm$ 3.7 & -20.09 $\pm$ 0.01 & -2.10 $\pm$ 0.09 \\

\end{longtable}

\vspace{0.3cm}
\noindent We report the observed EW$_0$ without correcting for \lya\ slit losses. M$_{\mathrm{UV}}$ values are derived from spectra and include the photometric correction. Symbols: $\ddag$ - source identified as AGN; $*$ - BLAGN also identified in \cite{Taylor2024, Hviding2025}; $\dag$ - source in an identified ionized bubble; 
$+$ - \lya\ emission discussed in \cite{Chen2025}; $\star$ - $\beta$ slope derived using the \cite{Dottorini2025} relation.

\begin{figure*}[ht!]
    \centering

    \begin{subfigure}[b]{0.22\textwidth}
        \includegraphics[width=\linewidth]{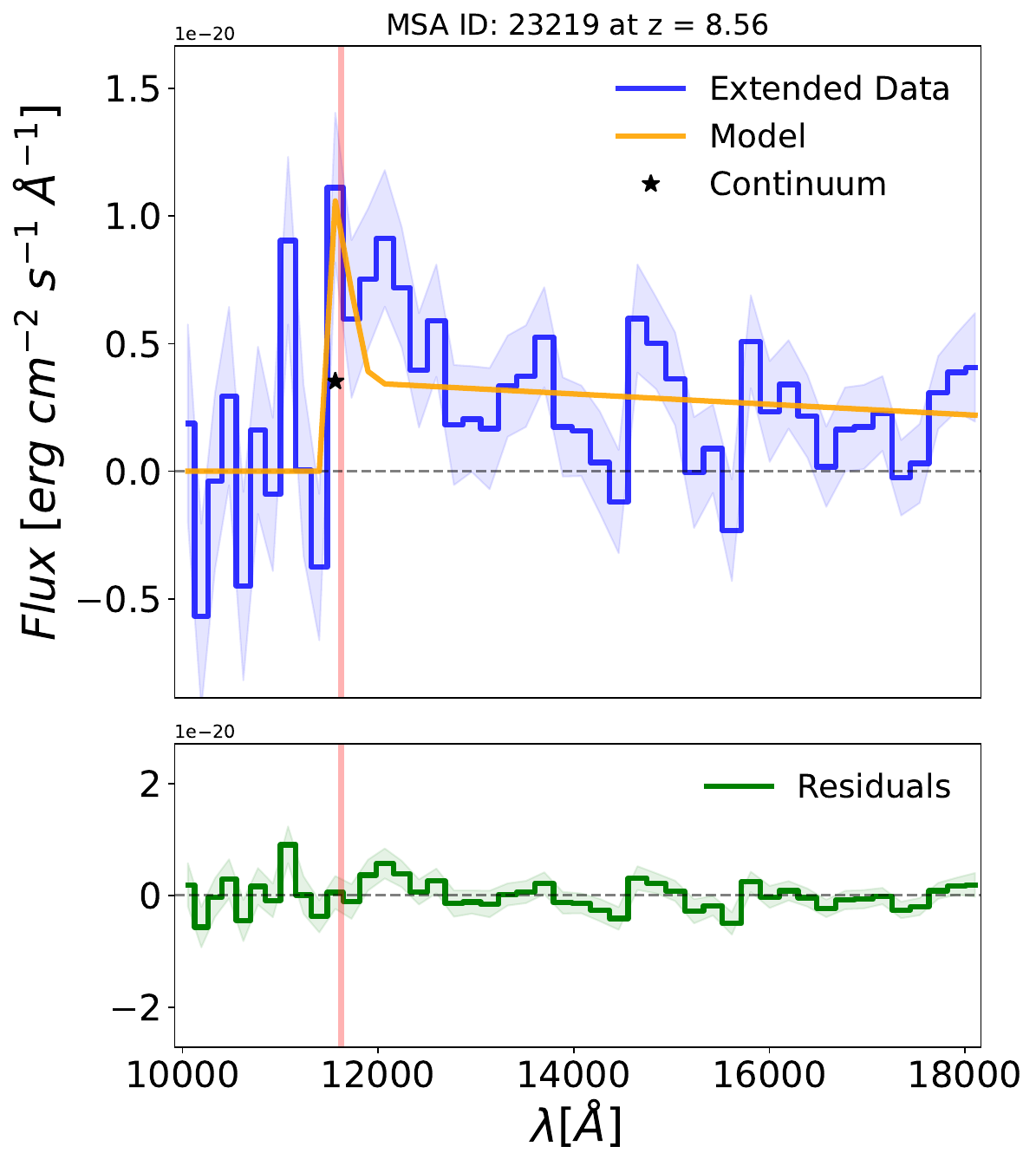}
    \end{subfigure}
    \begin{subfigure}[b]{0.22\textwidth}
        \includegraphics[width=\linewidth]{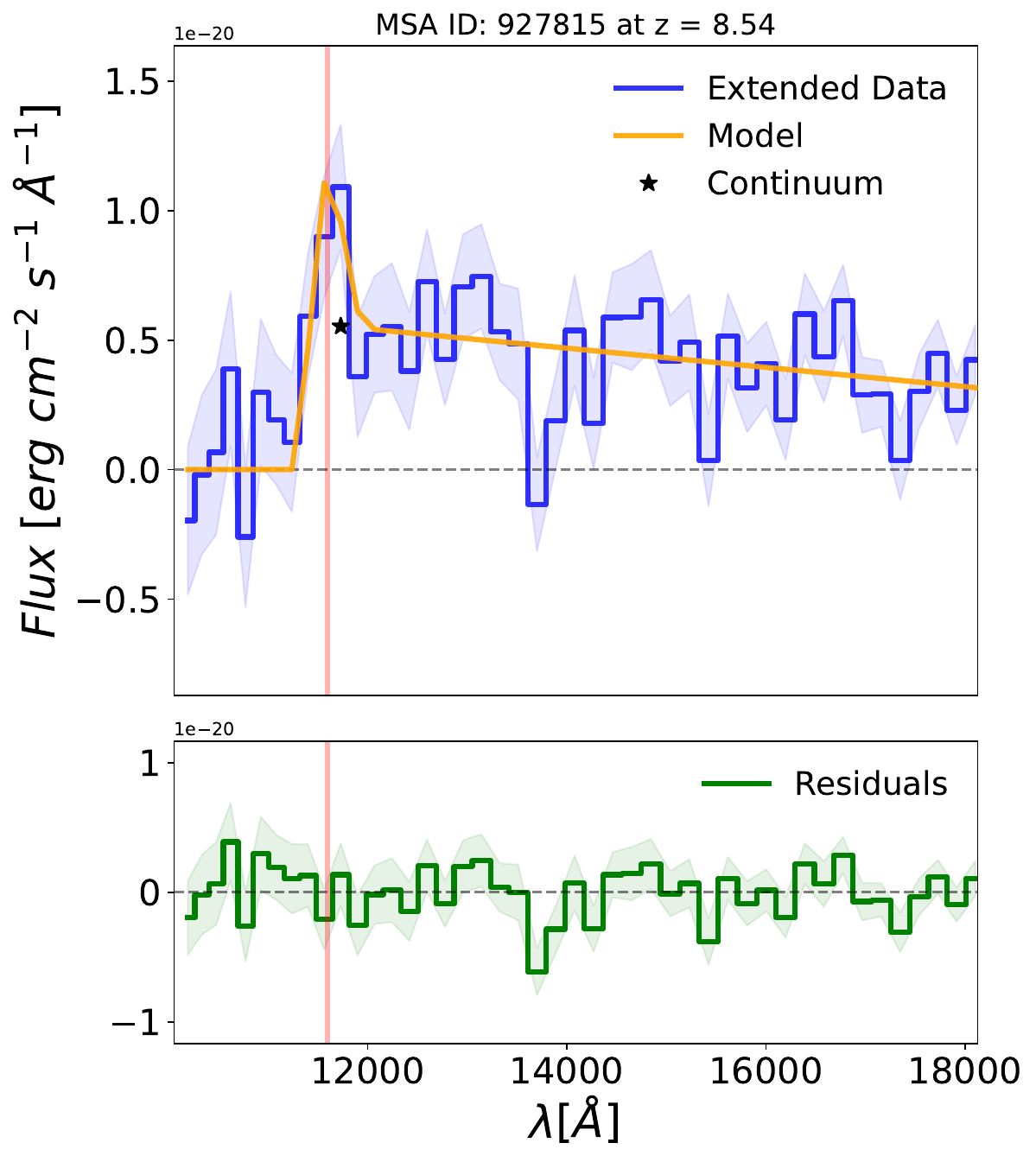}
    \end{subfigure}
    \begin{subfigure}[b]{0.22\textwidth}
        \includegraphics[width=\linewidth]{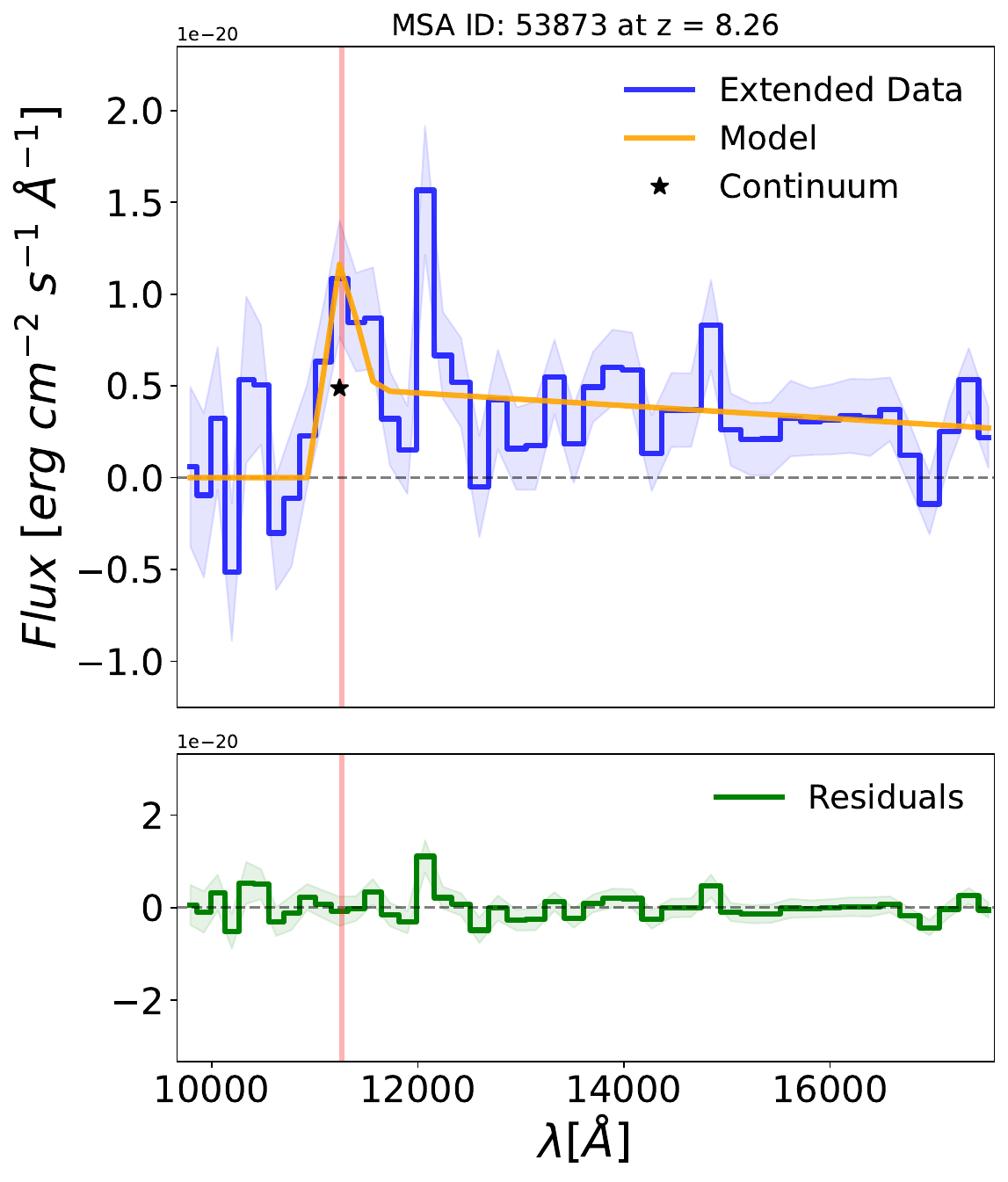}
    \end{subfigure}
    \begin{subfigure}[b]{0.22\textwidth}
        \includegraphics[width=\linewidth]{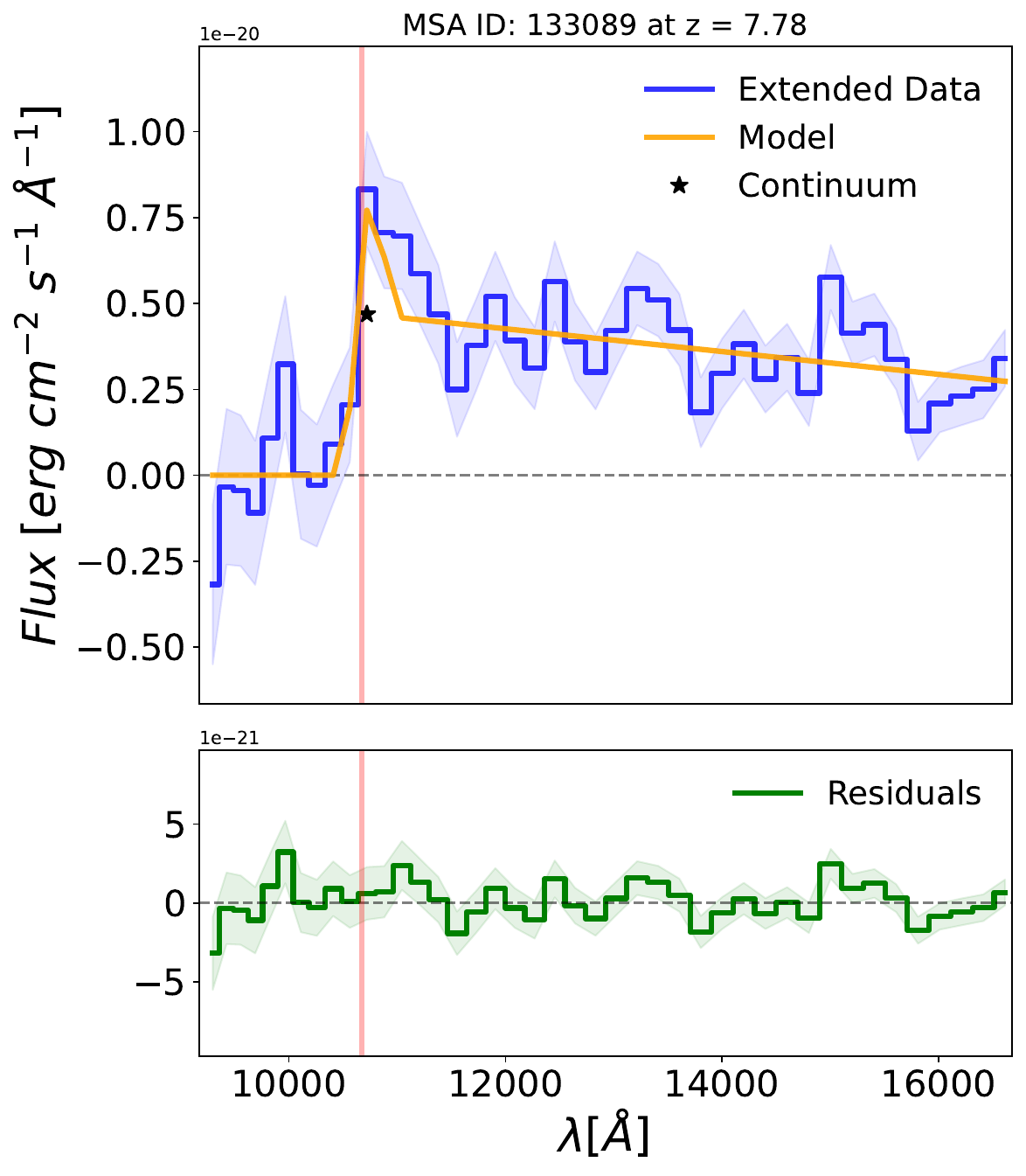}
    \end{subfigure}
    
    \vspace{0.1cm}

    \begin{subfigure}[b]{0.22\textwidth}
        \includegraphics[width=\linewidth]{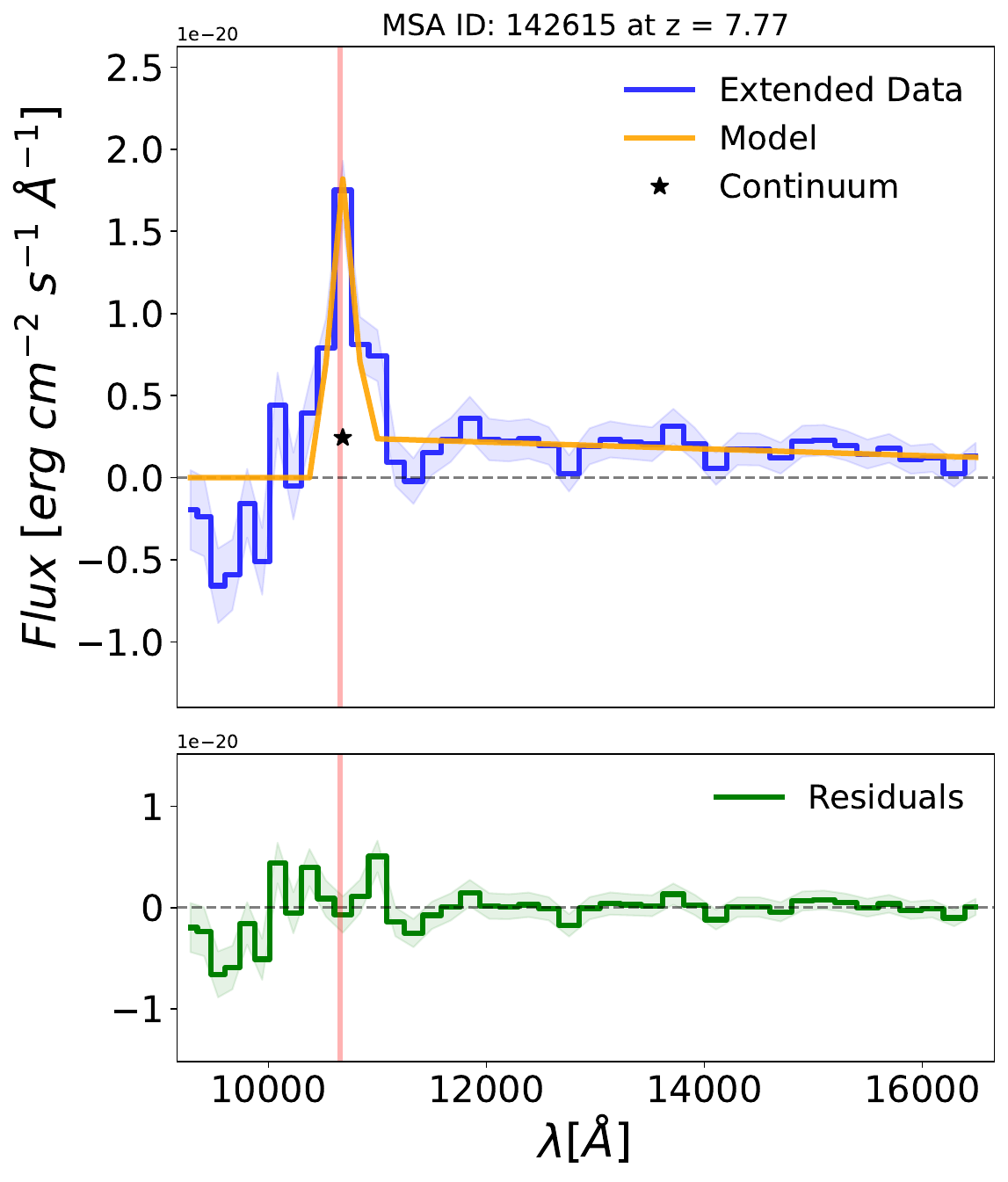}
    \end{subfigure}
    \begin{subfigure}[b]{0.22\textwidth}
        \includegraphics[width=\linewidth]{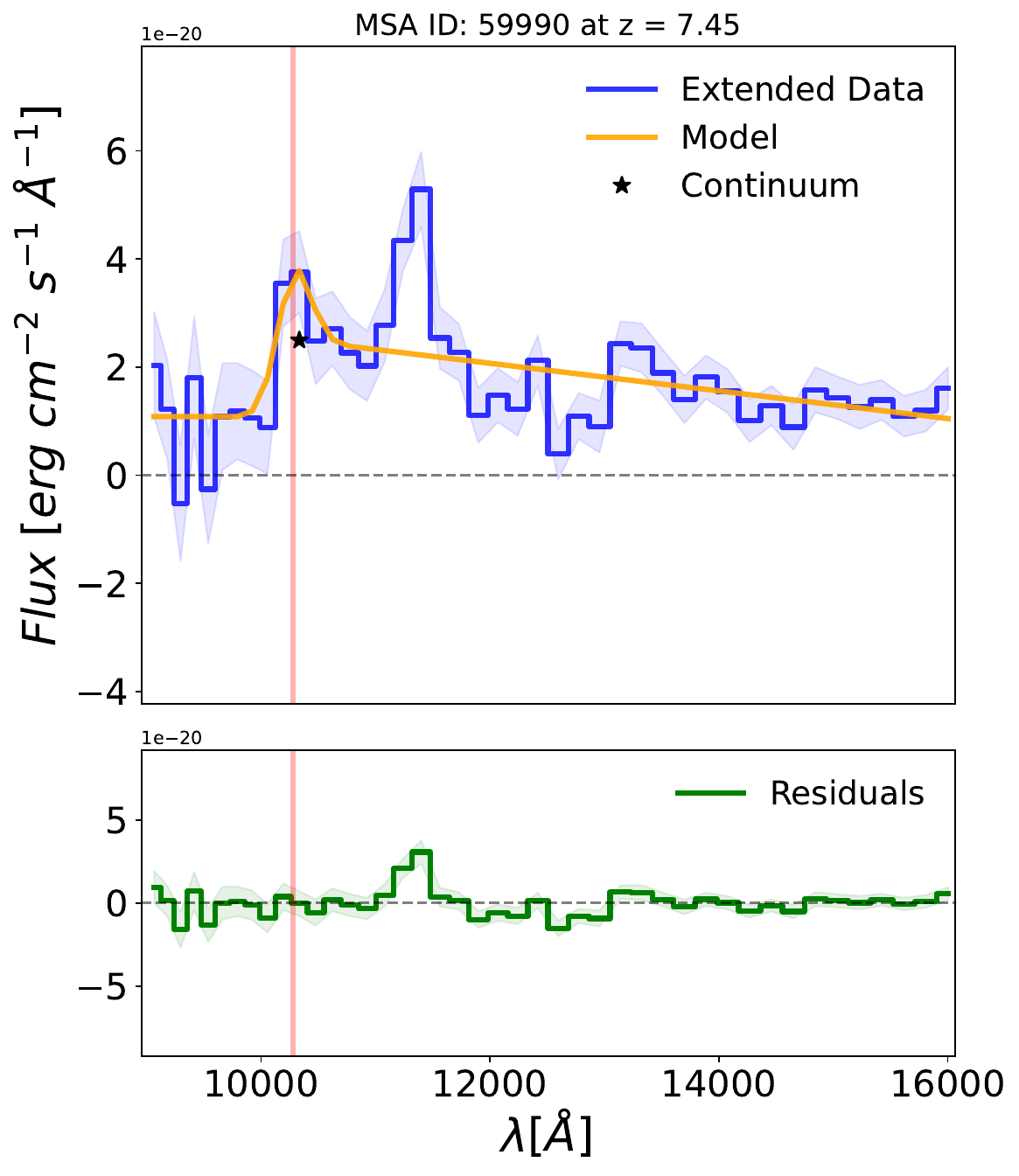}
    \end{subfigure}
    \begin{subfigure}[b]{0.22\textwidth}
        \includegraphics[width=\linewidth]{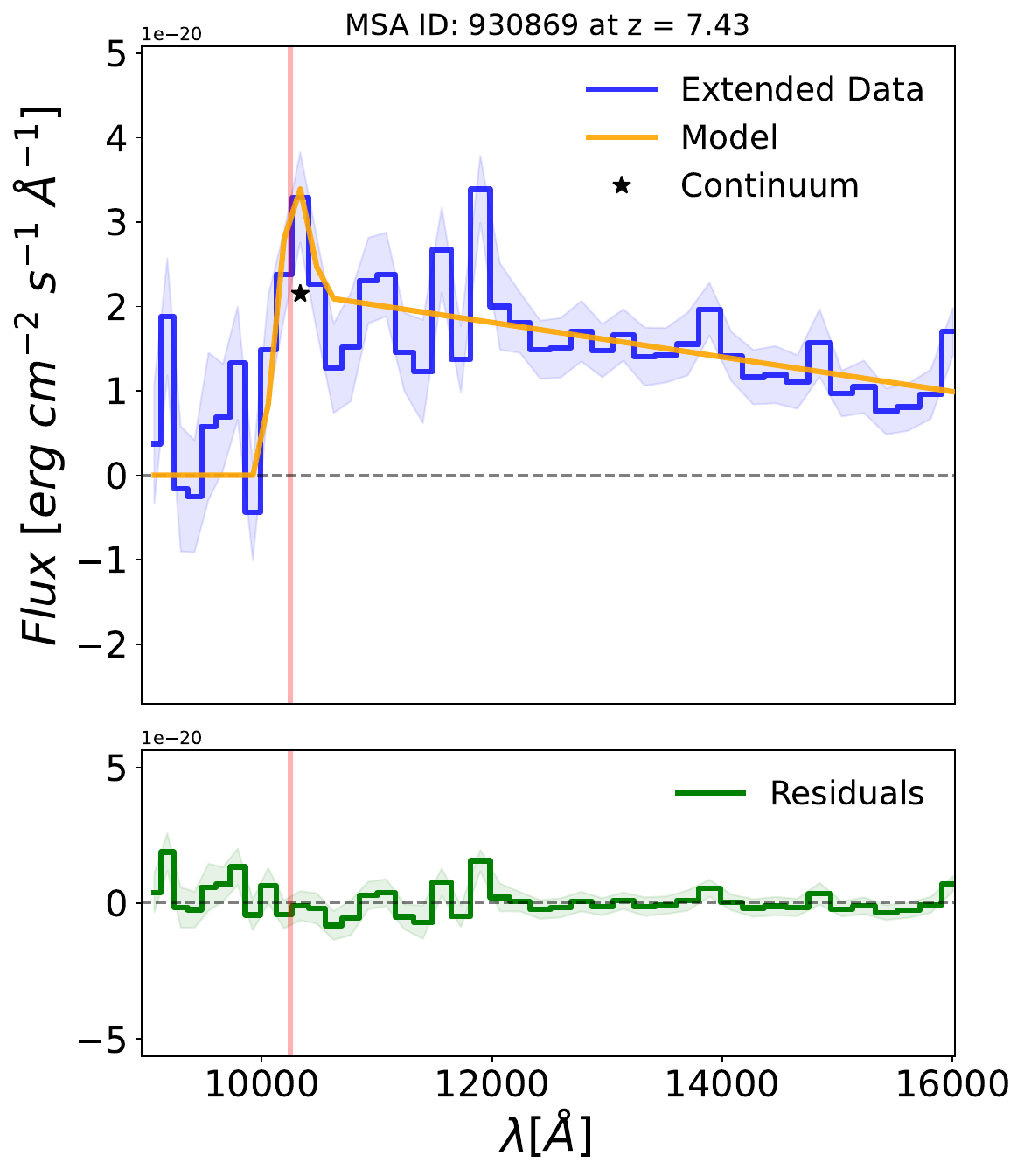}
    \end{subfigure}
    \begin{subfigure}[b]{0.22\textwidth}
        \includegraphics[width=\linewidth]{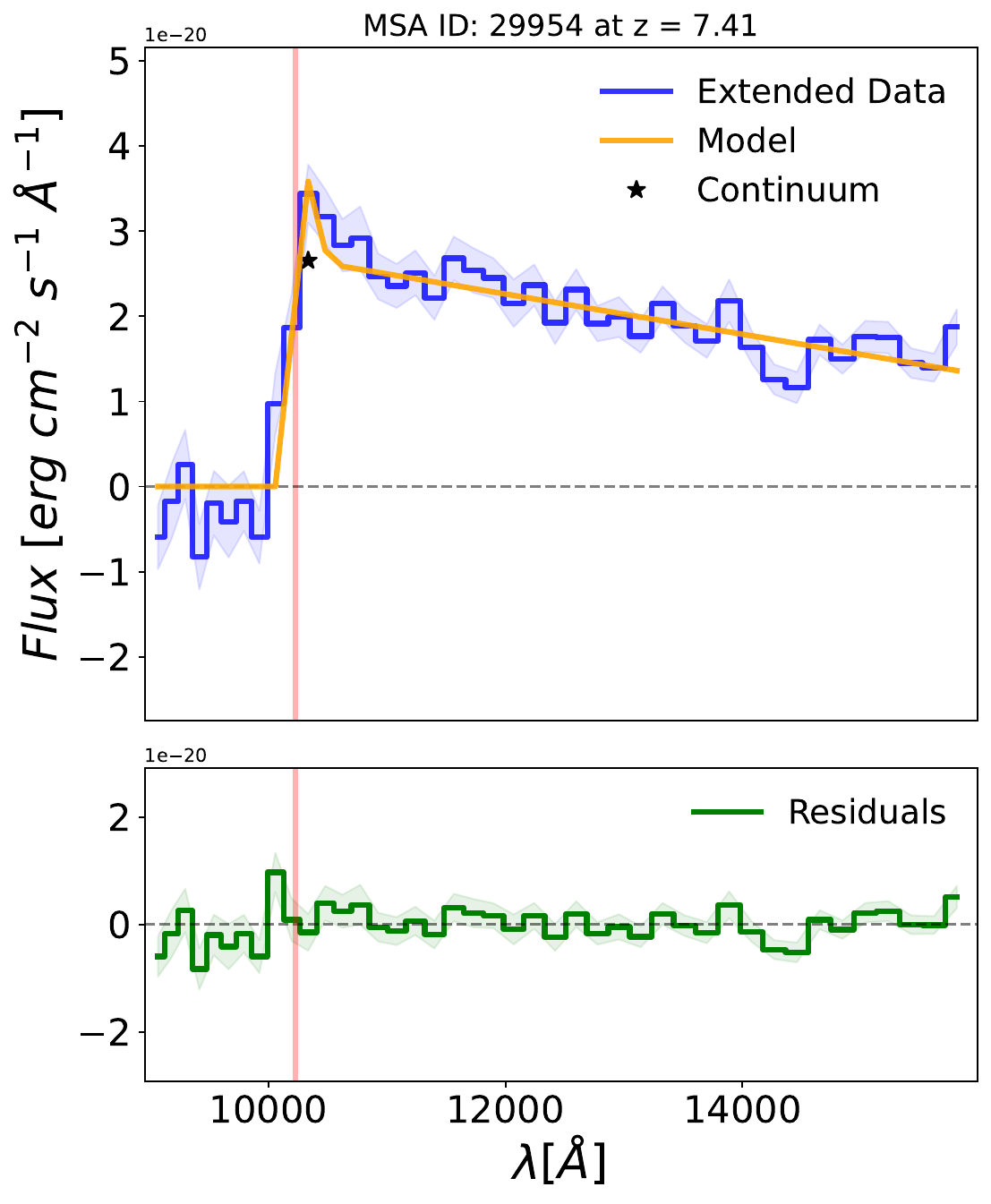}
    \end{subfigure}
    
    \vspace{0.1cm}

    \begin{subfigure}[b]{0.22\textwidth}
        \includegraphics[width=\linewidth]{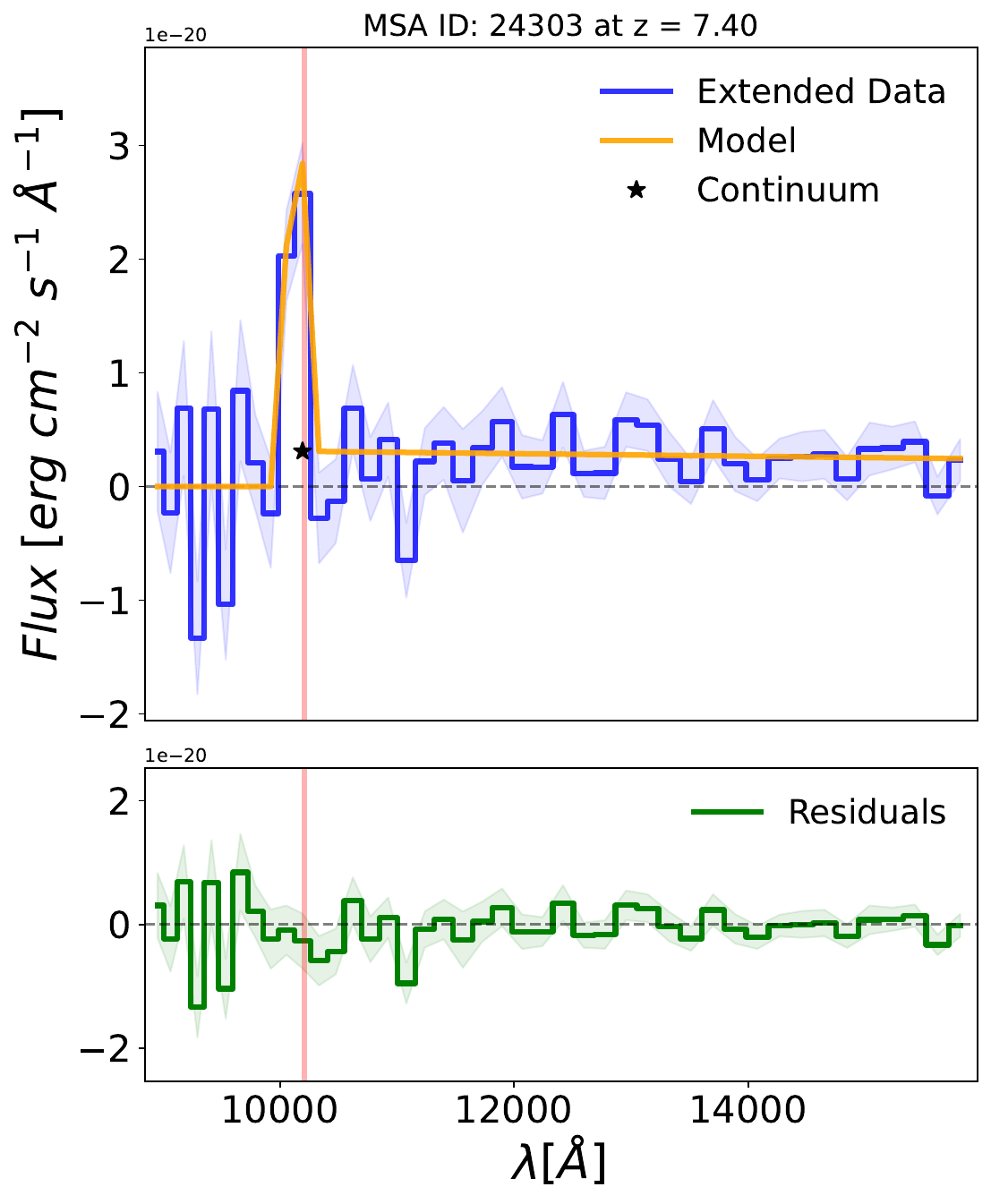}
    \end{subfigure}
    \begin{subfigure}[b]{0.22\textwidth}
        \includegraphics[width=\linewidth]{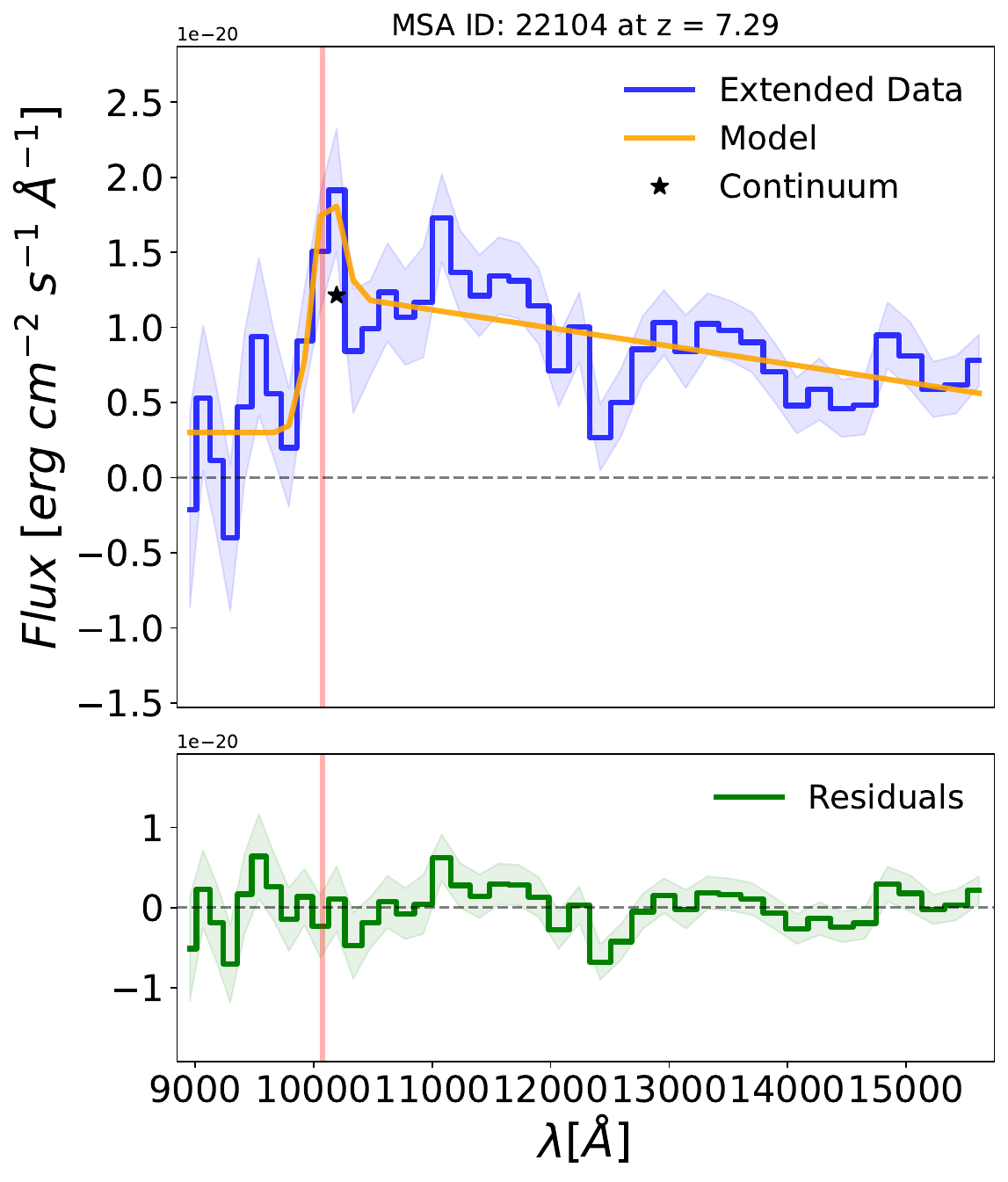}
    \end{subfigure}
    \begin{subfigure}[b]{0.22\textwidth}
        \includegraphics[width=\linewidth]{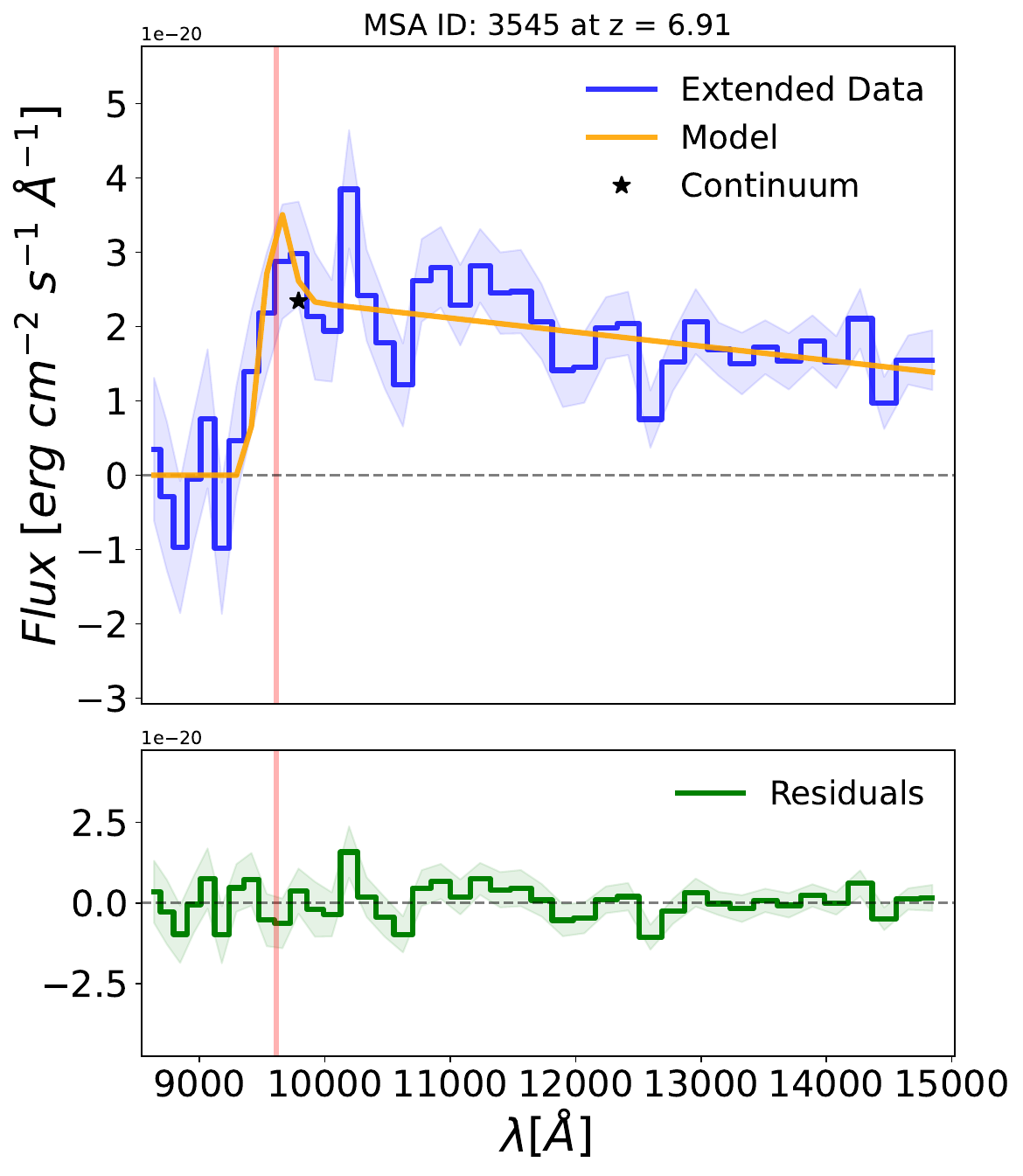}
    \end{subfigure}
    \begin{subfigure}[b]{0.22\textwidth}
        \includegraphics[width=\linewidth]{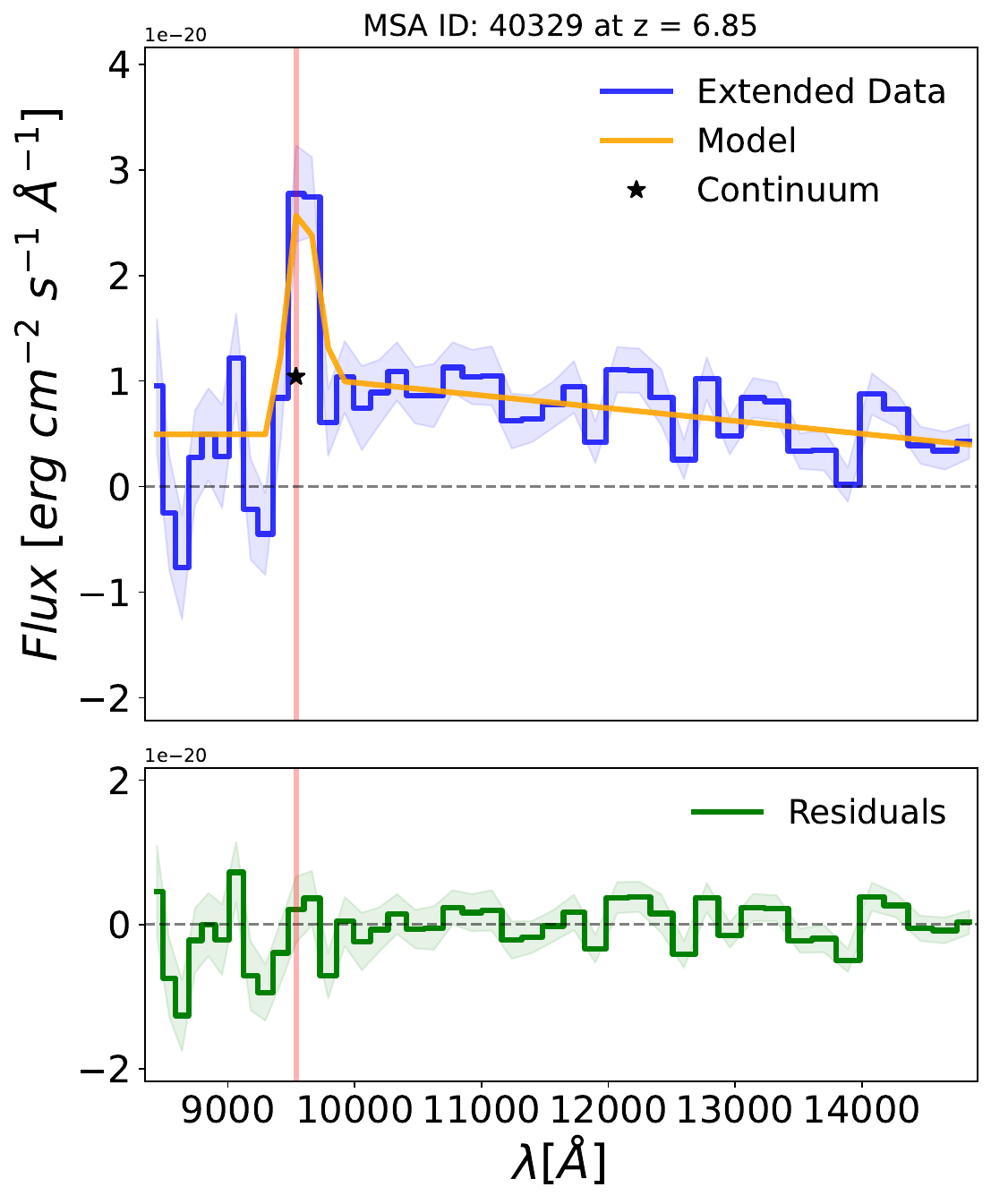}
    \end{subfigure}
    
    \caption{Fit \lya\ line profiles of the 73 galaxies with robust S/N > 3 \lya\ emission. The blue solid line and shaded area denote the flux and error measurement as a function of the observed wavelength. The best fit-model is represented by the orange solid line, while the fit continuum value at the \lya\ line peak is represented by the star symbol. The vertical red line indicates the \lya\ expected wavelength at the systemic redshift of the source. The lower panels show the residuals from the fit.}
    \label{fig:Lya_emitters} 
\end{figure*}

\twocolumn

\begin{figure*}[ht!]
    \ContinuedFloat
    
    \centering

    \begin{subfigure}[b]{0.22\textwidth}
        \includegraphics[width=\linewidth]{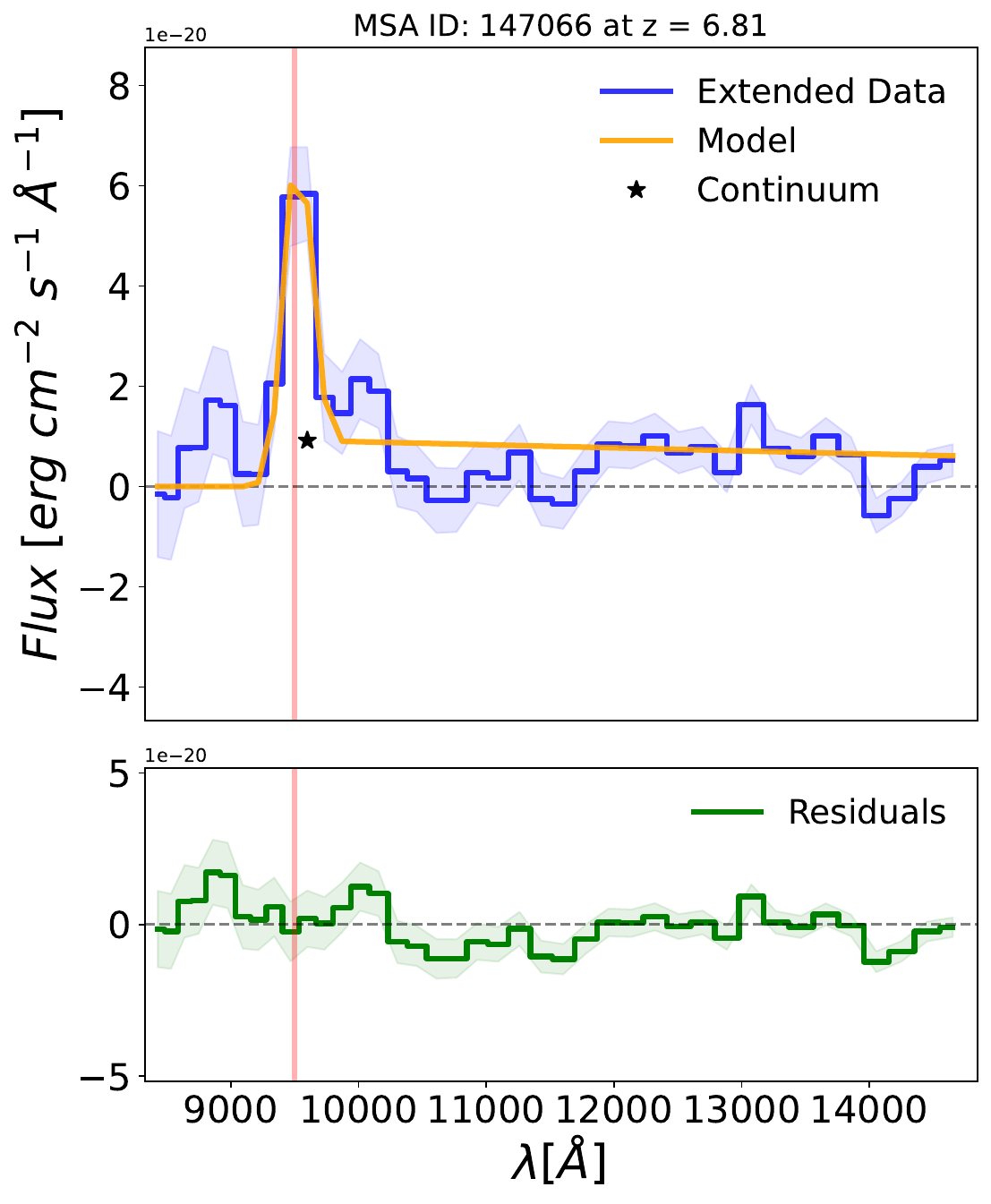}
    \end{subfigure}
    \begin{subfigure}[b]{0.22\textwidth}
        \includegraphics[width=\linewidth]{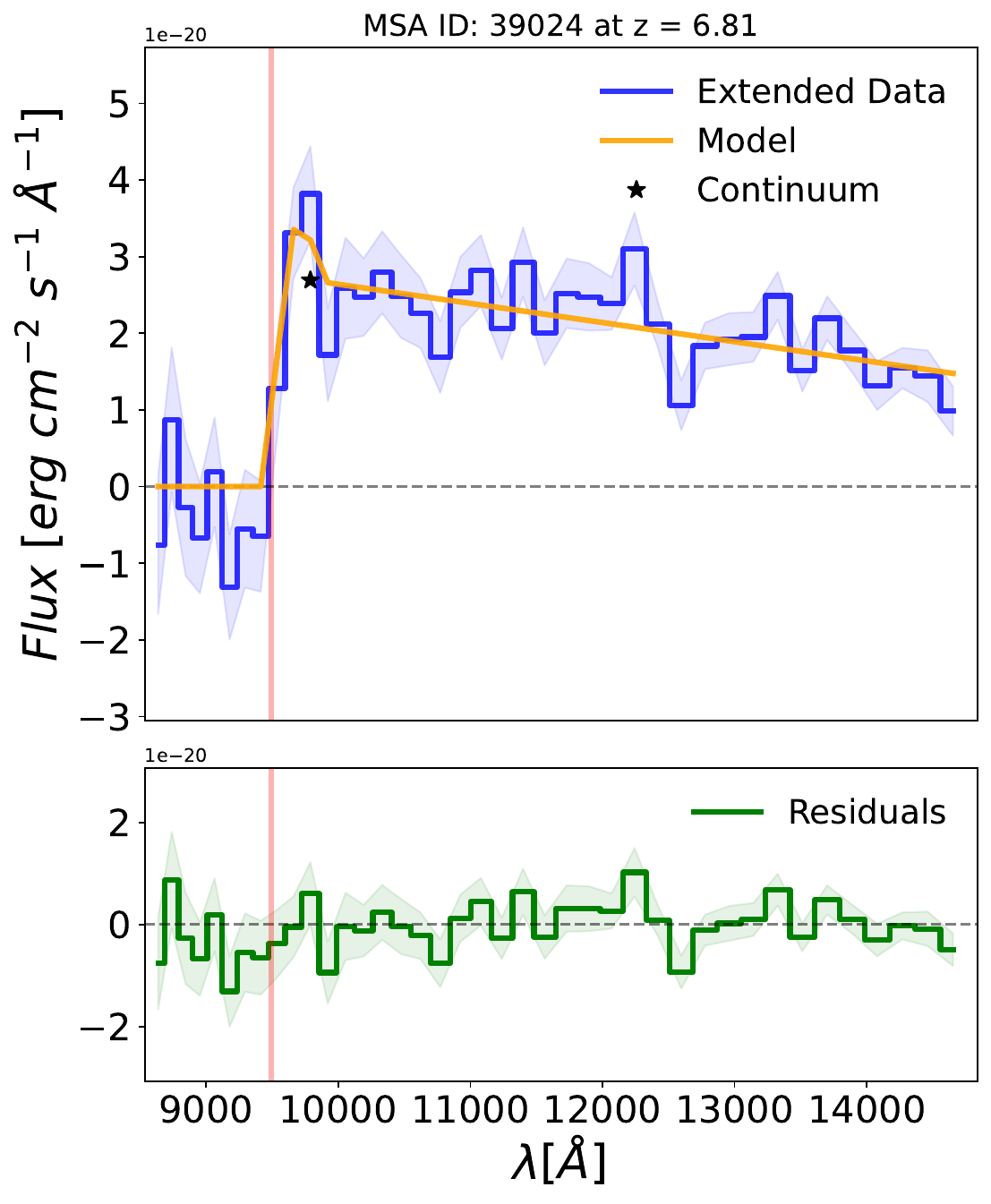}
    \end{subfigure}
    \begin{subfigure}[b]{0.22\textwidth}
        \includegraphics[width=\linewidth]{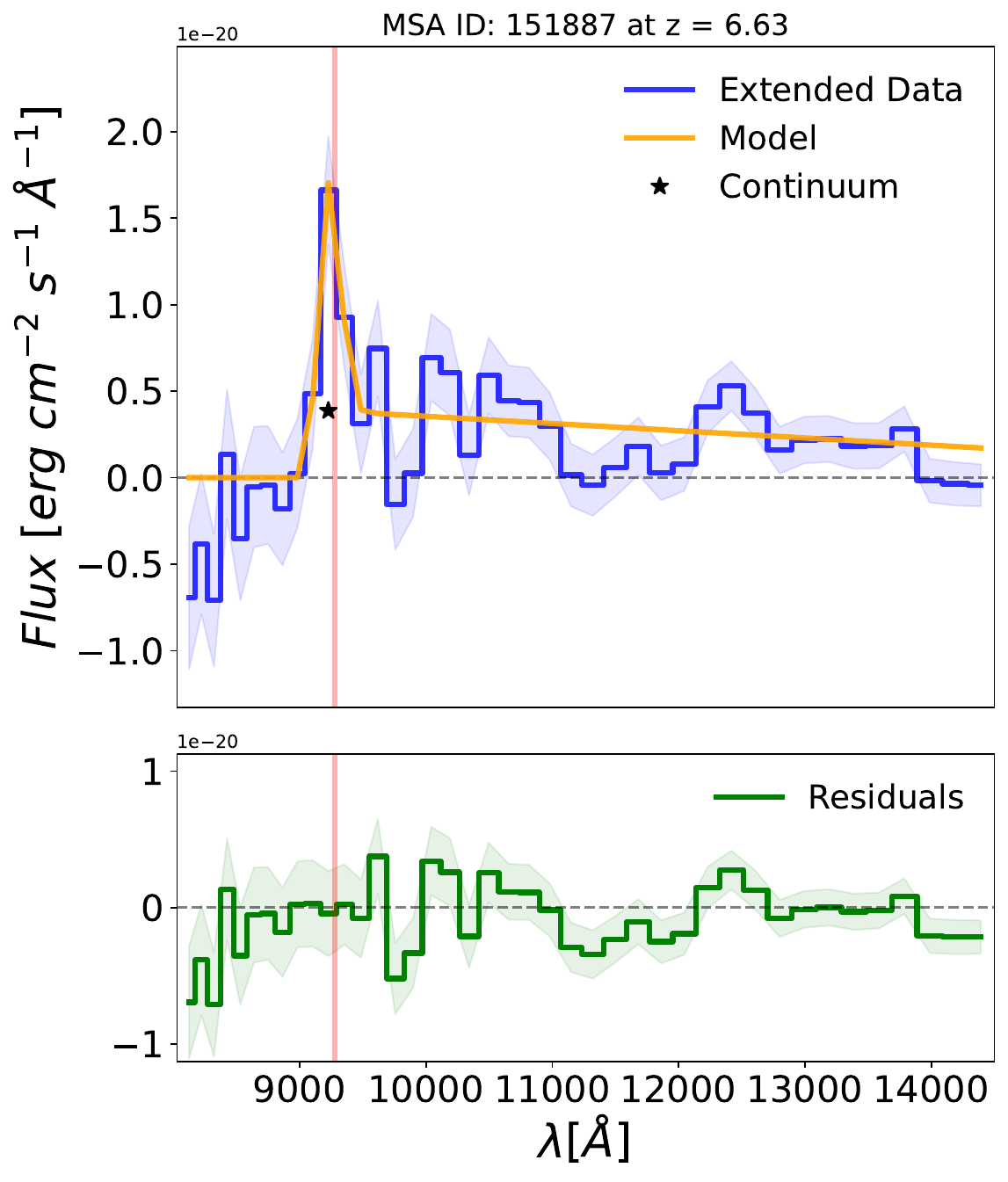}
    \end{subfigure}
    \begin{subfigure}[b]{0.22\textwidth}
        \includegraphics[width=\linewidth]{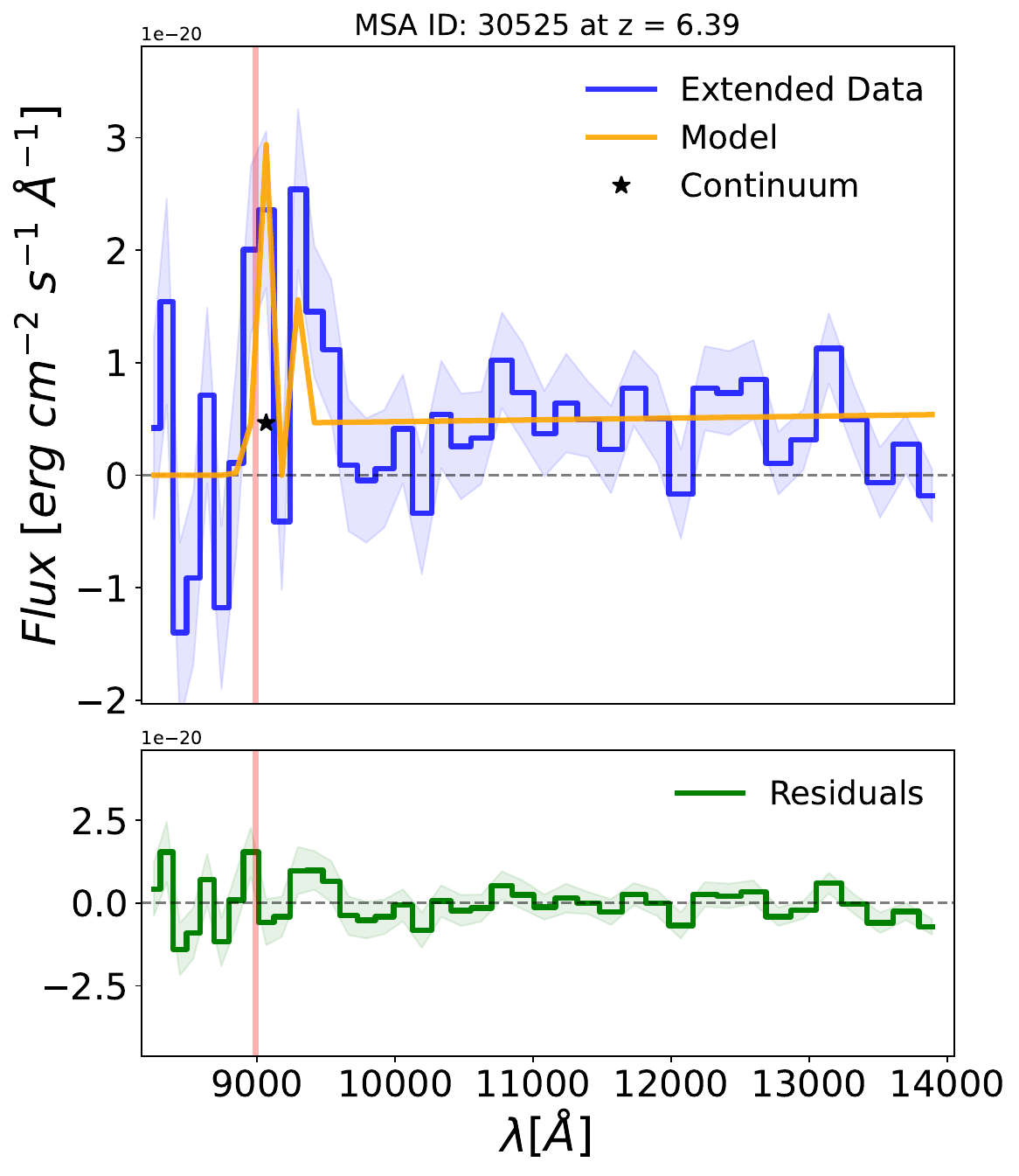}
    \end{subfigure}

    \vspace{0.1cm}

    \begin{subfigure}[b]{0.22\textwidth}
        \includegraphics[width=\linewidth]{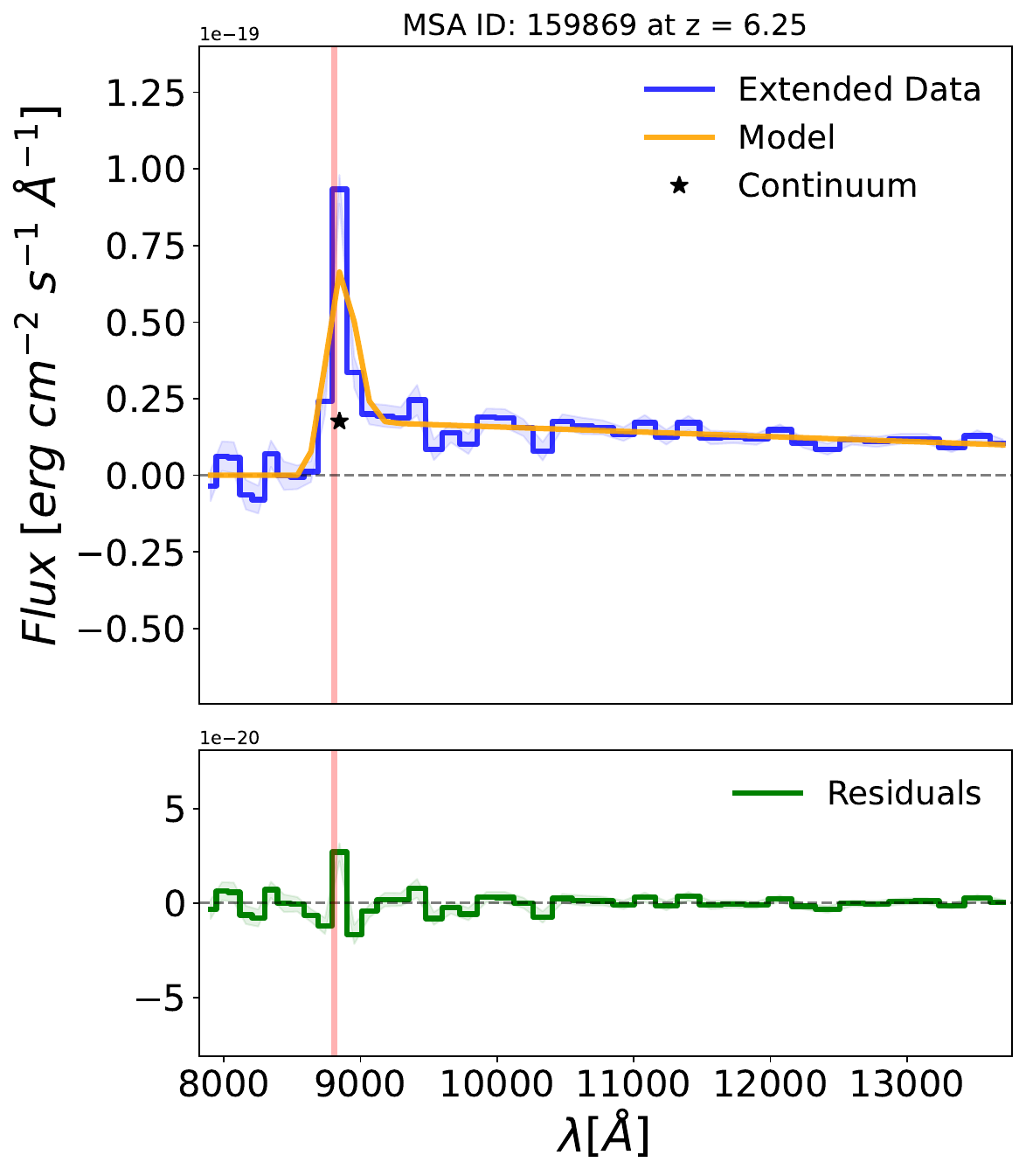}
    \end{subfigure}
    \begin{subfigure}[b]{0.22\textwidth}
        \includegraphics[width=\linewidth]{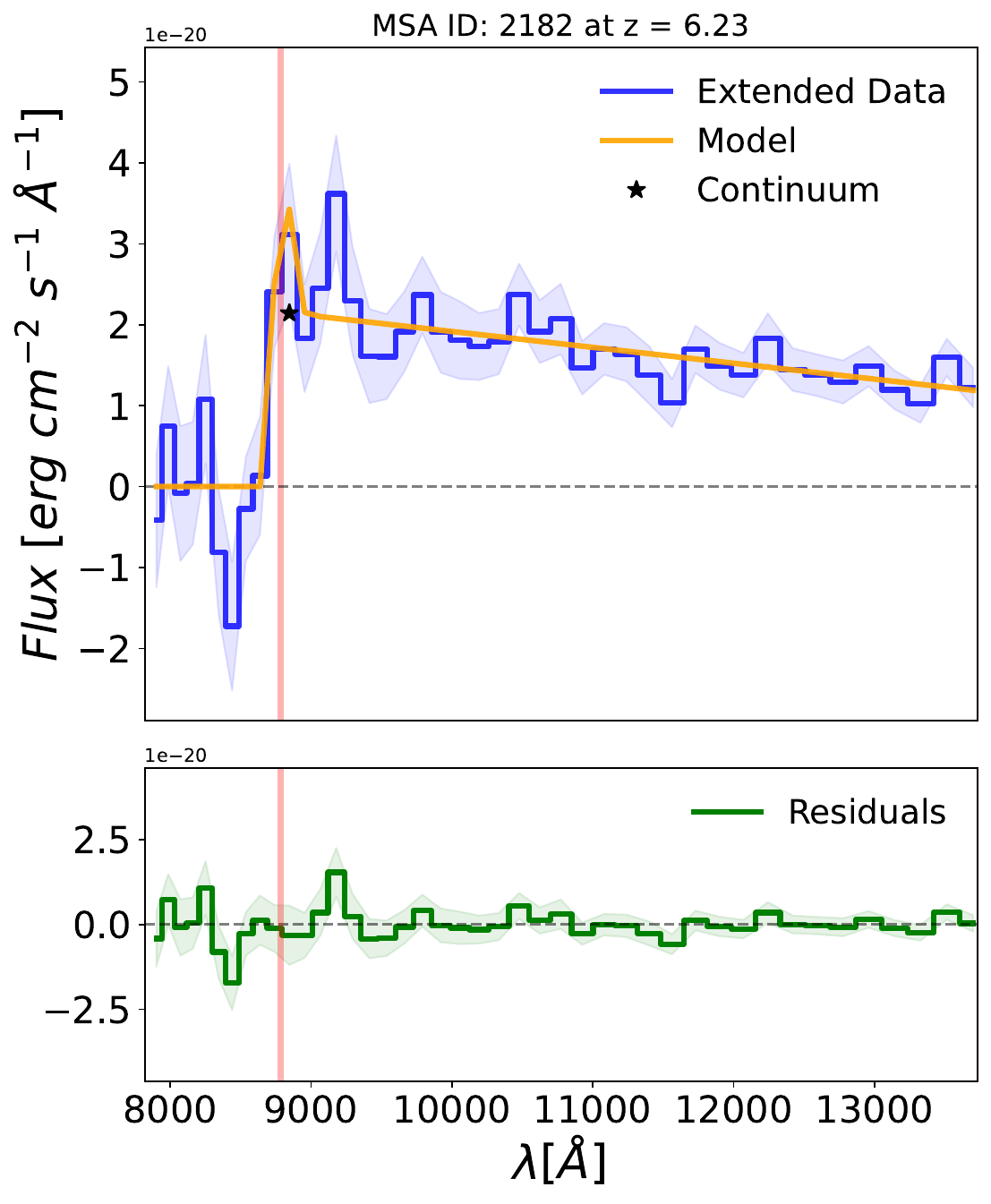}
    \end{subfigure}
    \begin{subfigure}[b]{0.22\textwidth}
        \includegraphics[width=\linewidth]{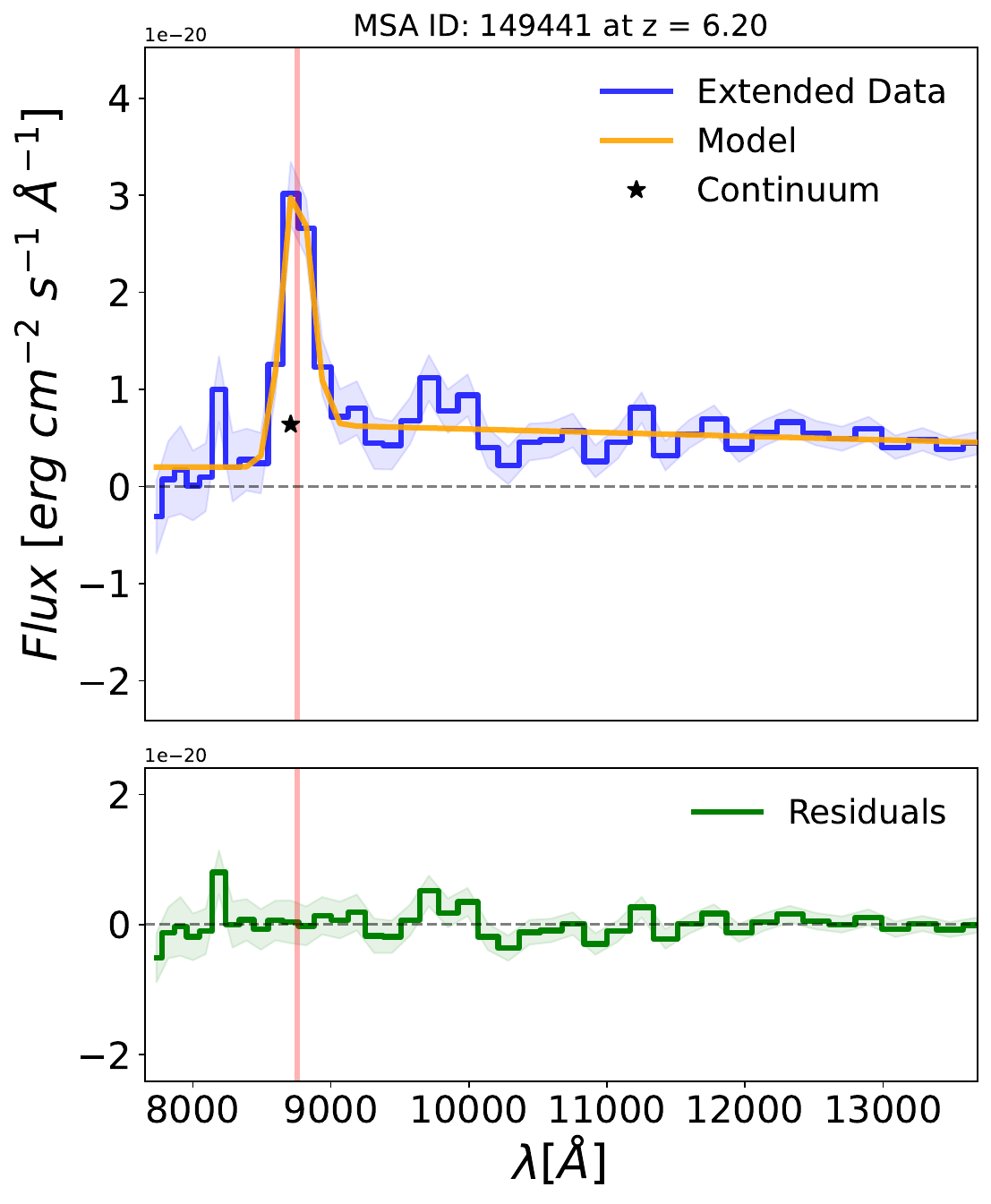}
    \end{subfigure}
    \begin{subfigure}[b]{0.22\textwidth}
        \includegraphics[width=\linewidth]{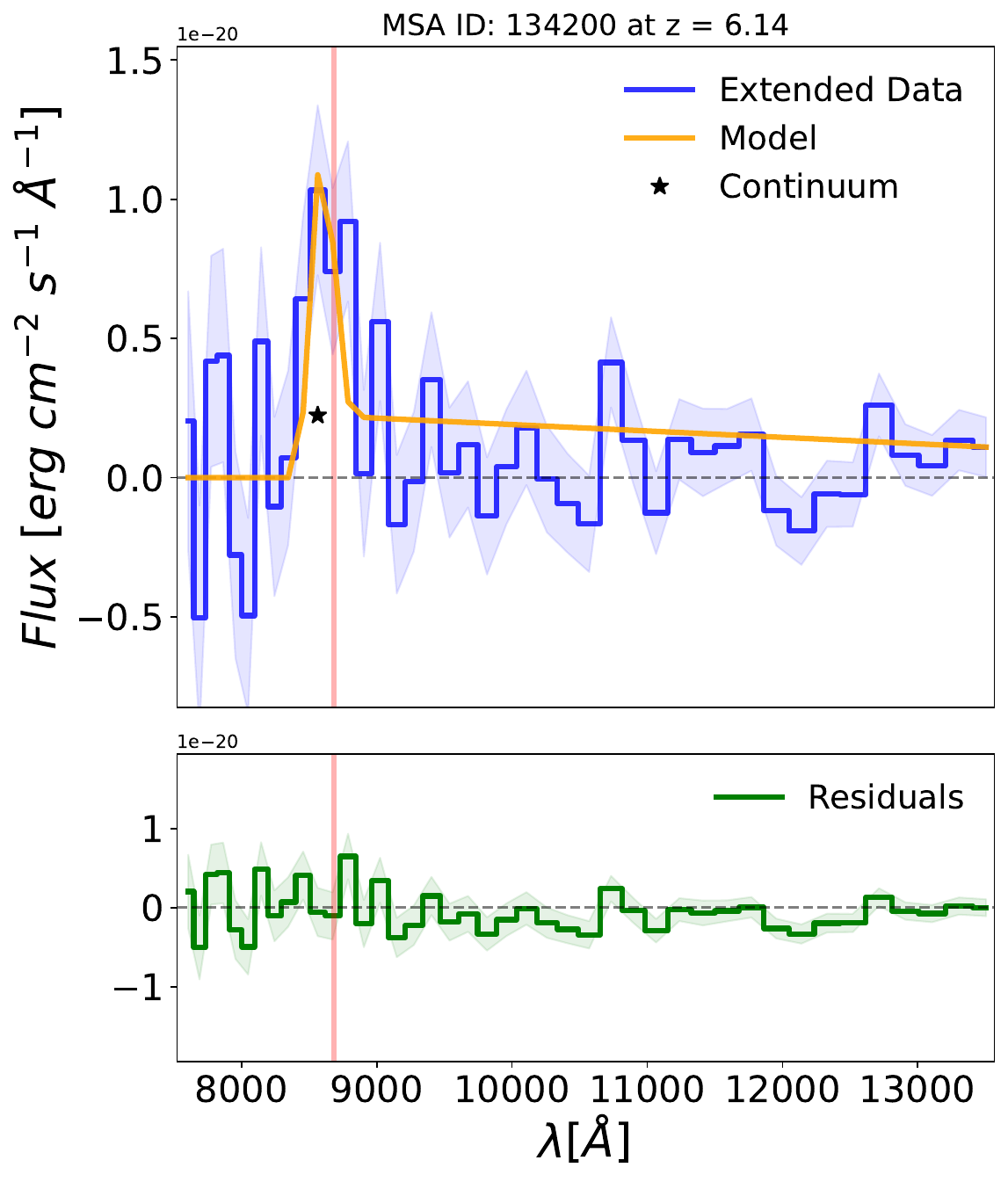}
    \end{subfigure}

    \vspace{0.1cm}

    \begin{subfigure}[b]{0.22\textwidth}
        \includegraphics[width=\linewidth]{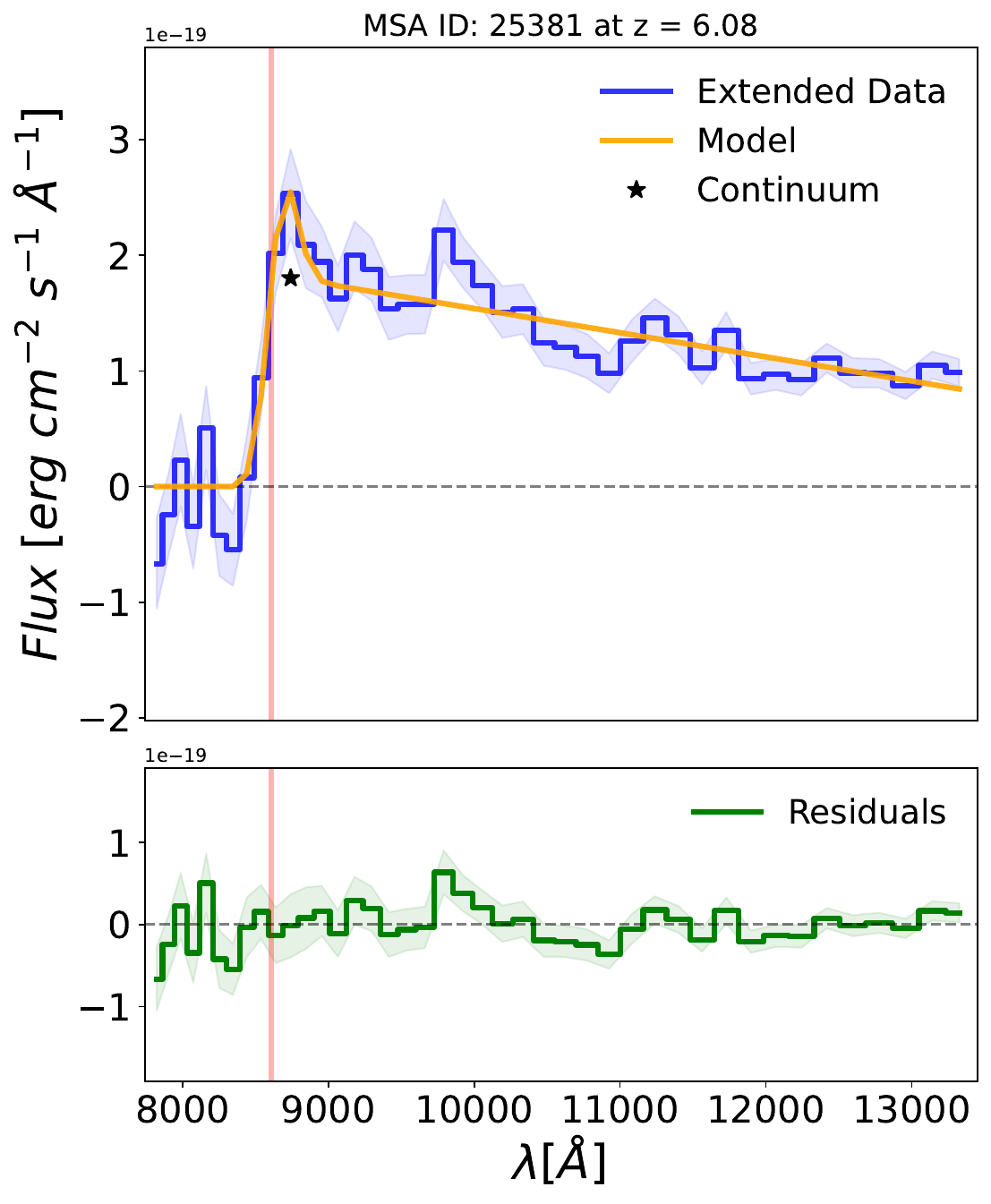}
    \end{subfigure}
    \begin{subfigure}[b]{0.22\textwidth}
        \includegraphics[width=\linewidth]{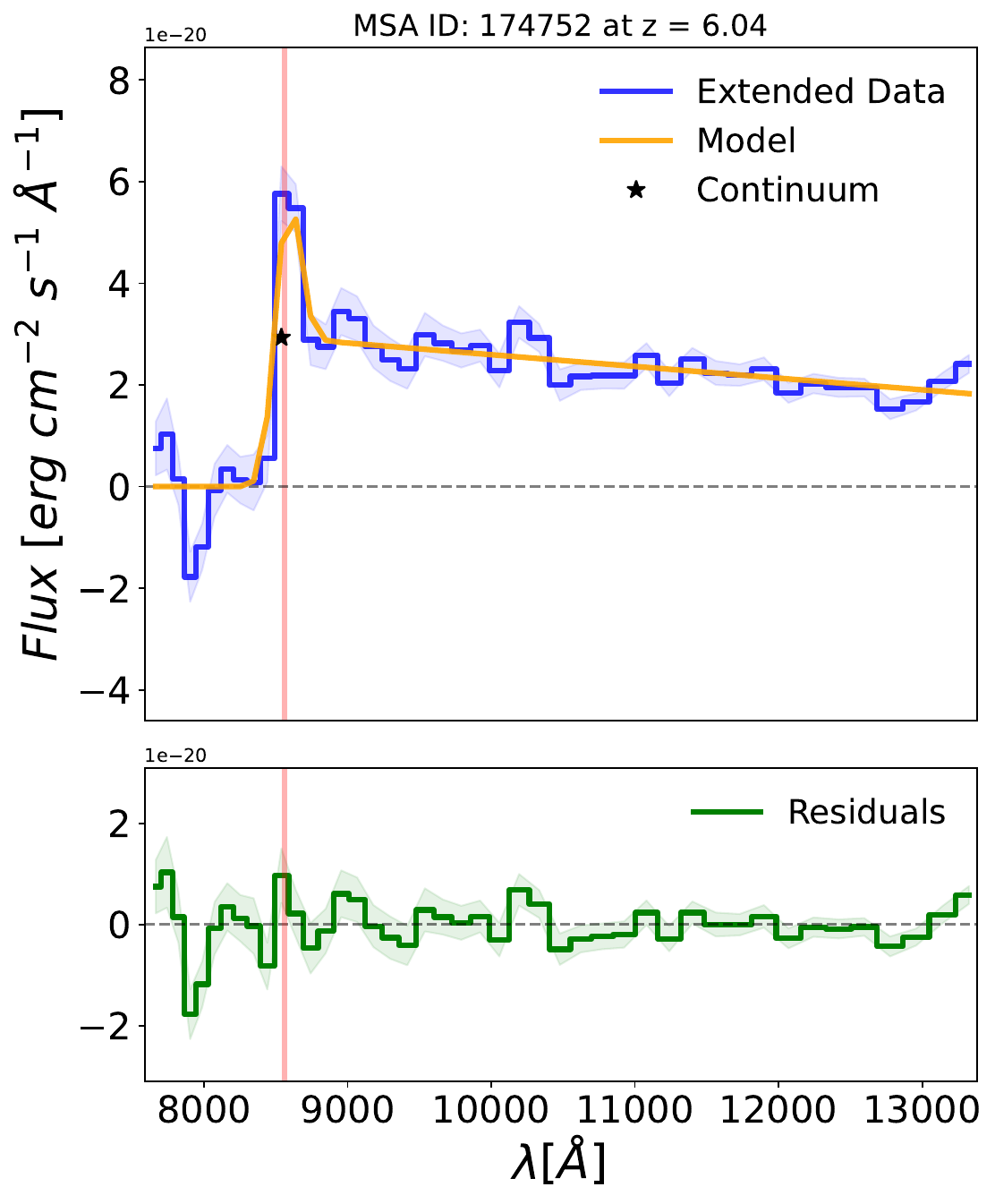}
    \end{subfigure}
    \begin{subfigure}[b]{0.22\textwidth}
        \includegraphics[width=\linewidth]{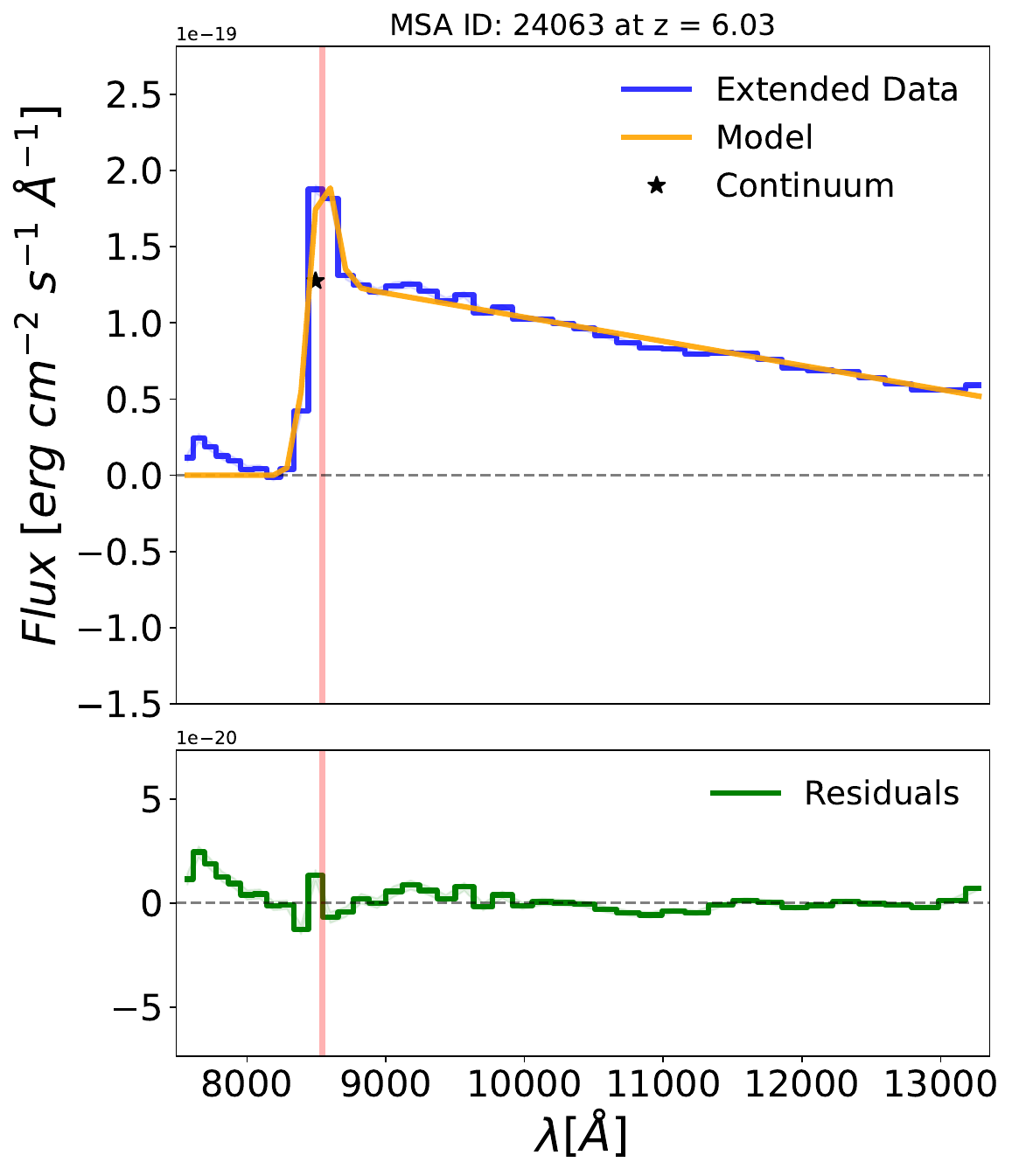}
    \end{subfigure}
    \begin{subfigure}[b]{0.22\textwidth}
        \includegraphics[width=\linewidth]{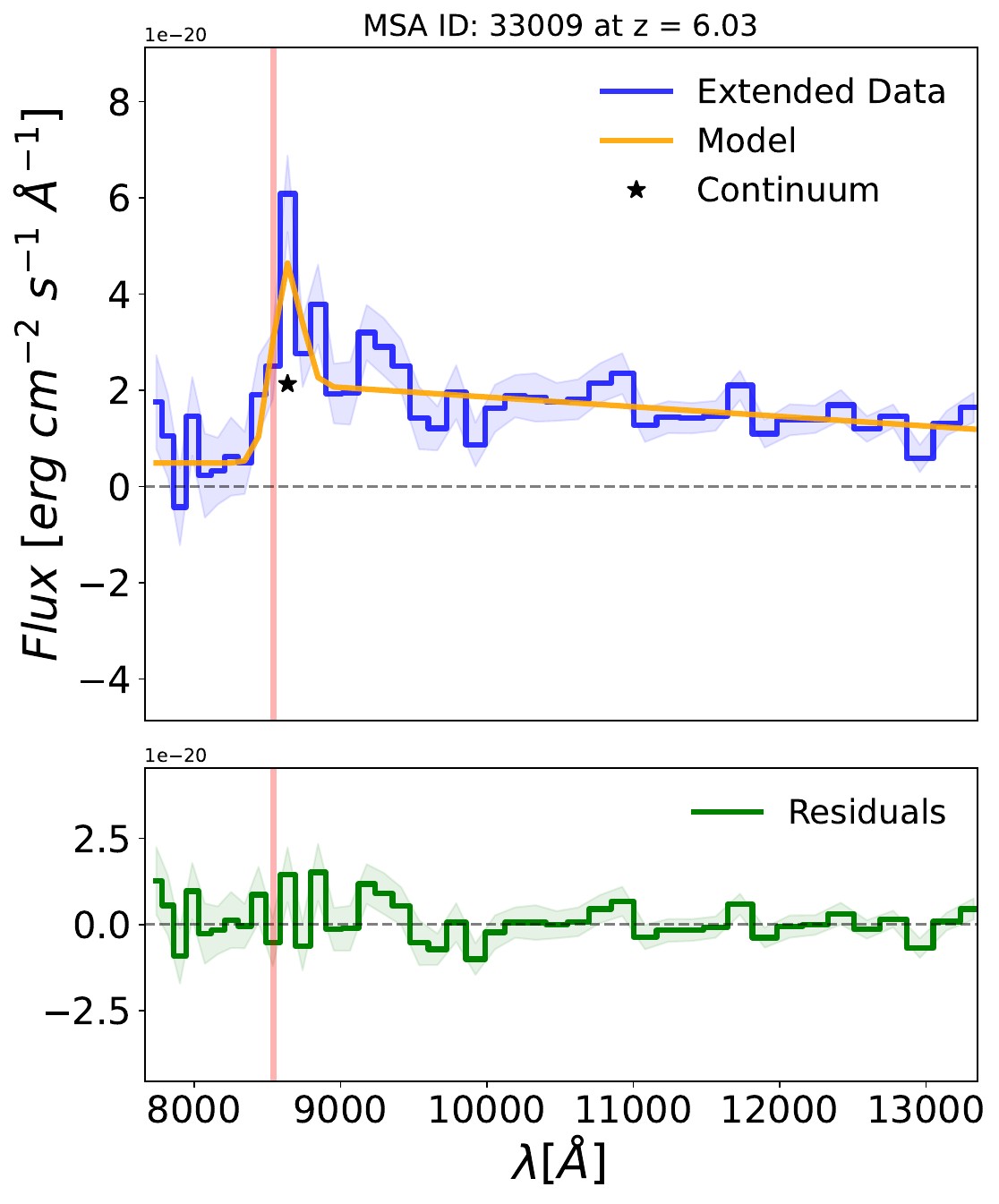}
    \end{subfigure}

    \vspace{0.1cm}

    \begin{subfigure}[b]{0.22\textwidth}
        \includegraphics[width=\linewidth]{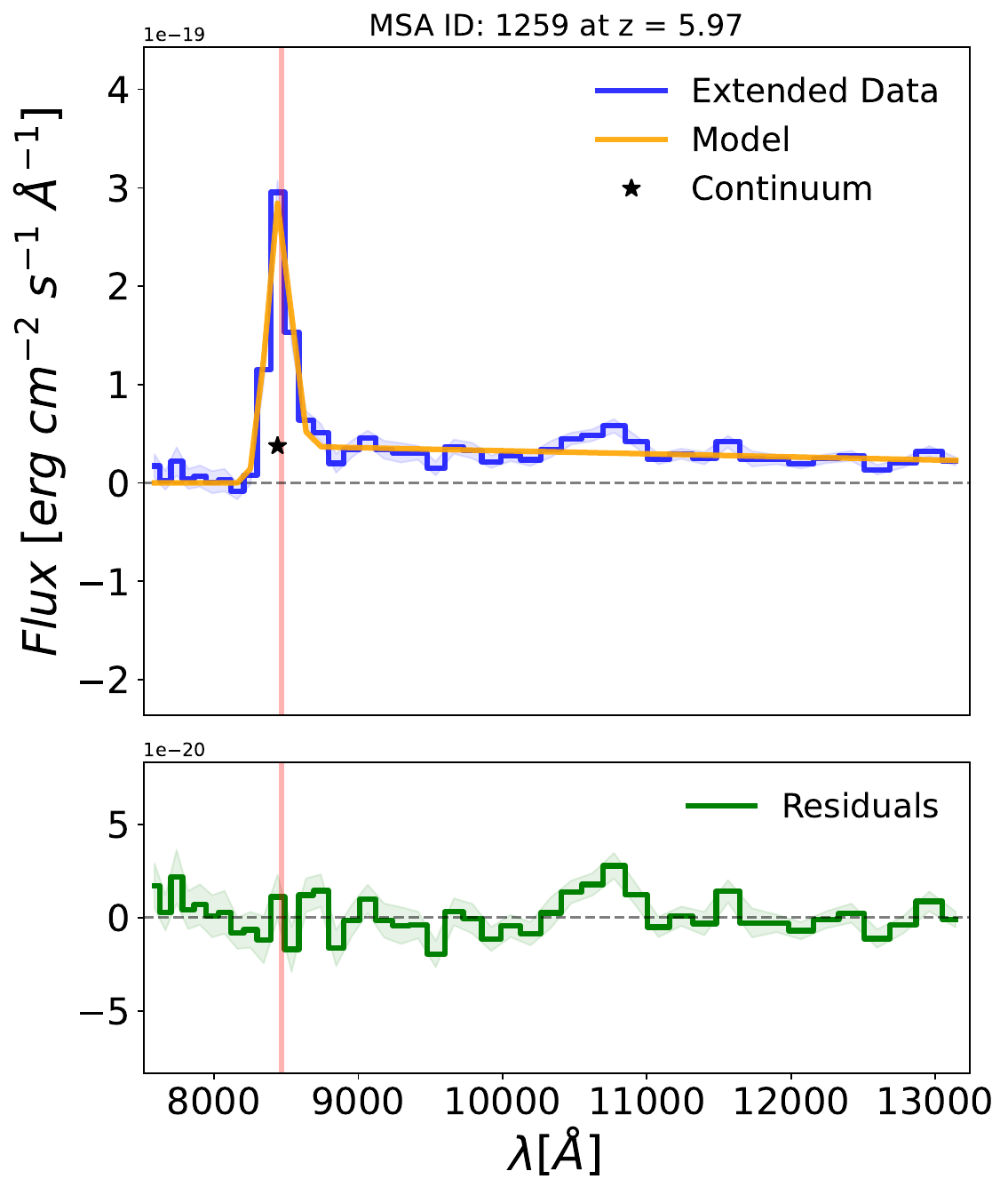}
    \end{subfigure}
    \begin{subfigure}[b]{0.22\textwidth}
        \includegraphics[width=\linewidth]{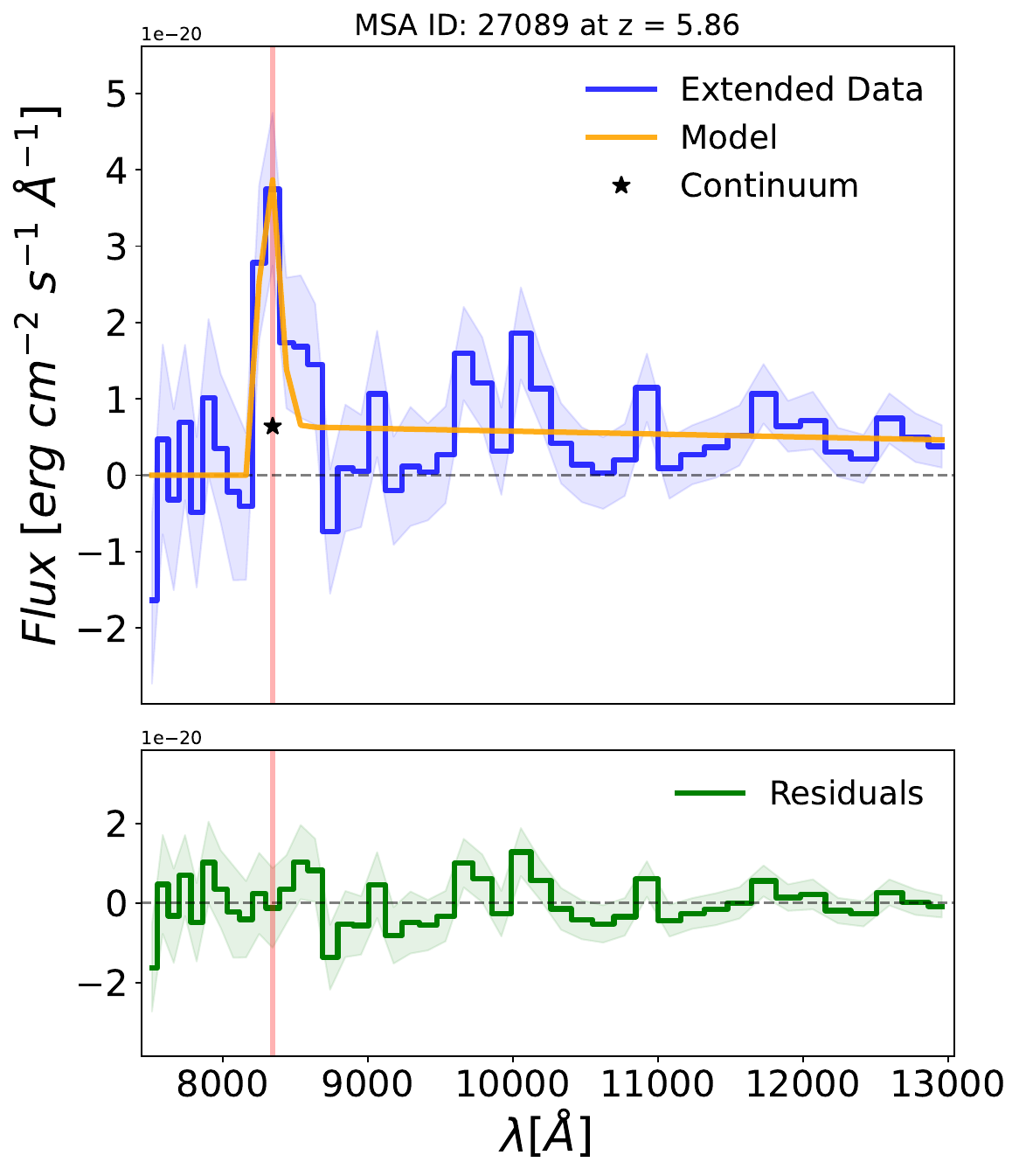}
    \end{subfigure}
    \begin{subfigure}[b]{0.22\textwidth}
        \includegraphics[width=\linewidth]{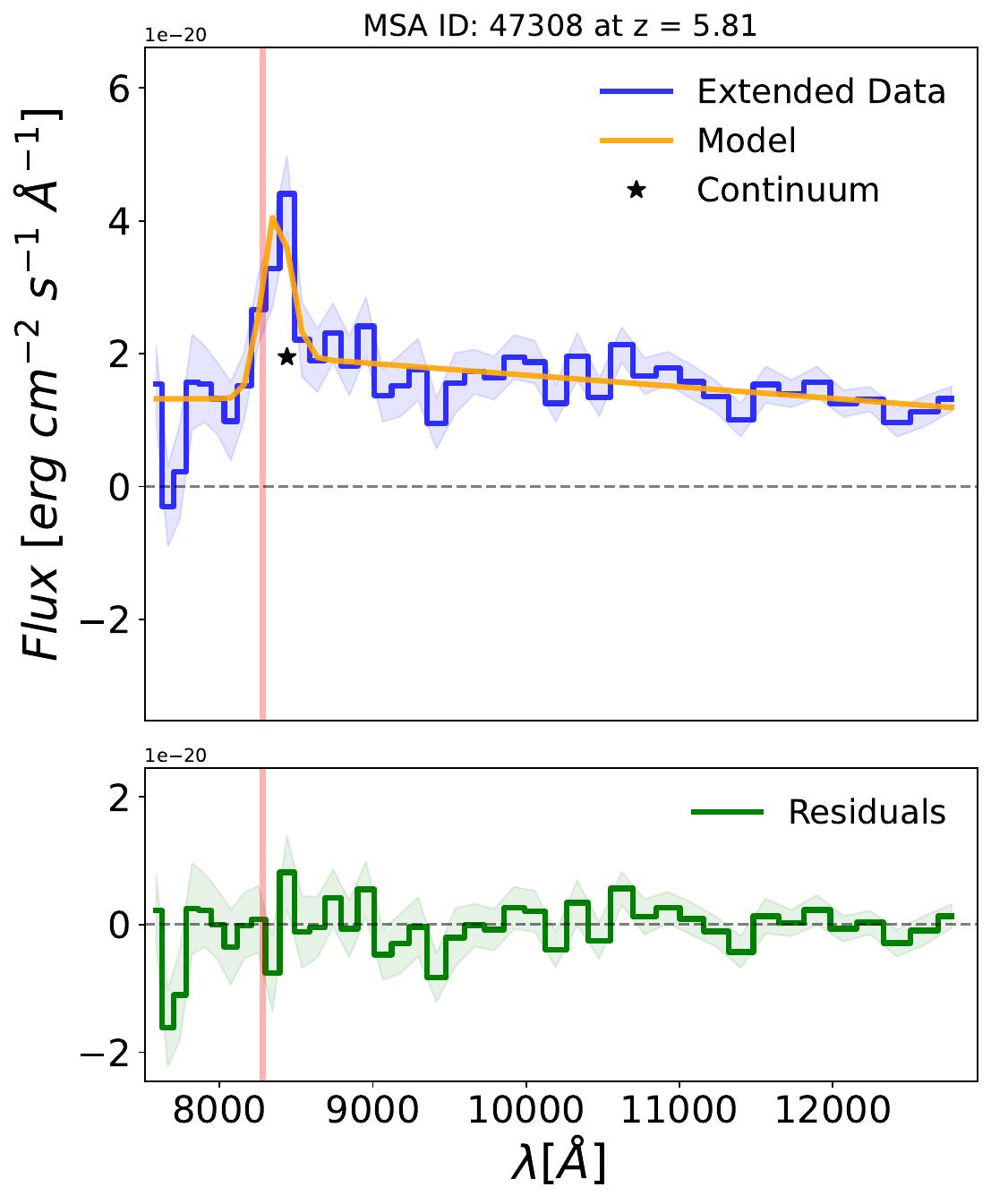}
    \end{subfigure}
    \begin{subfigure}[b]{0.22\textwidth}
        \includegraphics[width=\linewidth]{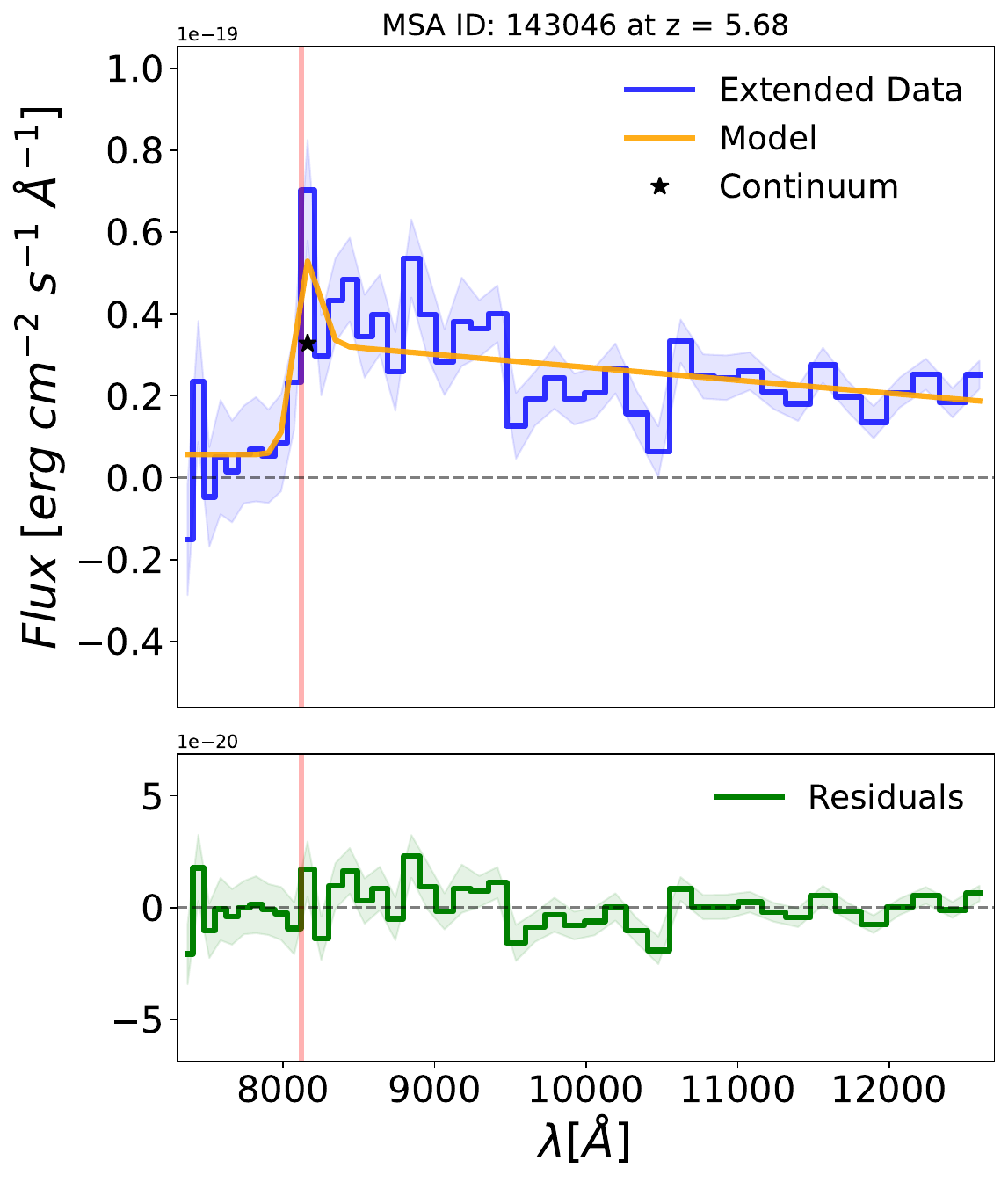}
    \end{subfigure}

    \caption{continued.}
    \label{fig:Lya_emitters_continued1}
\end{figure*}

\begin{figure*}[ht!]
    \ContinuedFloat

    \centering

    \begin{subfigure}[b]{0.22\textwidth}
        \includegraphics[width=\linewidth]{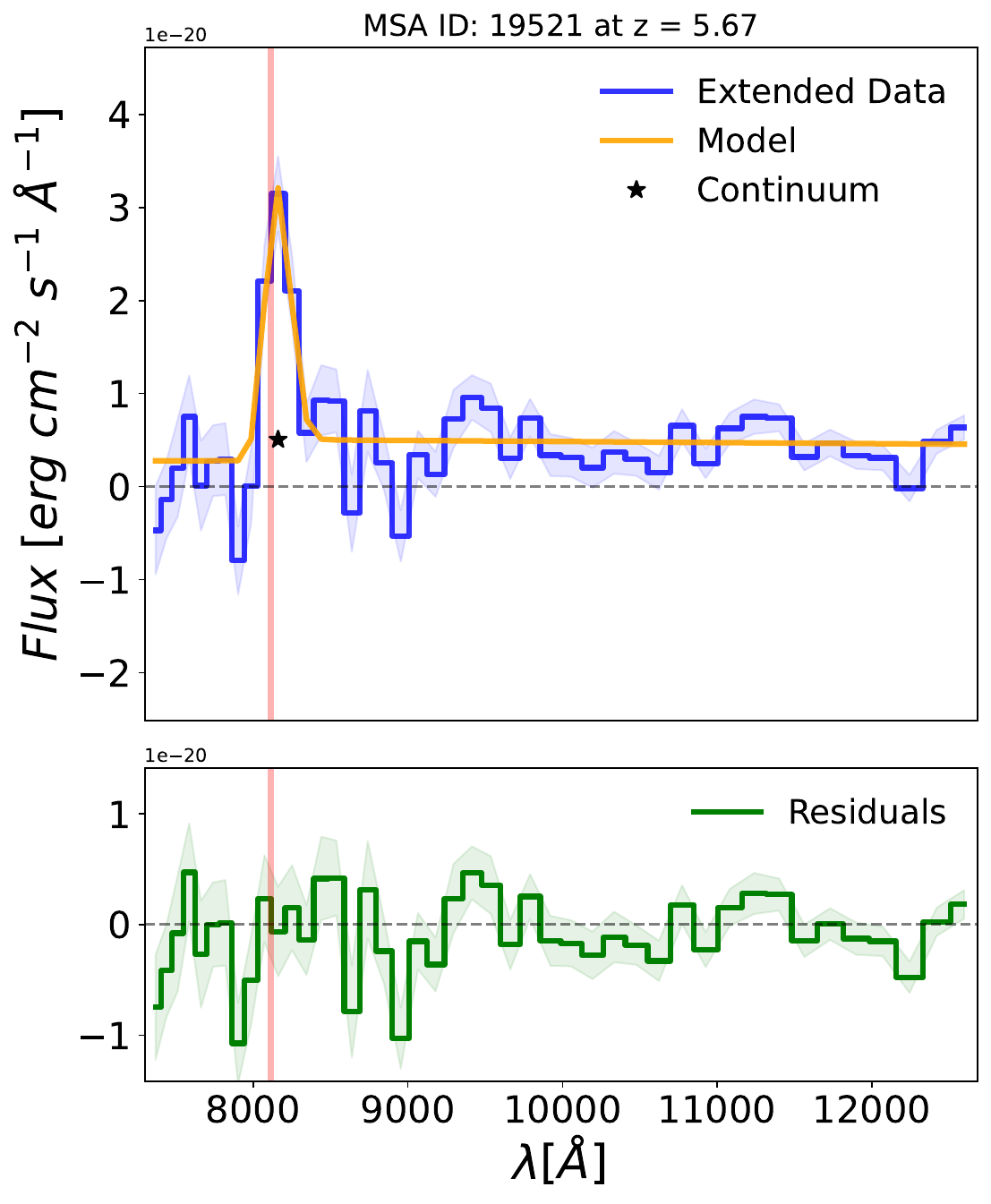}
    \end{subfigure}
    \begin{subfigure}[b]{0.22\textwidth}
        \includegraphics[width=\linewidth]{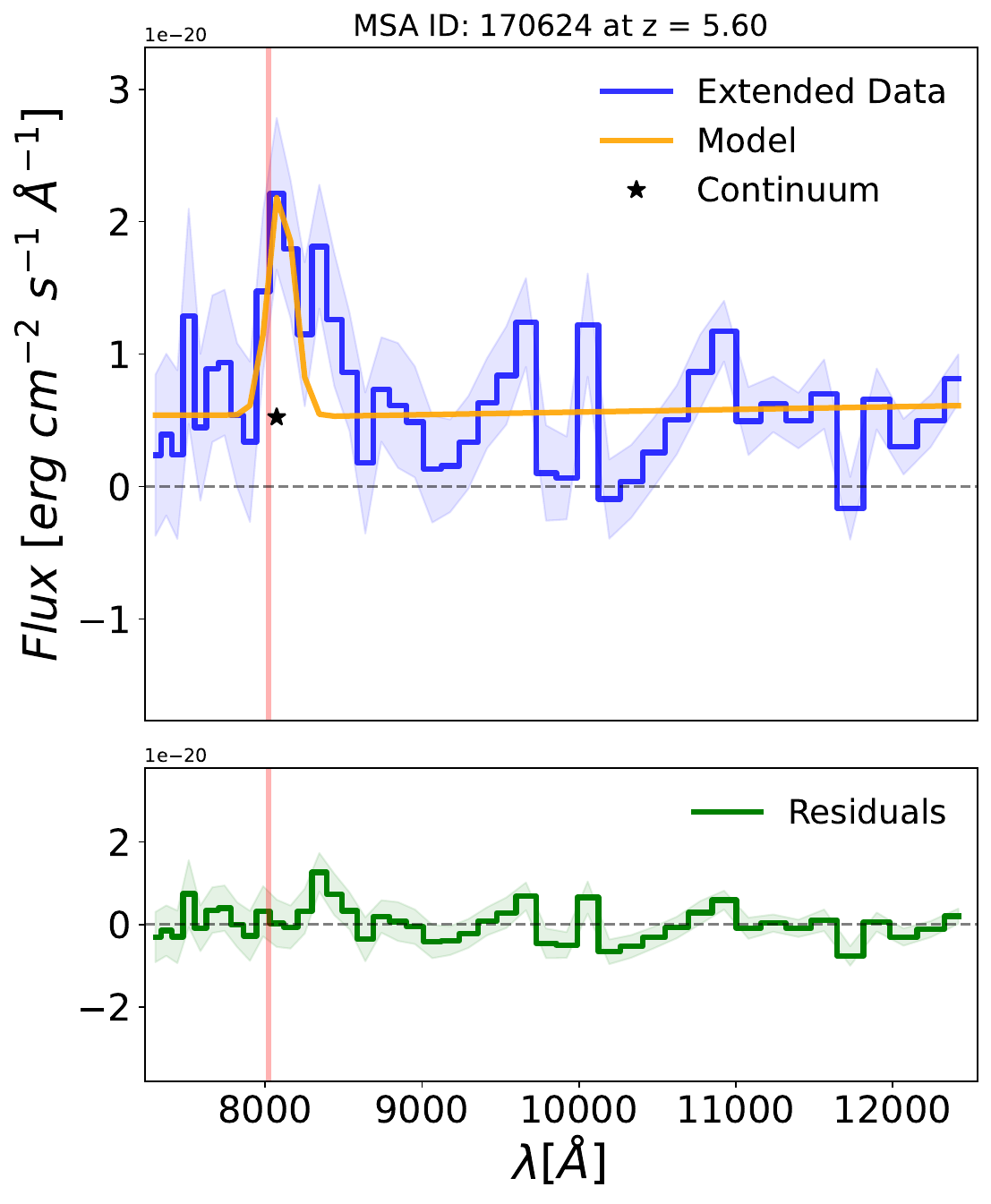}
    \end{subfigure}
    \begin{subfigure}[b]{0.22\textwidth}
        \includegraphics[width=\linewidth]{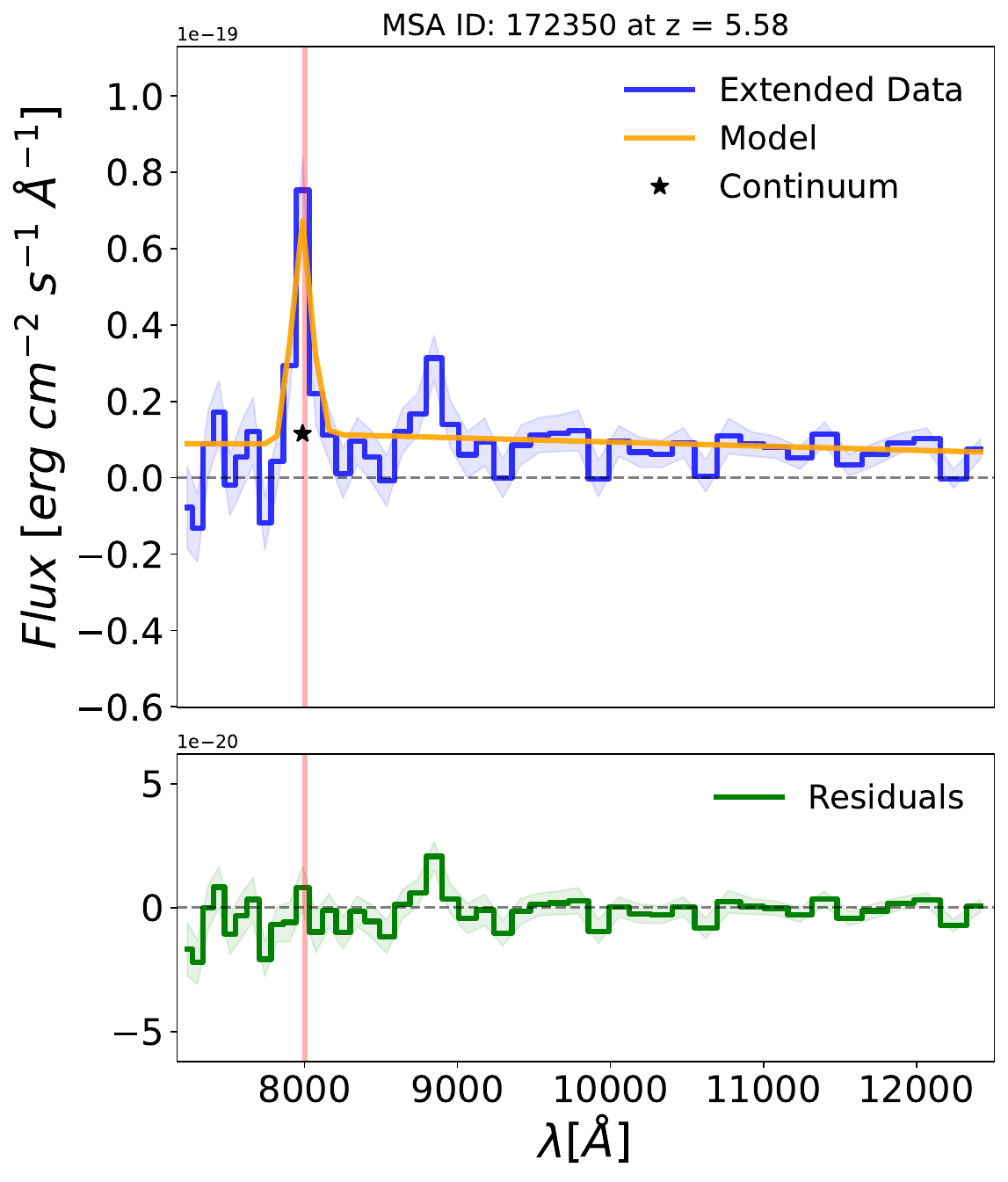}
    \end{subfigure}
    \begin{subfigure}[b]{0.22\textwidth}
        \includegraphics[width=\linewidth]{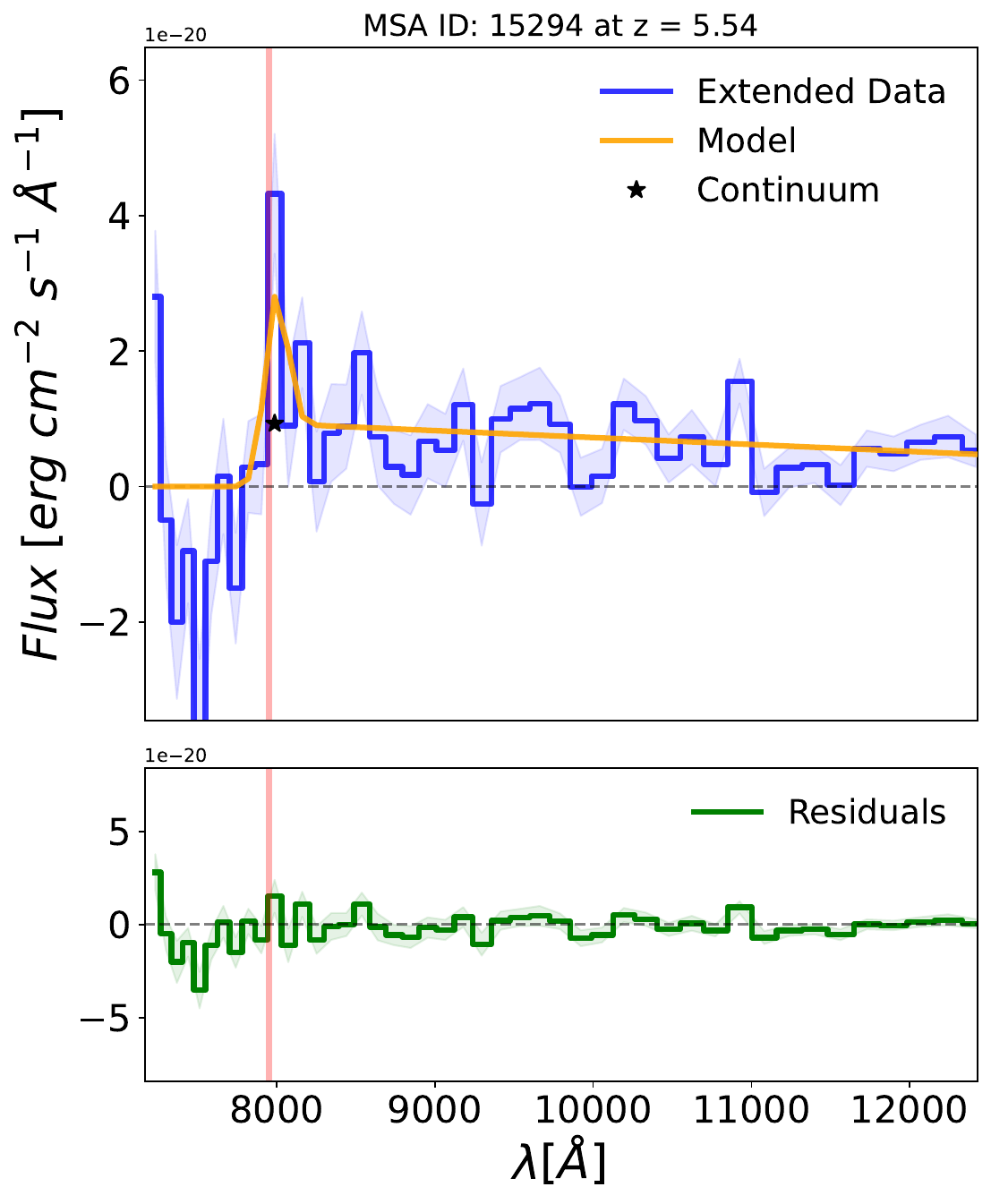}
    \end{subfigure}

    \vspace{0.1cm}

    \begin{subfigure}[b]{0.22\textwidth}
        \includegraphics[width=\linewidth]{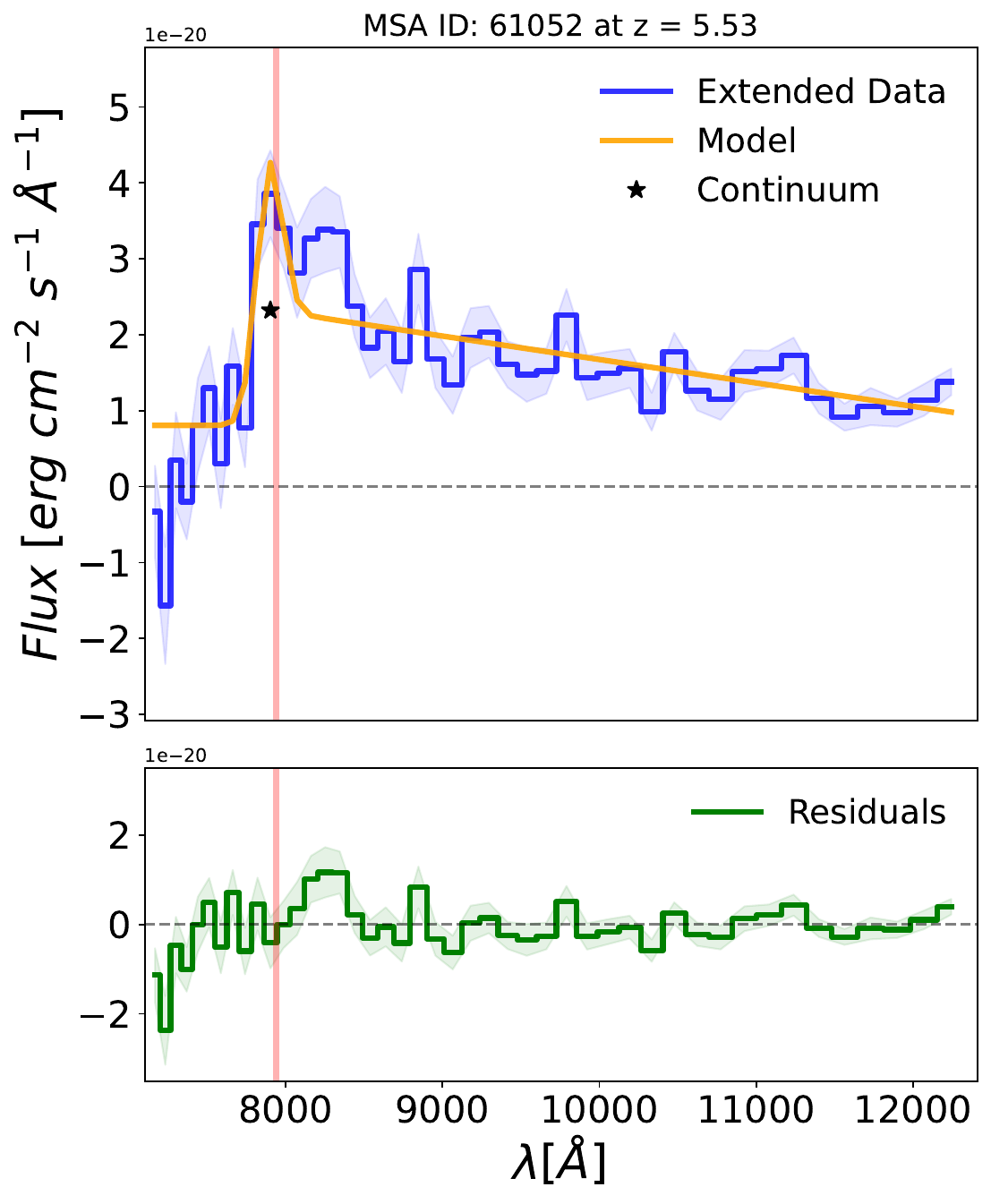}
    \end{subfigure}
    \begin{subfigure}[b]{0.22\textwidth}
        \includegraphics[width=\linewidth]{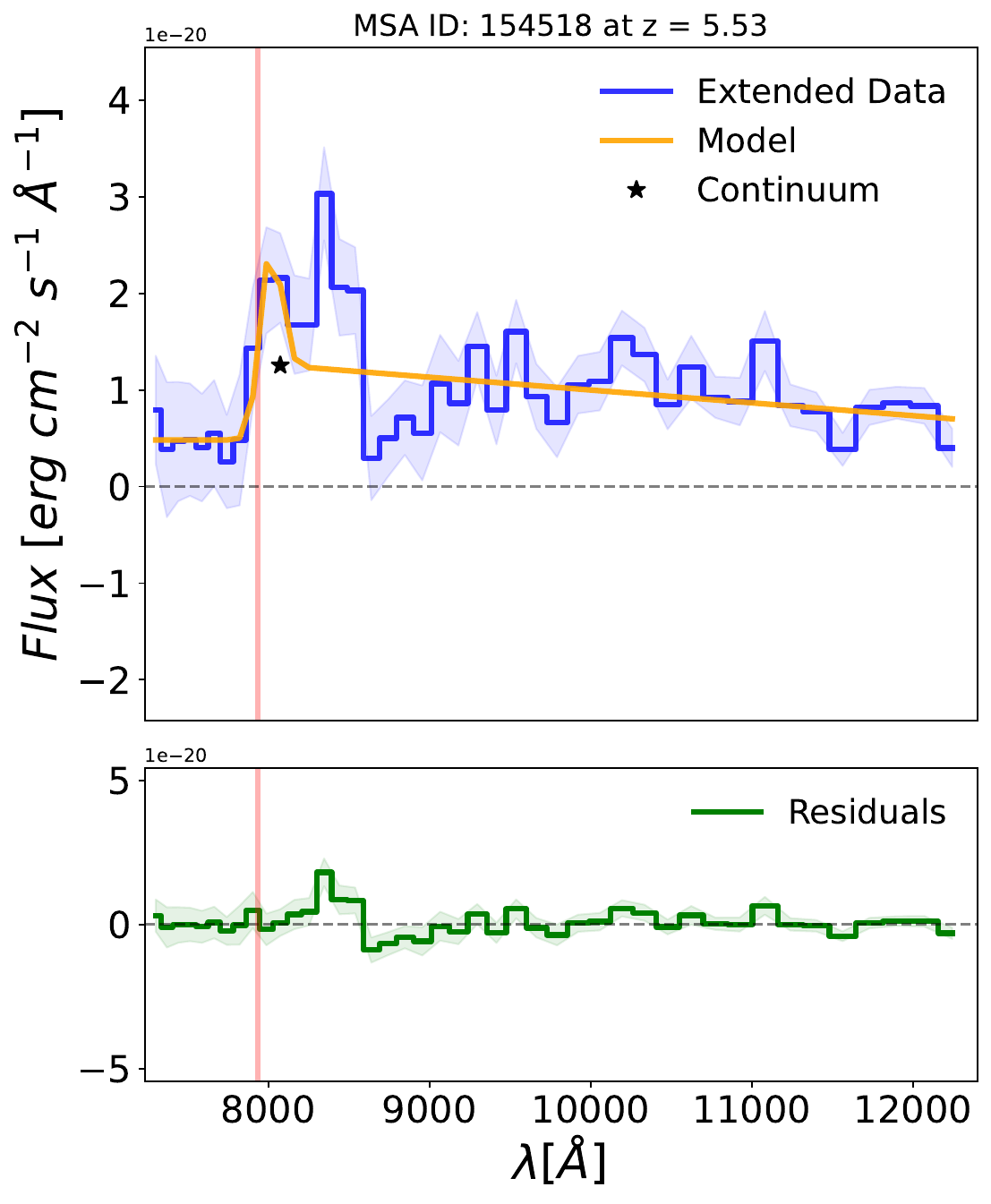}
    \end{subfigure}
    \begin{subfigure}[b]{0.22\textwidth}
        \includegraphics[width=\linewidth]{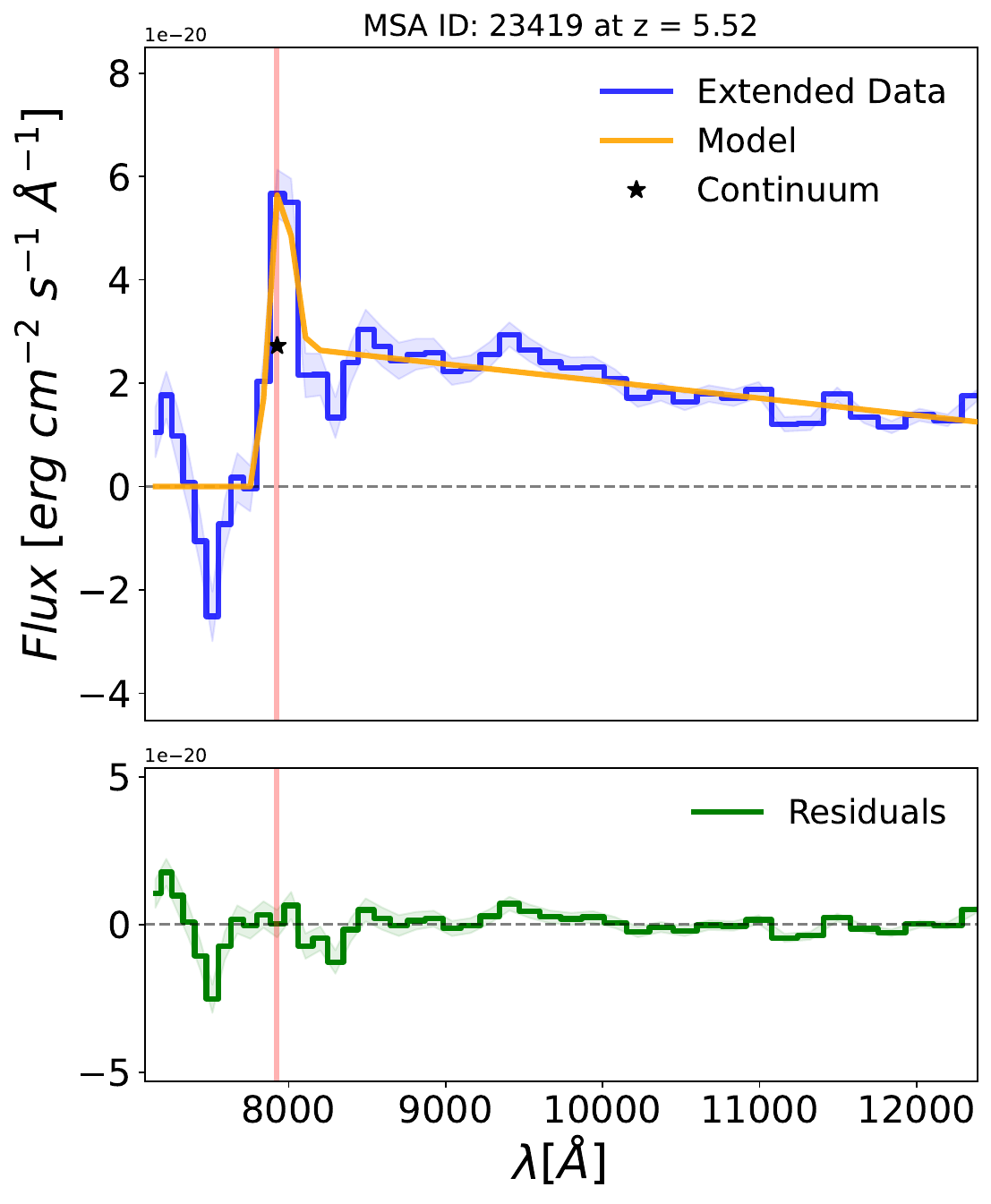}
    \end{subfigure}
    \begin{subfigure}[b]{0.22\textwidth}
        \includegraphics[width=\linewidth]{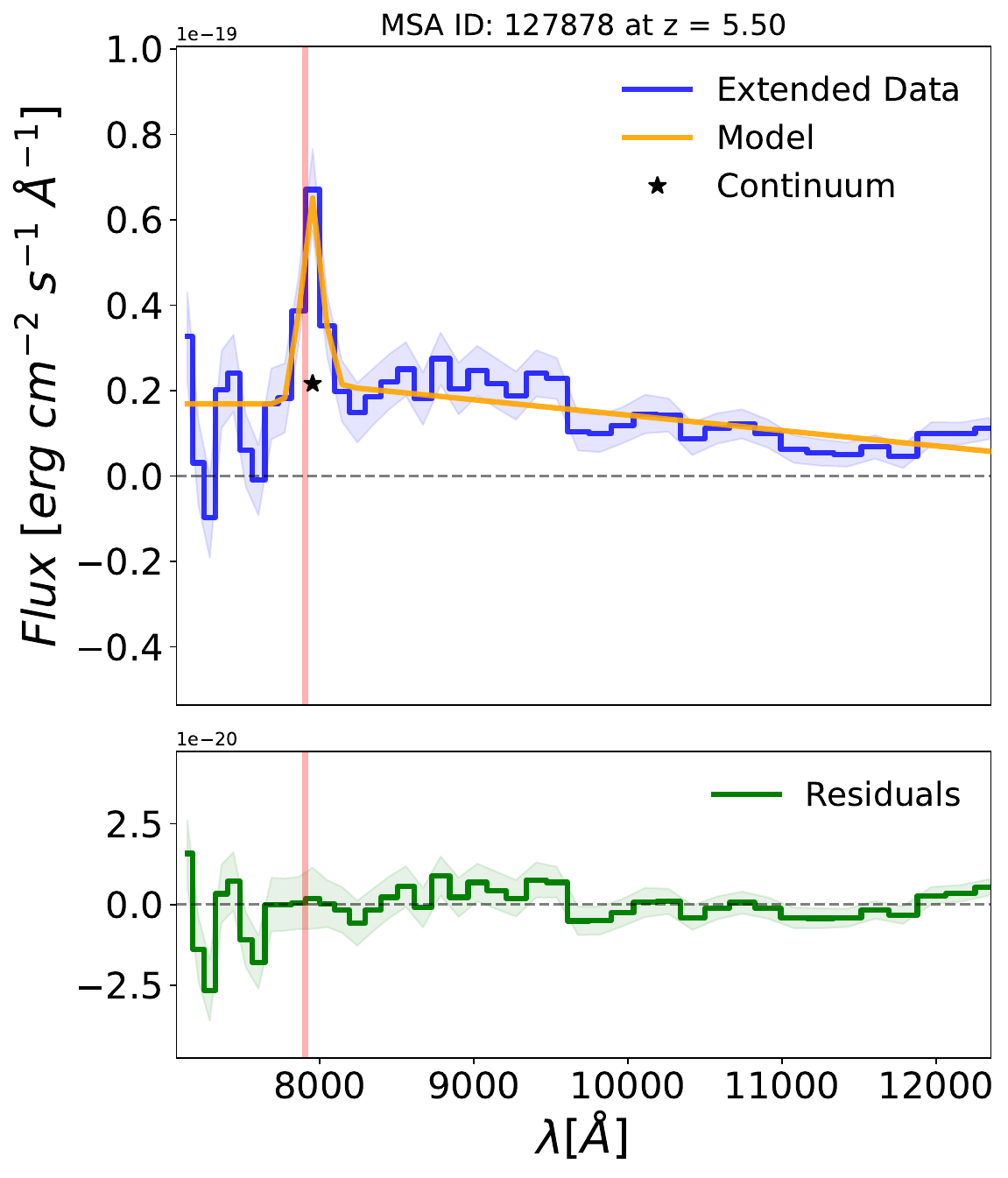}
    \end{subfigure}

    \vspace{0.1cm}

    \begin{subfigure}[b]{0.22\textwidth}
        \includegraphics[width=\linewidth]{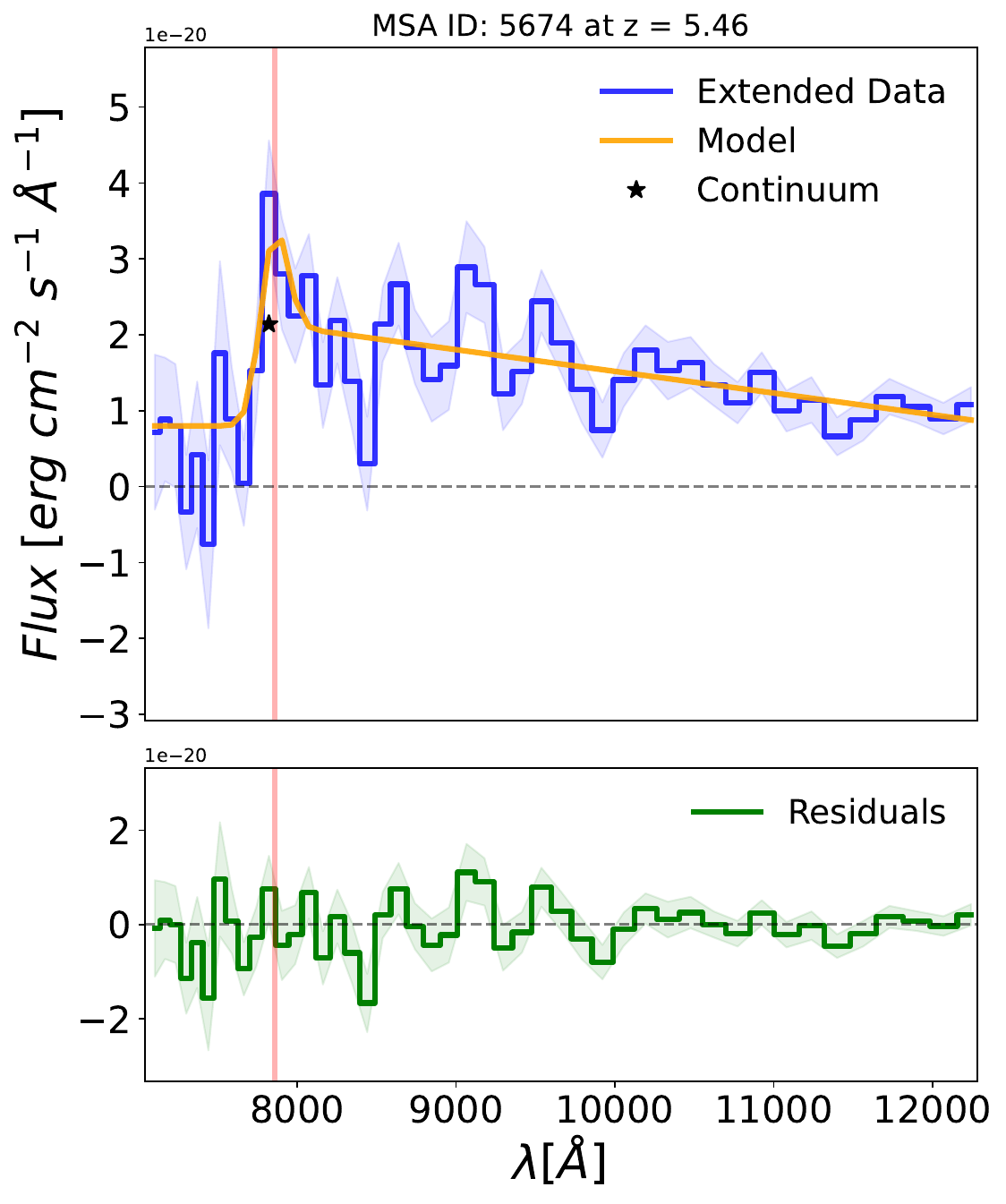}
    \end{subfigure}
    \begin{subfigure}[b]{0.22\textwidth}
        \includegraphics[width=\linewidth]{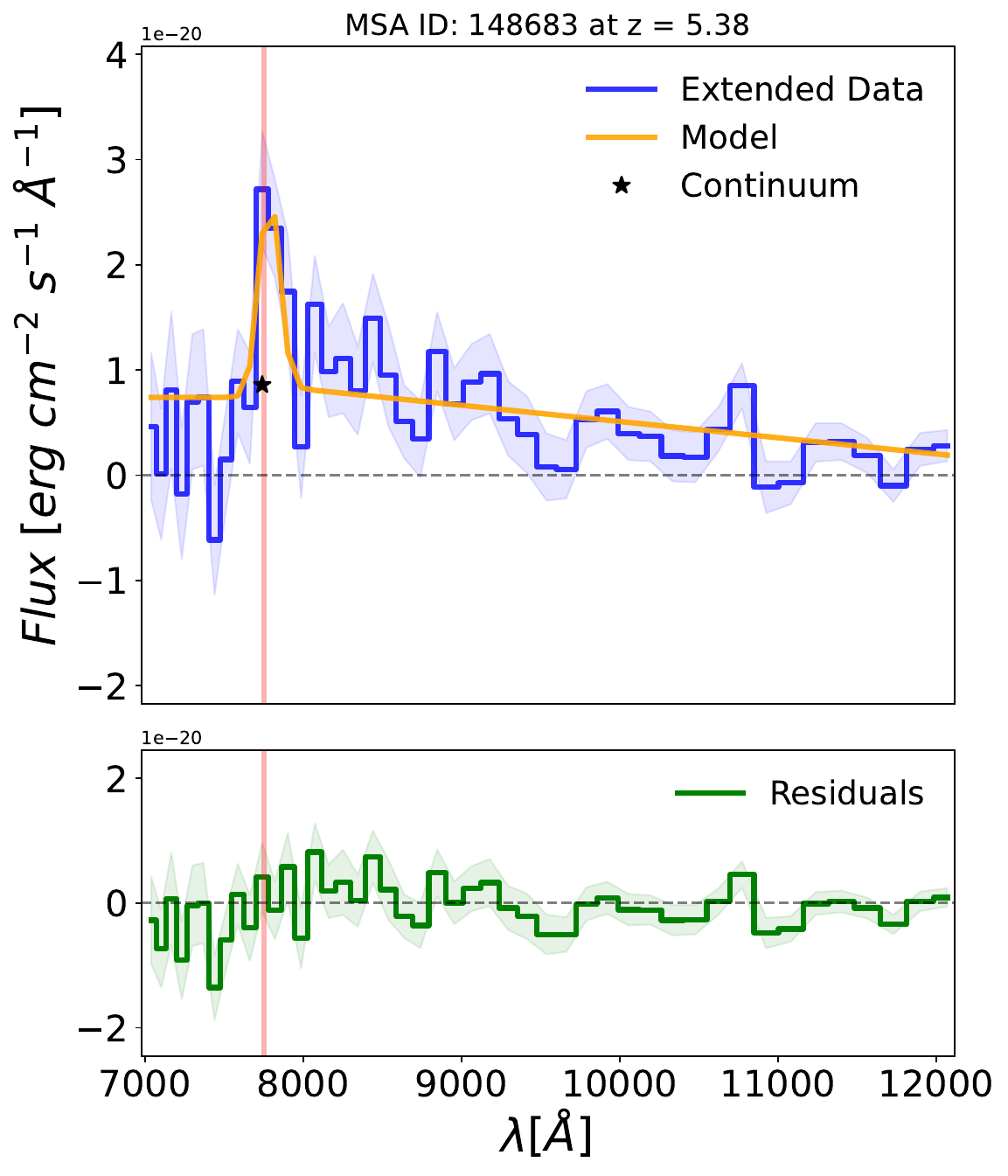}
    \end{subfigure}
    \begin{subfigure}[b]{0.22\textwidth}
        \includegraphics[width=\linewidth]{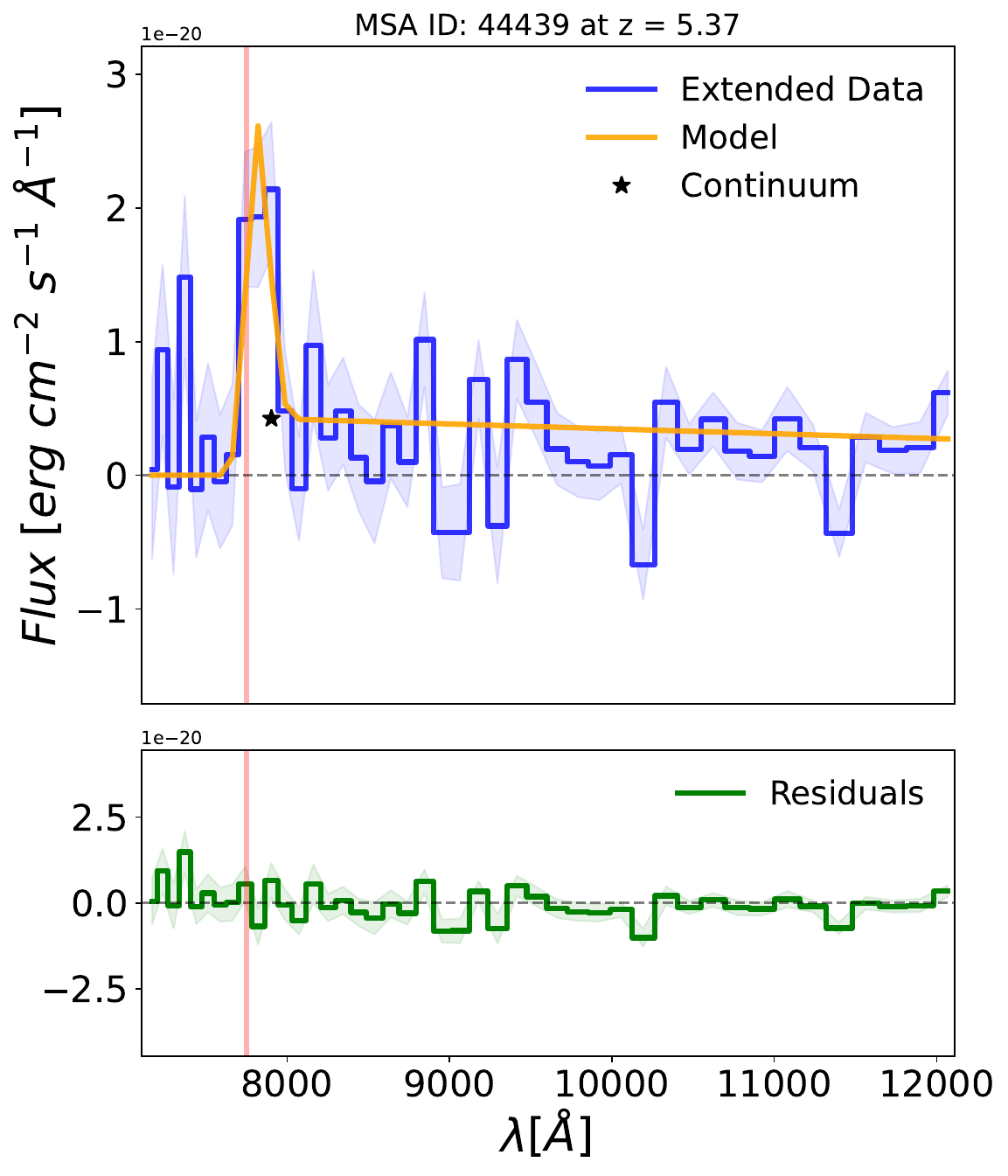}
    \end{subfigure}
    \begin{subfigure}[b]{0.22\textwidth}
        \includegraphics[width=\linewidth]{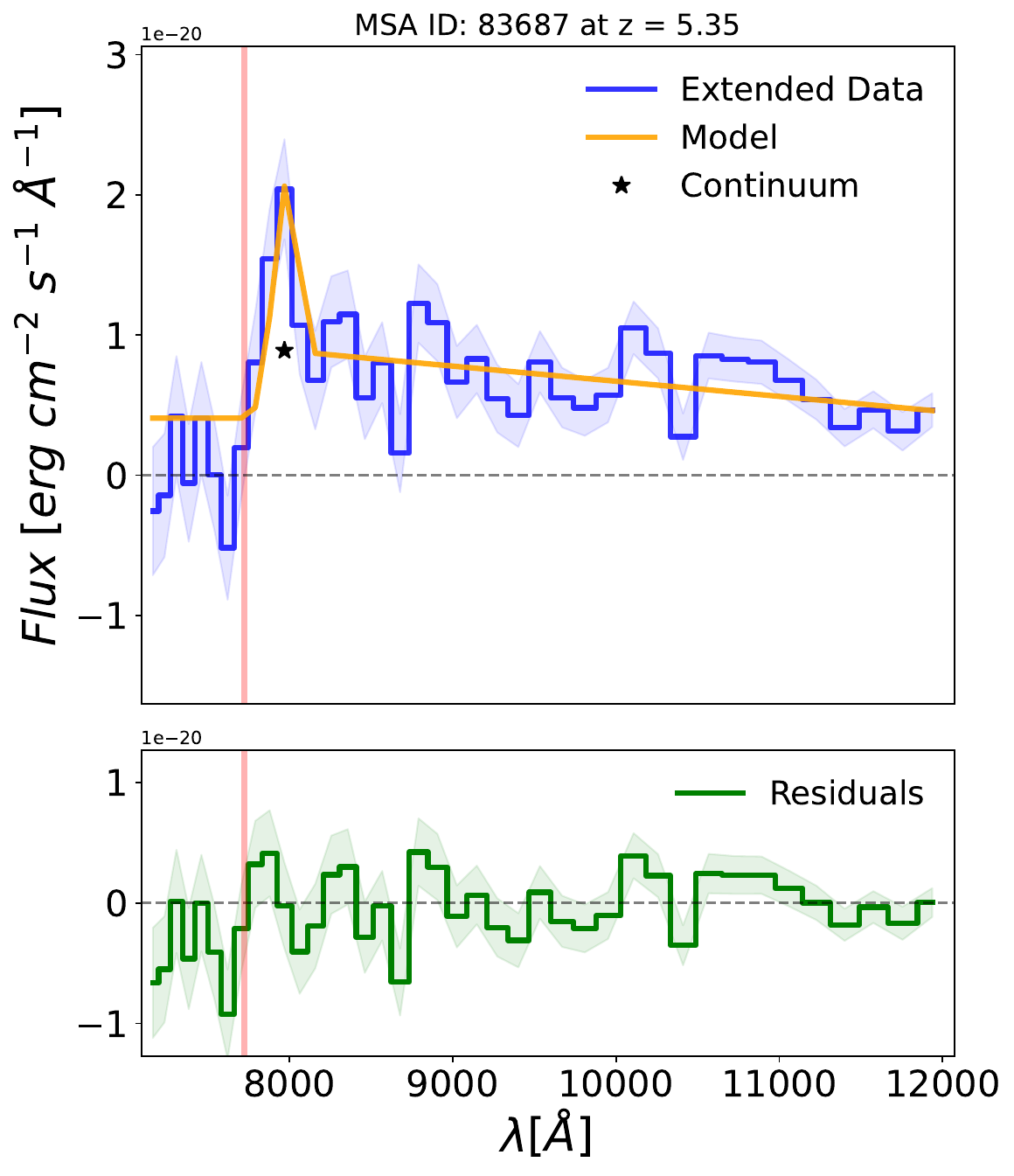}
    \end{subfigure}

    \vspace{0.1cm}

    \begin{subfigure}[b]{0.22\textwidth}
        \includegraphics[width=\linewidth]{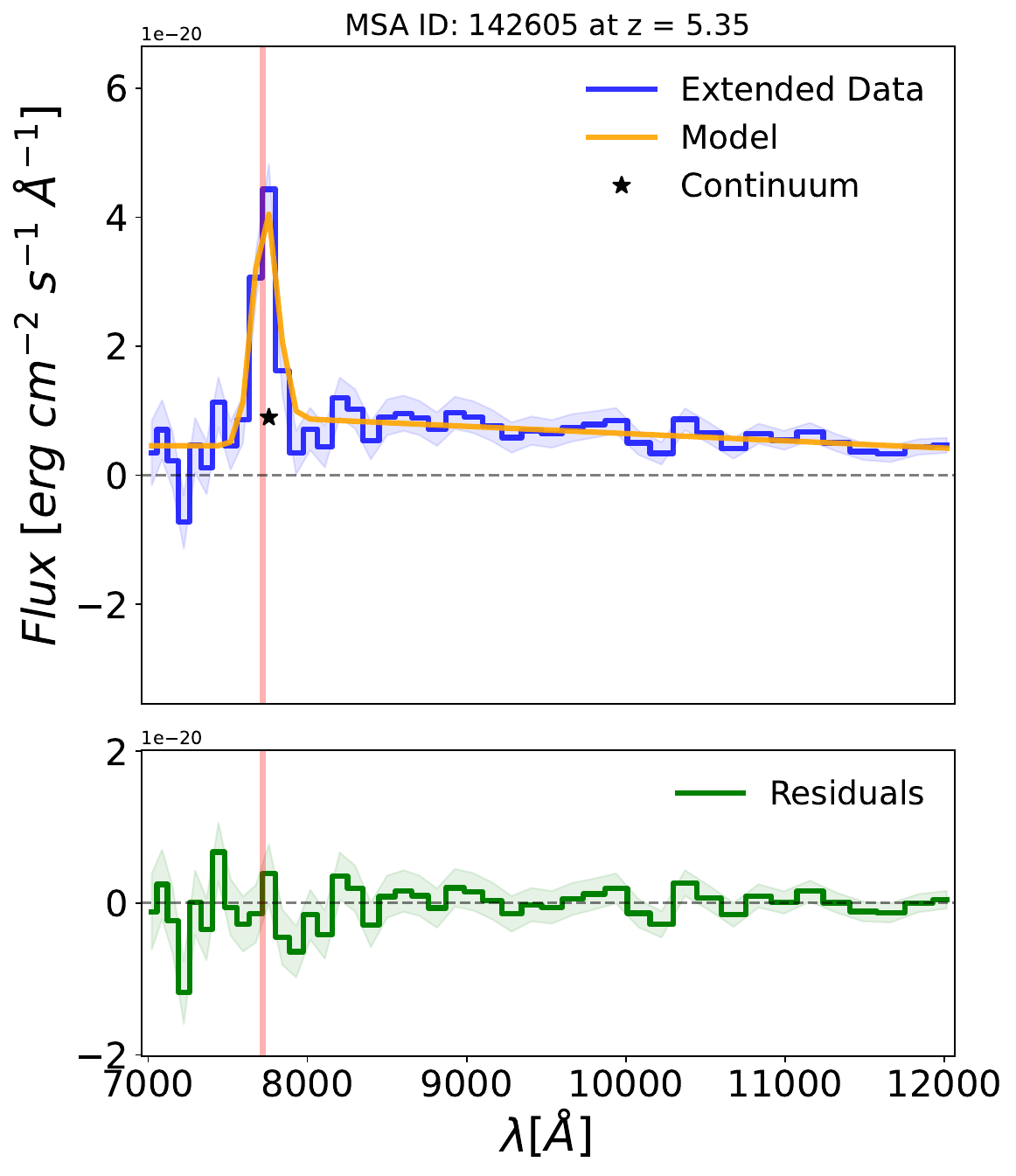}
    \end{subfigure}
    \begin{subfigure}[b]{0.22\textwidth}
        \includegraphics[width=\linewidth]{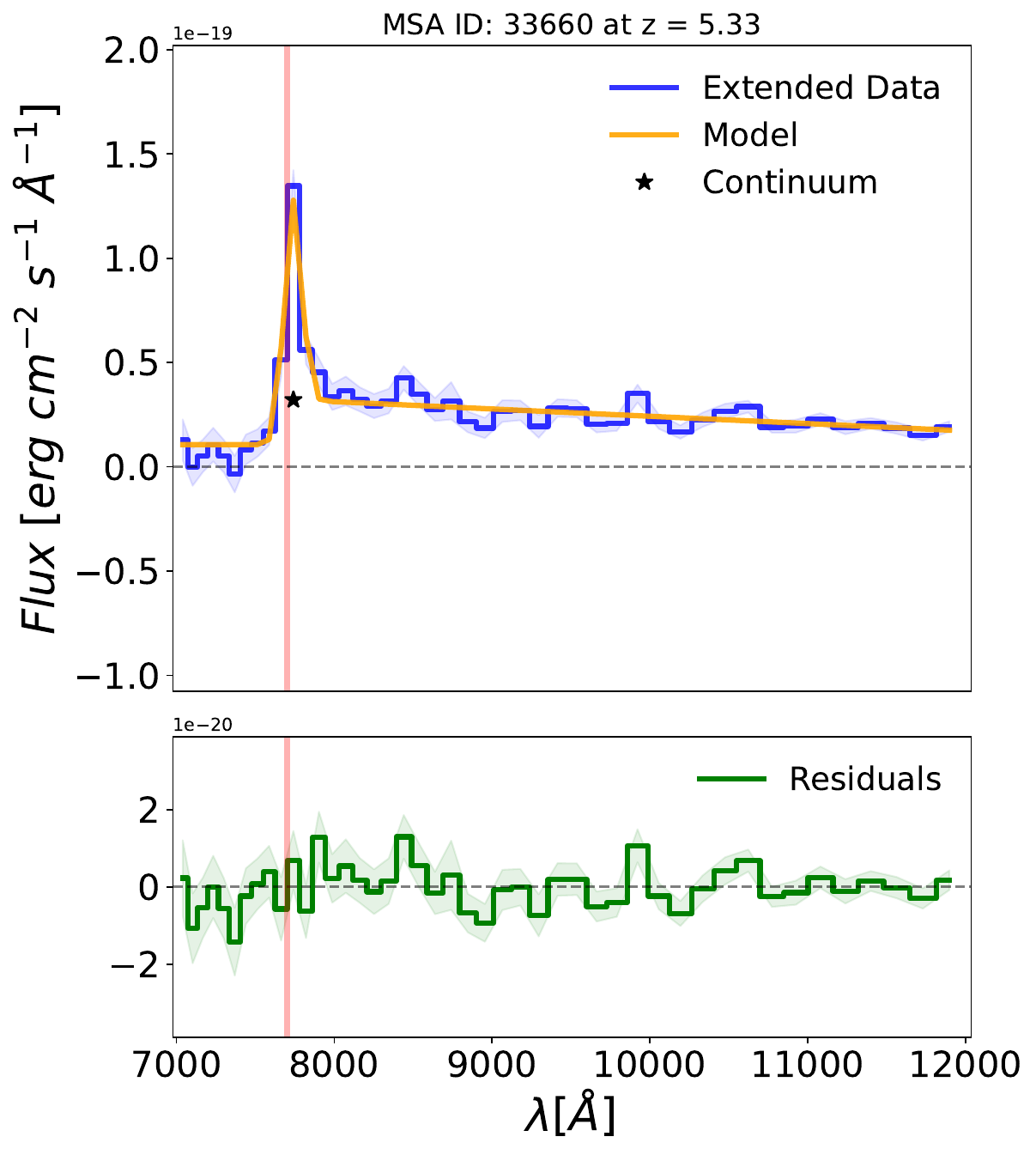}
    \end{subfigure}
    \begin{subfigure}[b]{0.22\textwidth}
        \includegraphics[width=\linewidth]{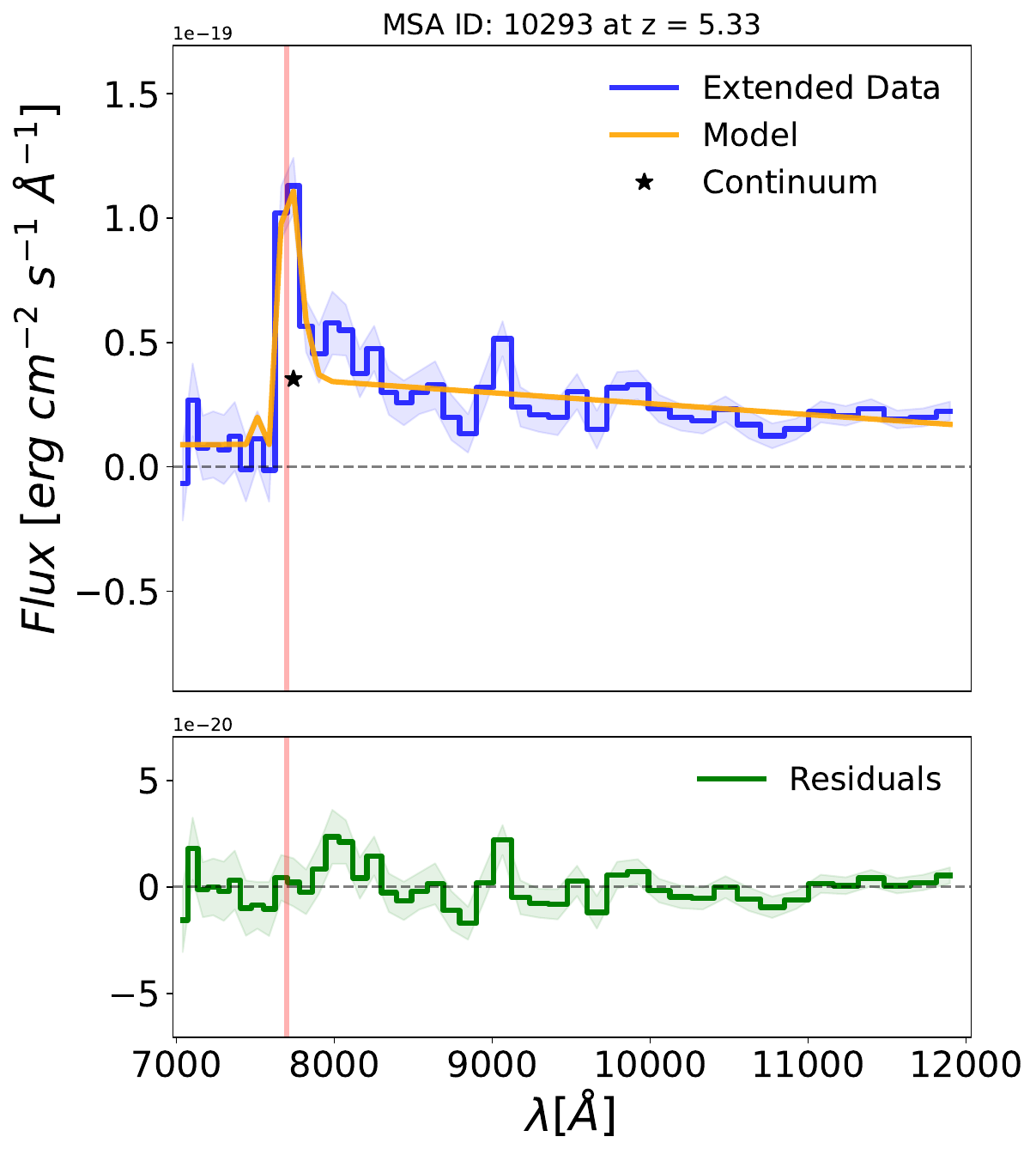}
    \end{subfigure}
    \begin{subfigure}[b]{0.22\textwidth}
        \includegraphics[width=\linewidth]{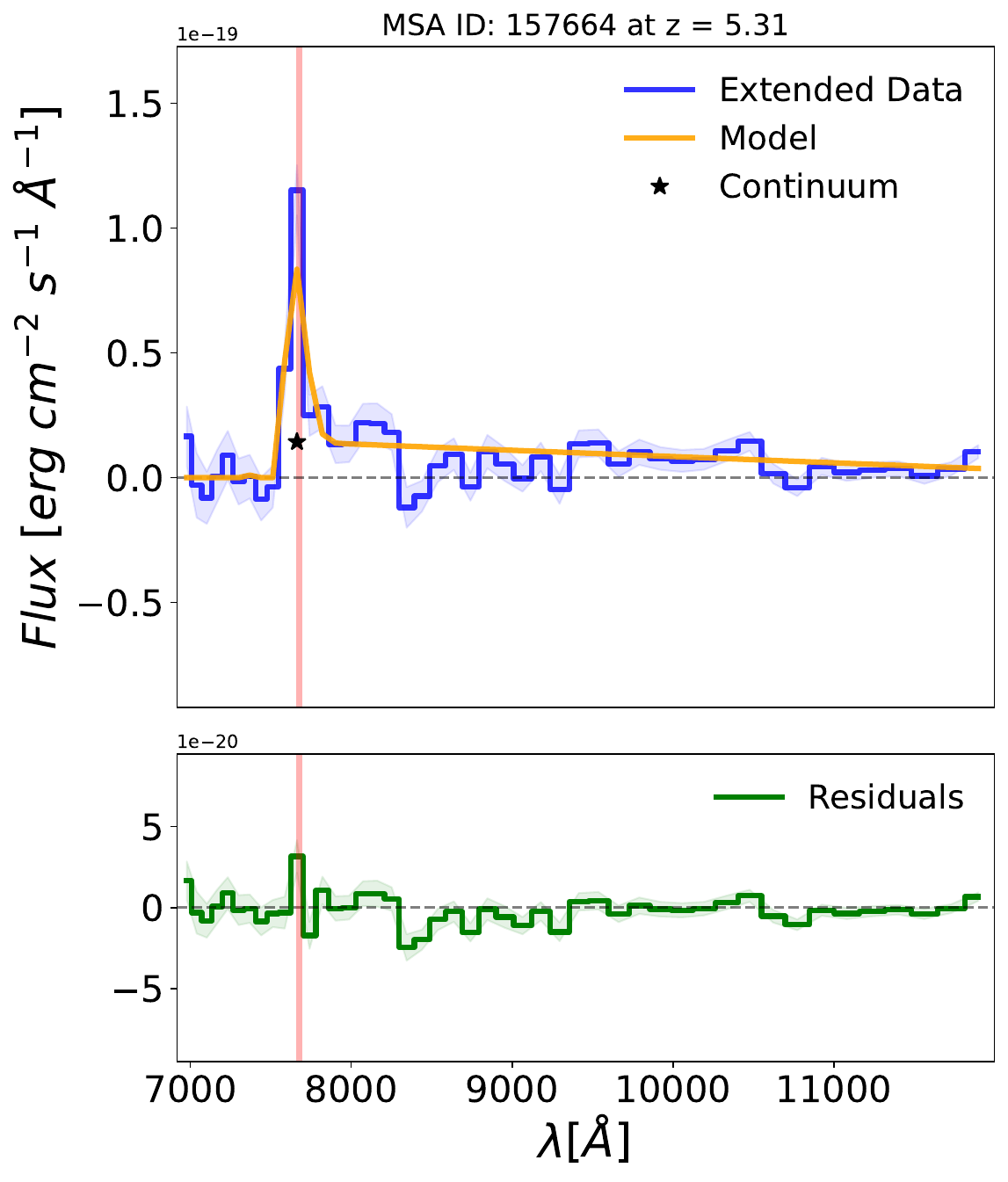}
    \end{subfigure}

    \caption{continued.}
    \label{fig:Lya_emitters_continued2}
\end{figure*}

\begin{figure*}[ht!]
    \ContinuedFloat

    \centering

    \begin{subfigure}[b]{0.22\textwidth}
        \includegraphics[width=\linewidth]{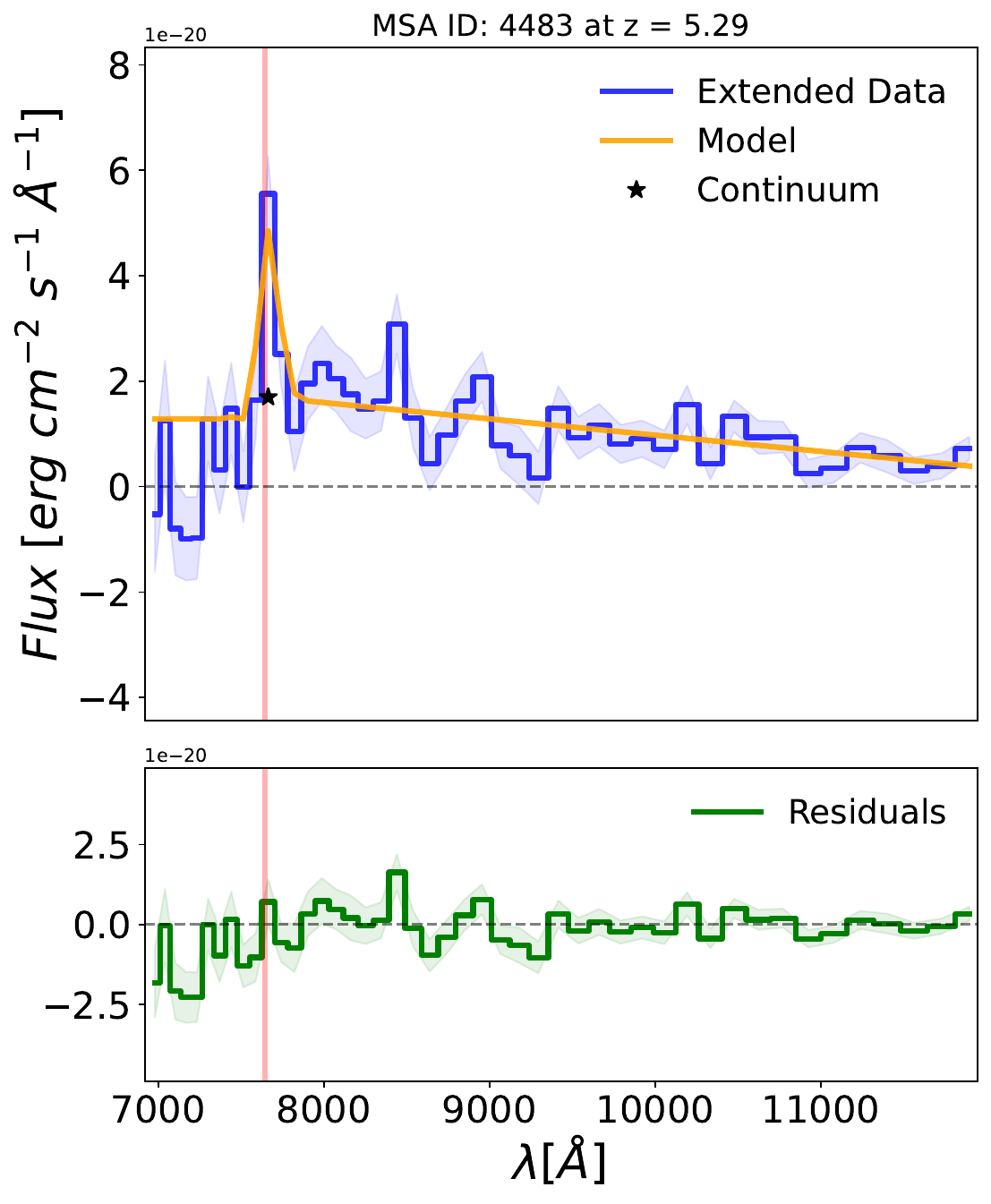}
    \end{subfigure}
    \begin{subfigure}[b]{0.22\textwidth}
        \includegraphics[width=\linewidth]{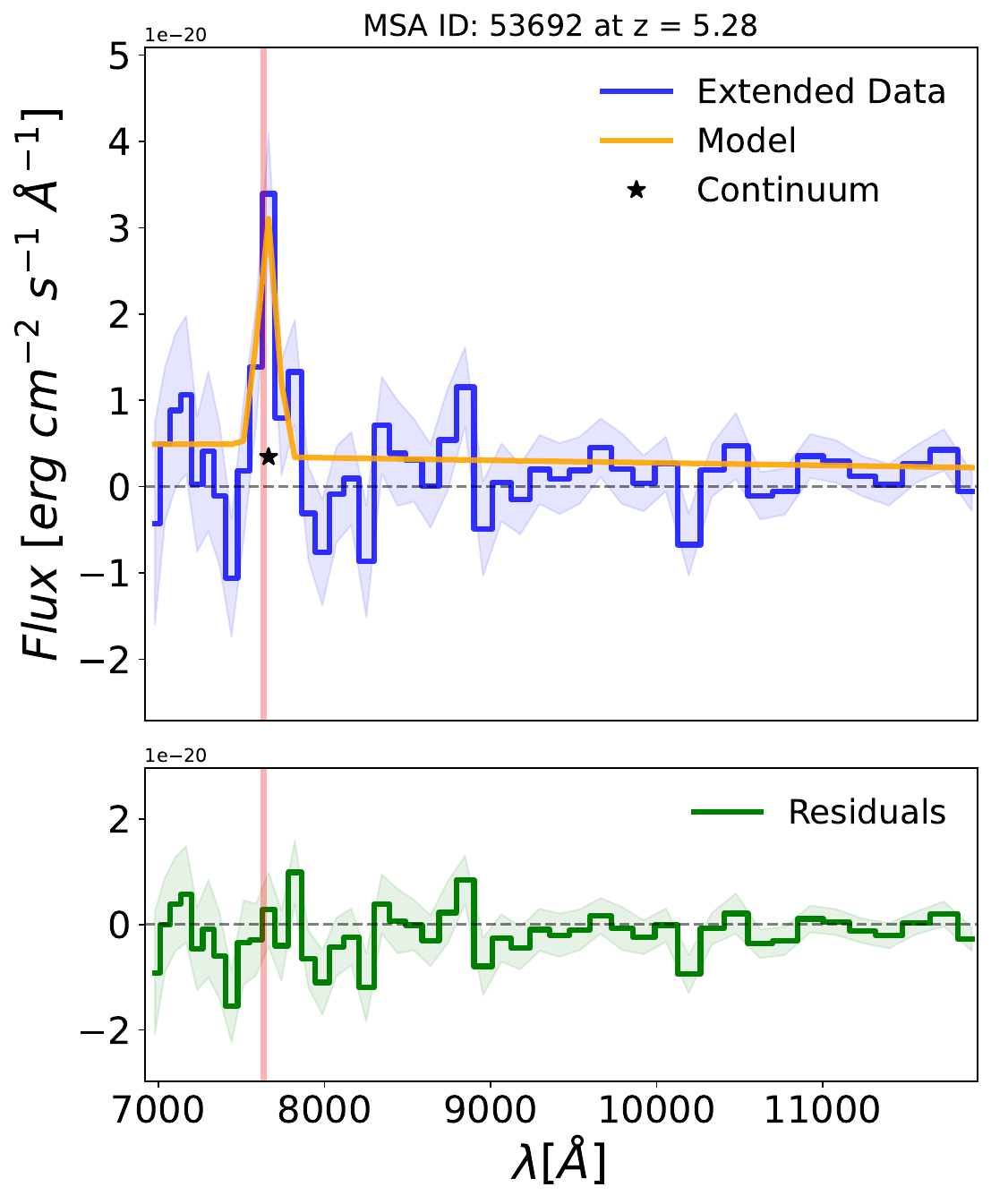}
    \end{subfigure}
    \begin{subfigure}[b]{0.22\textwidth}
        \includegraphics[width=\linewidth]{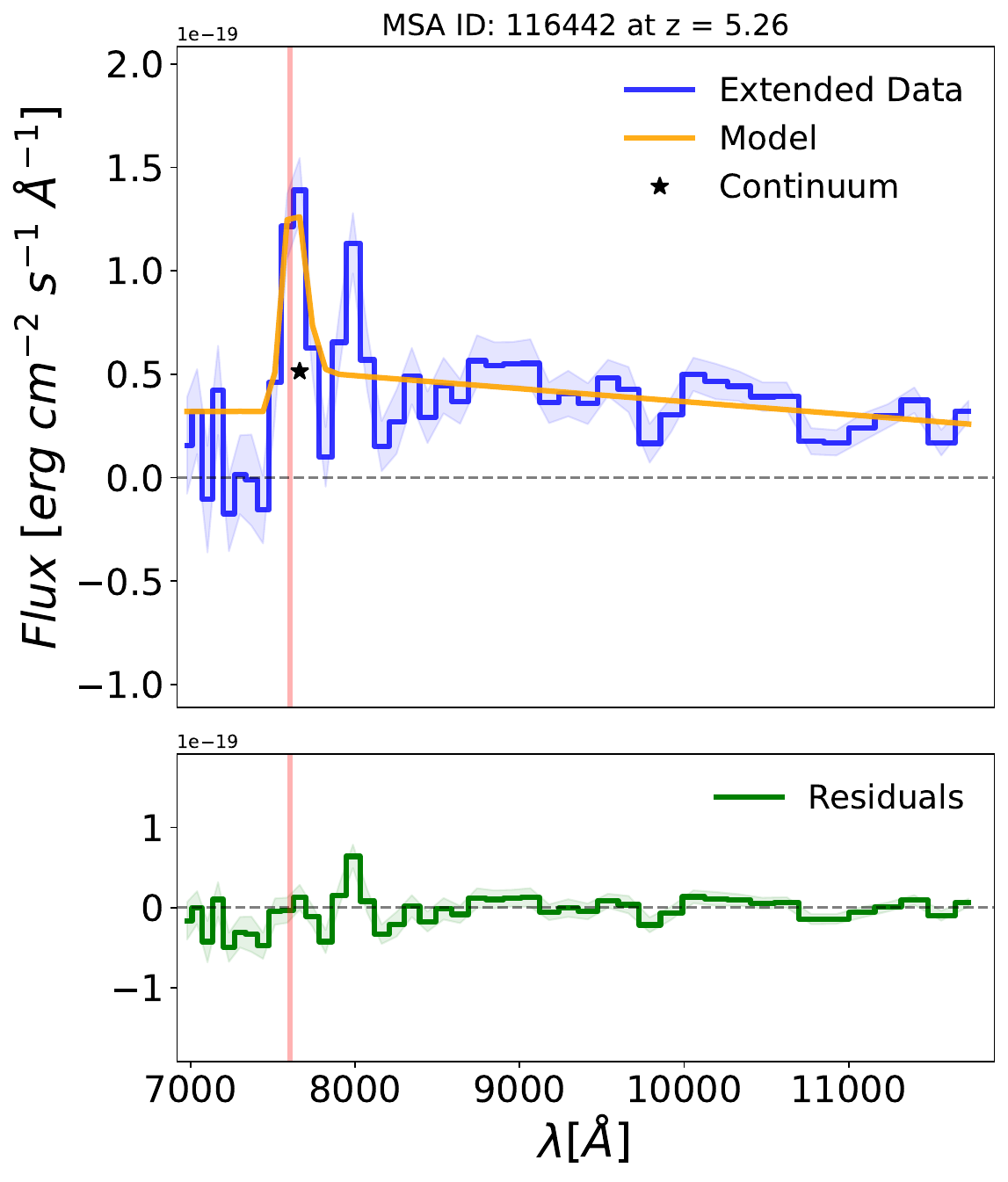}
    \end{subfigure}
    \begin{subfigure}[b]{0.22\textwidth}
        \includegraphics[width=\linewidth]{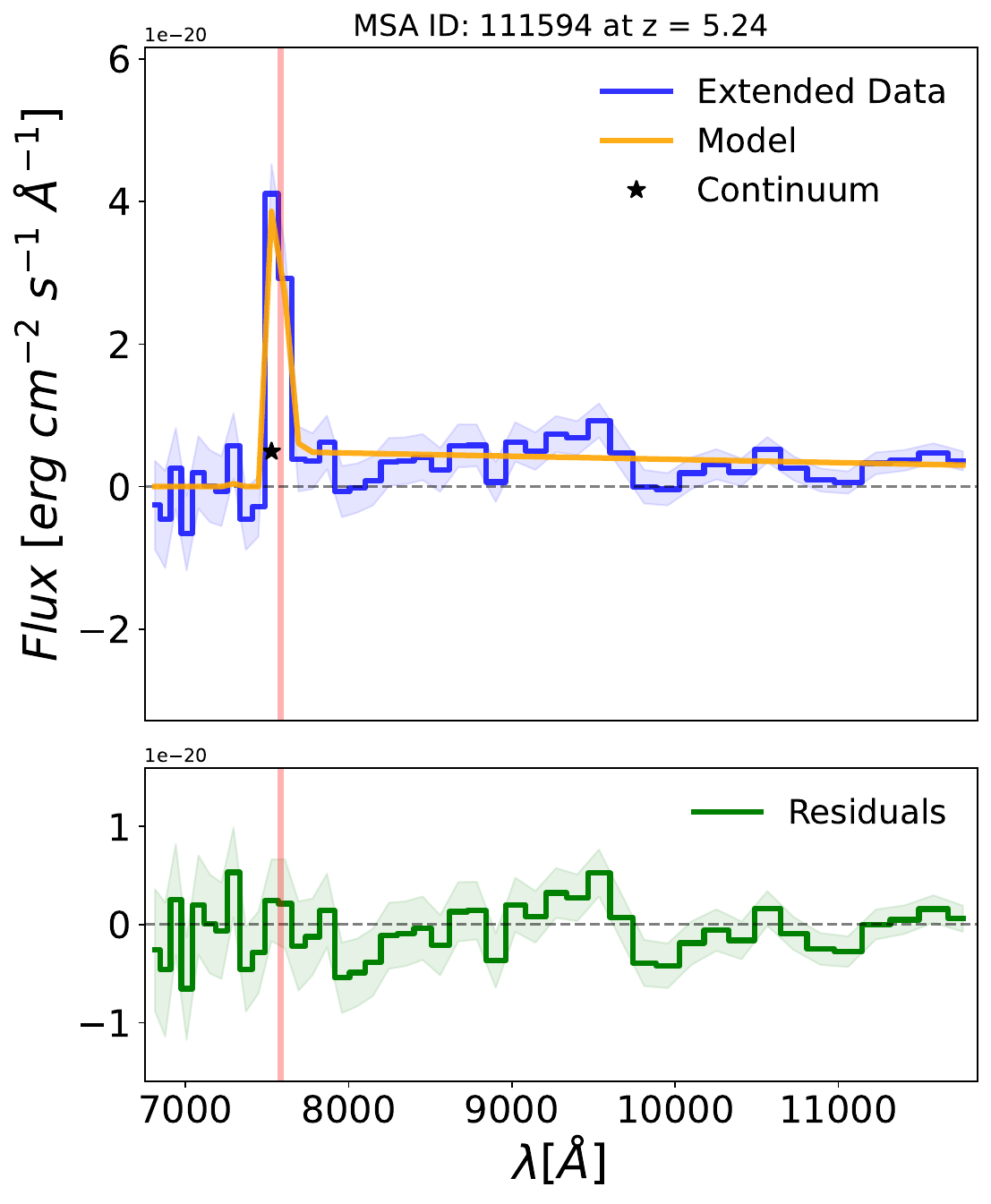}
    \end{subfigure}

    \vspace{0.1cm}

    \begin{subfigure}[b]{0.22\textwidth}
        \includegraphics[width=\linewidth]{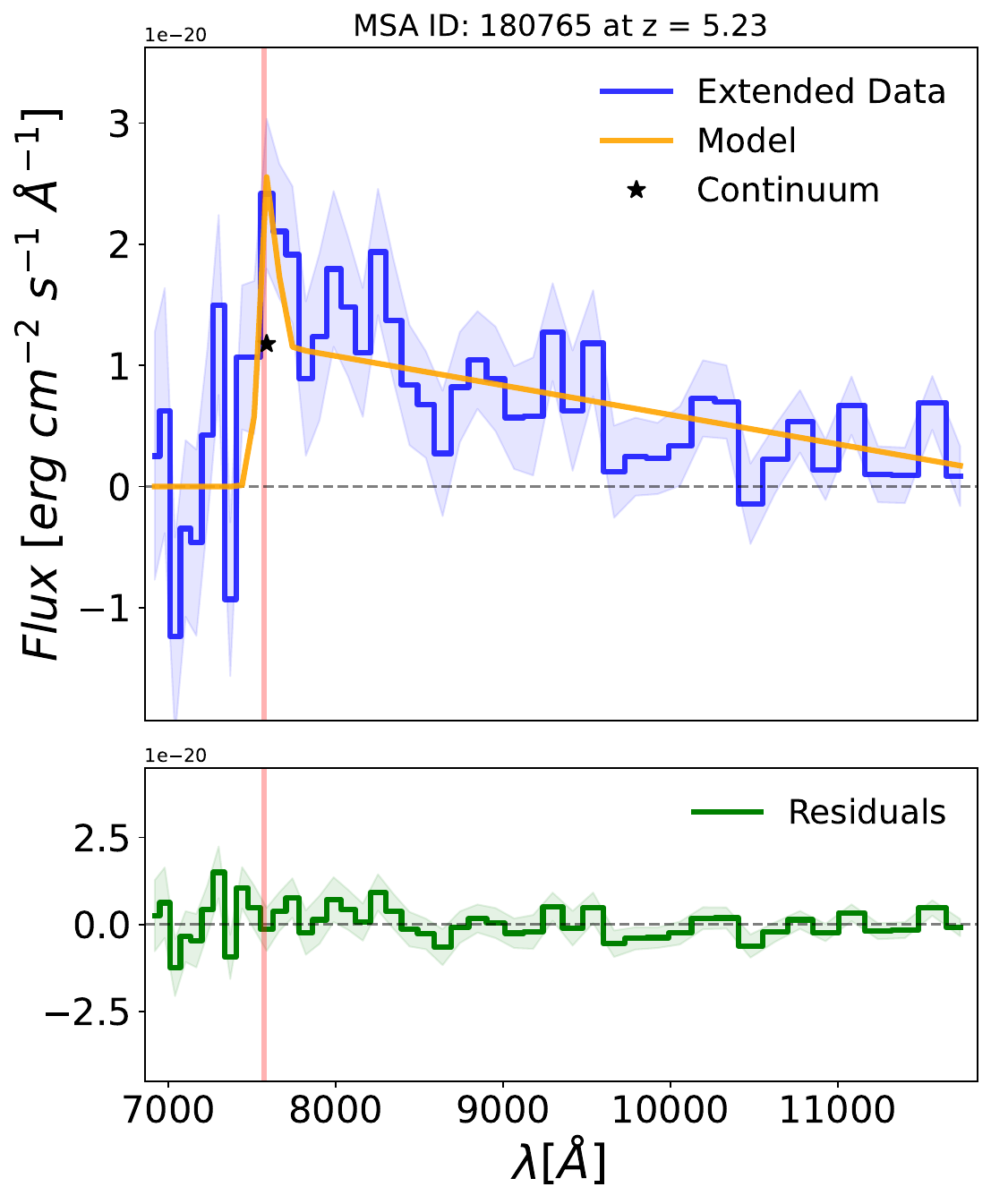}
    \end{subfigure}
    \begin{subfigure}[b]{0.22\textwidth}
        \includegraphics[width=\linewidth]{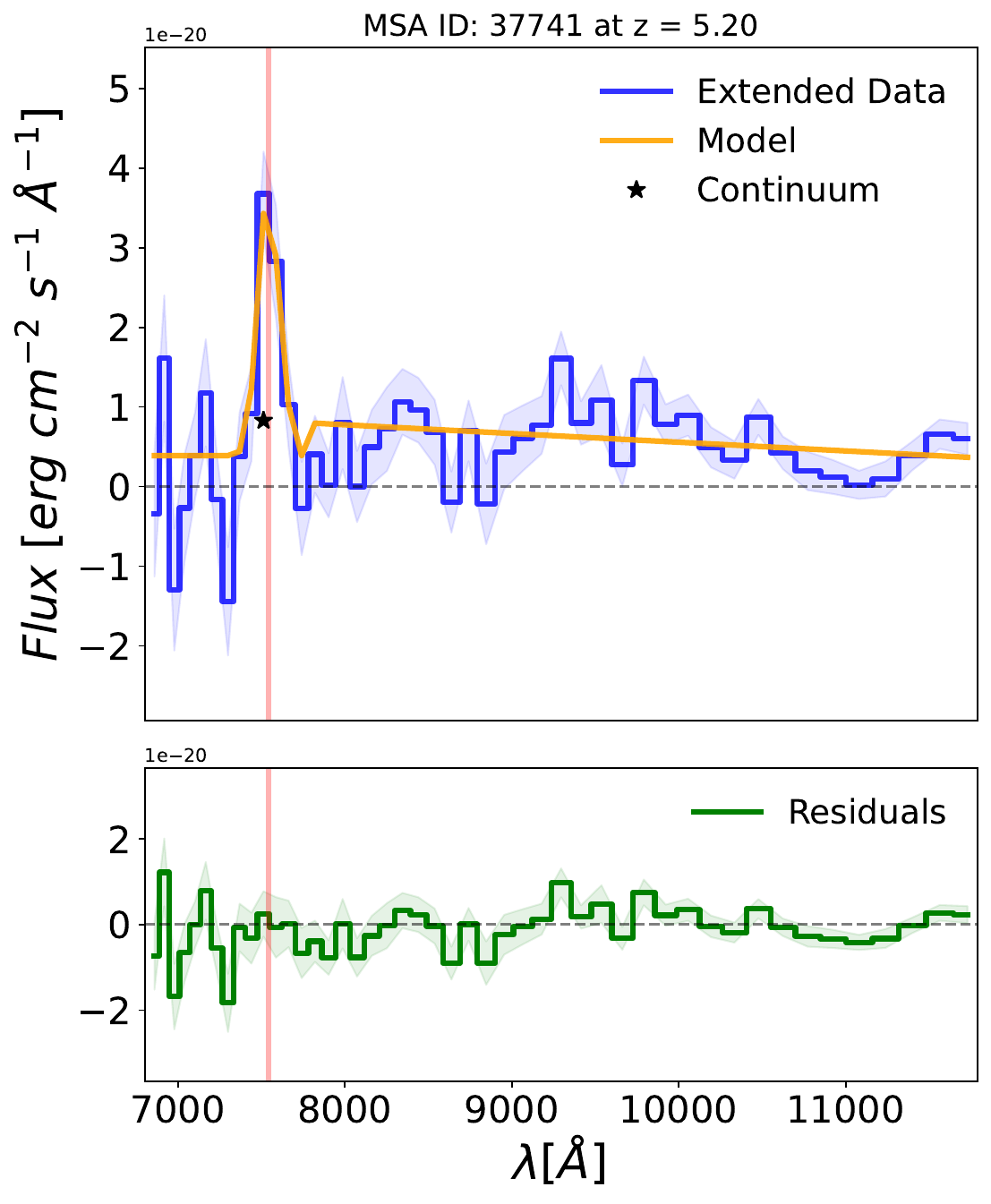}
    \end{subfigure}
    \begin{subfigure}[b]{0.22\textwidth}
        \includegraphics[width=\linewidth]{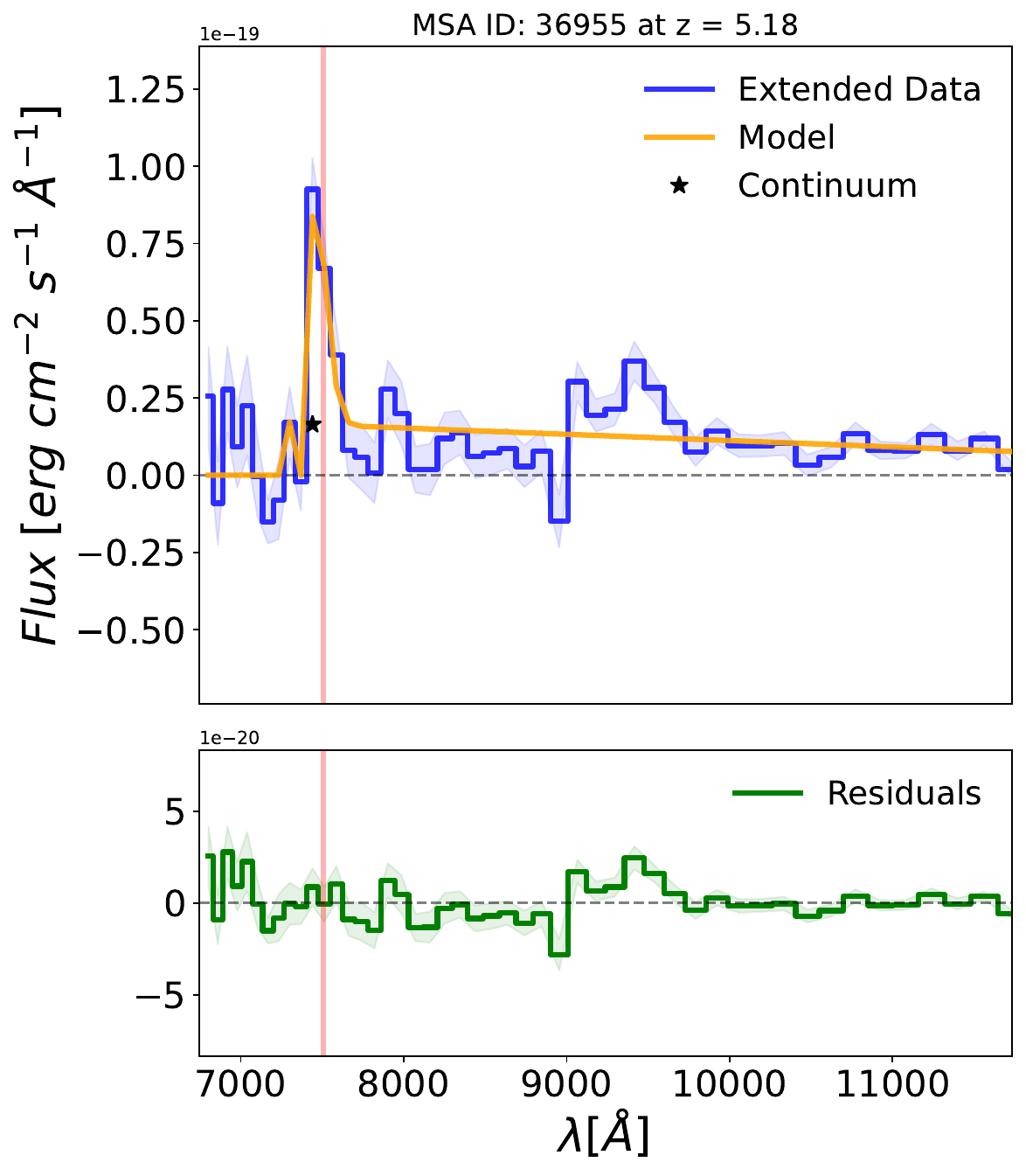}
    \end{subfigure}
    \begin{subfigure}[b]{0.22\textwidth}
        \includegraphics[width=\linewidth]{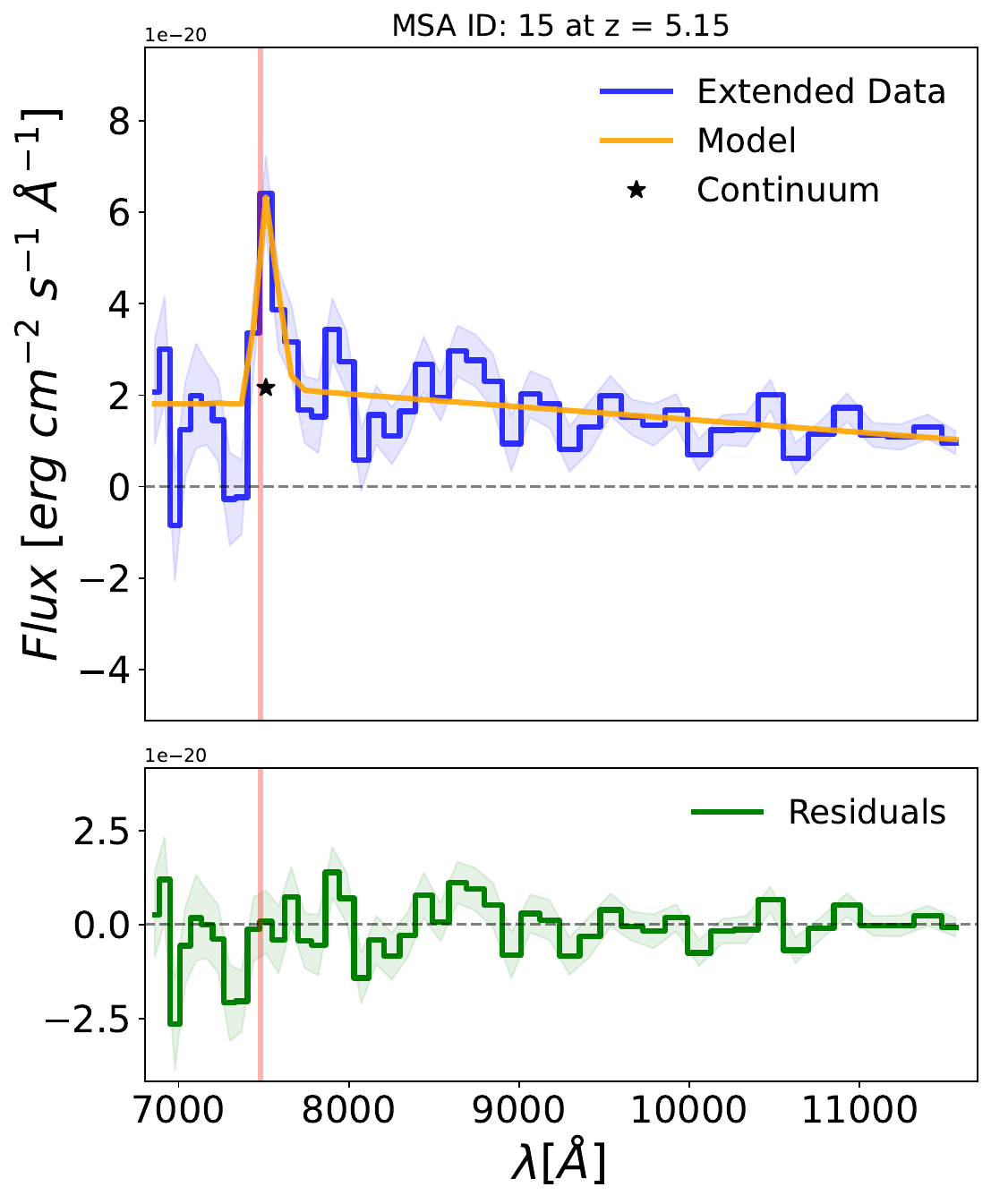}
    \end{subfigure}

    \vspace{0.1cm}

    \begin{subfigure}[b]{0.22\textwidth}
        \includegraphics[width=\linewidth]{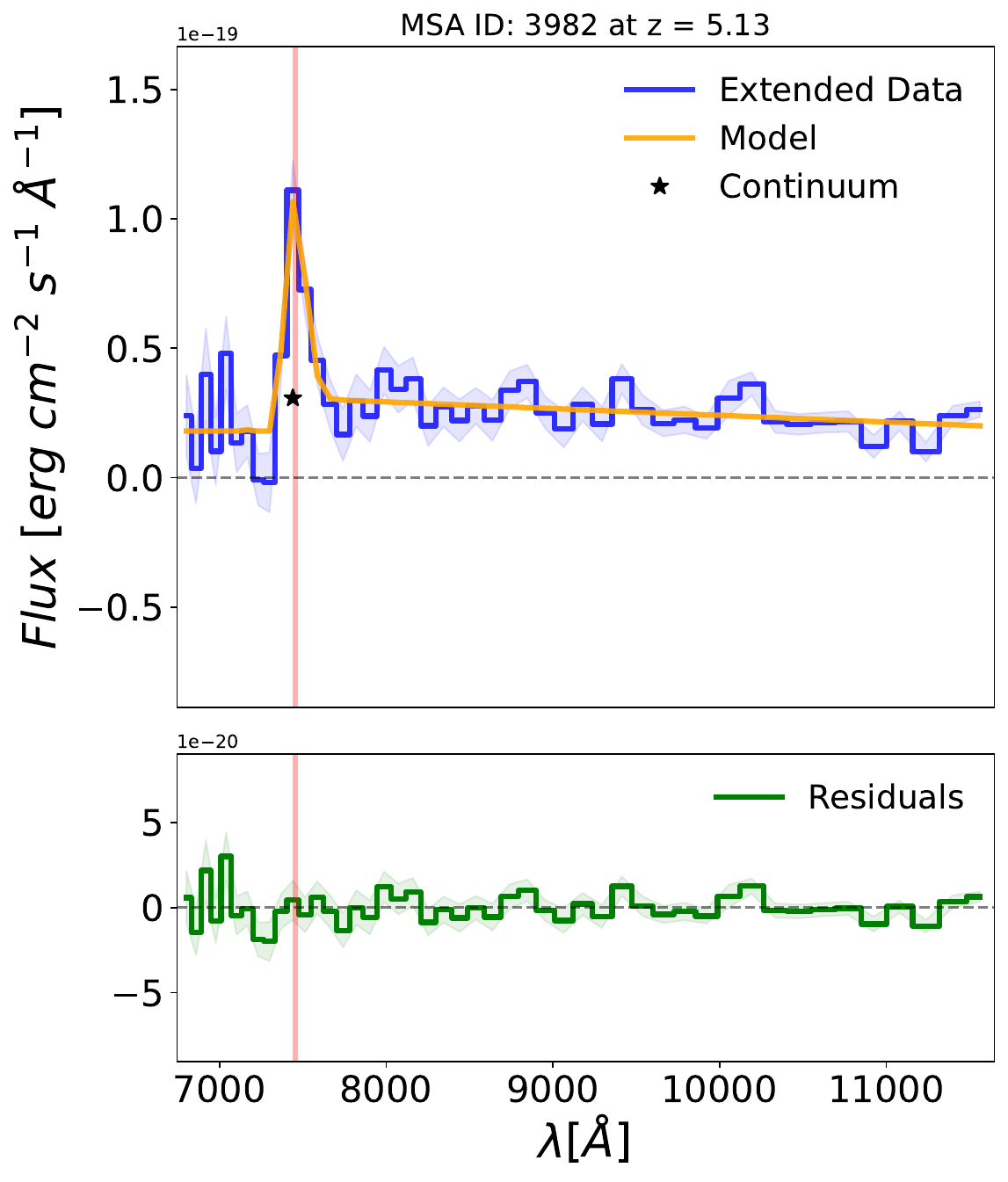}
    \end{subfigure}
    \begin{subfigure}[b]{0.22\textwidth}
        \includegraphics[width=\linewidth]{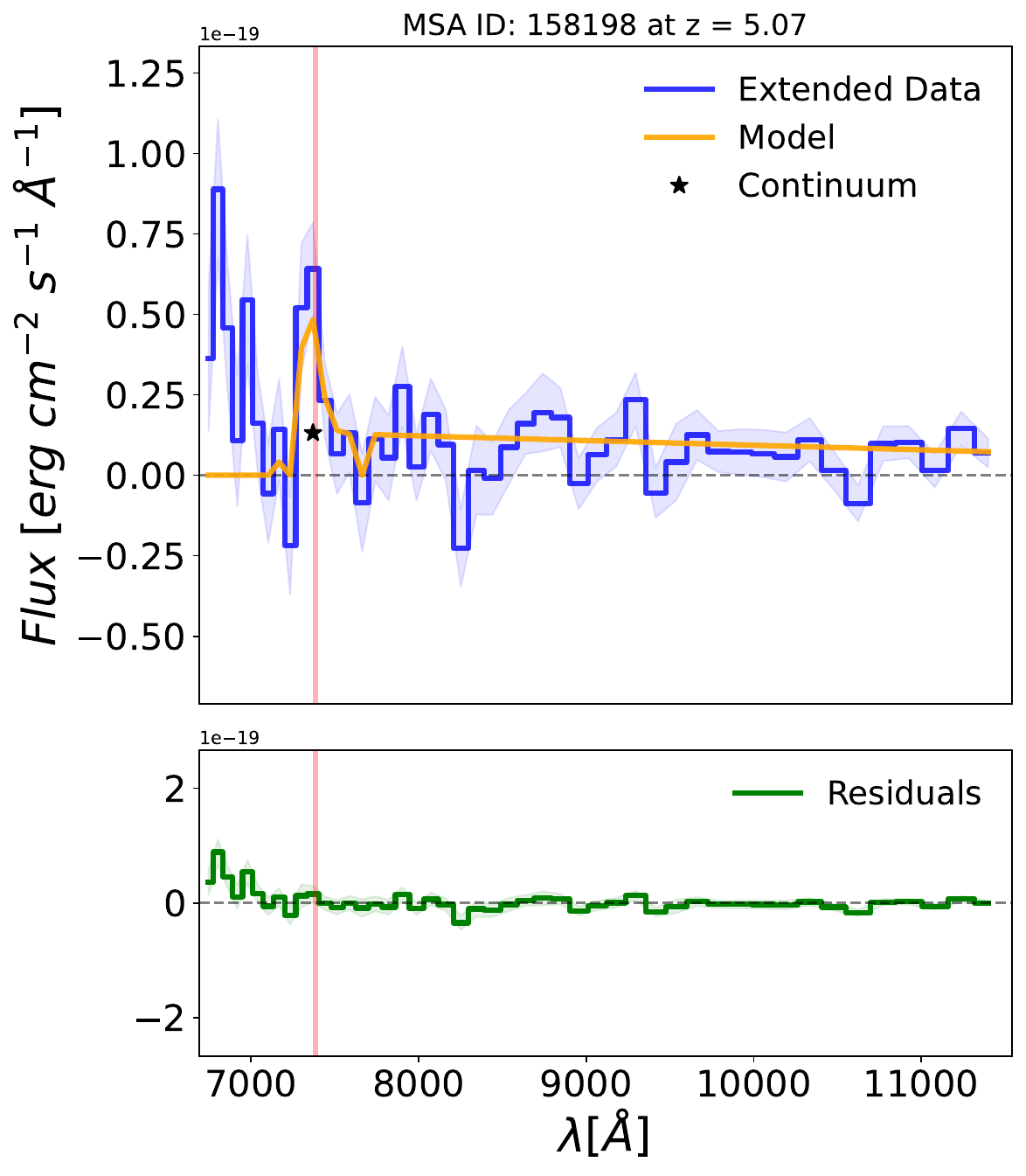}
    \end{subfigure}
    \begin{subfigure}[b]{0.22\textwidth}
        \includegraphics[width=\linewidth]{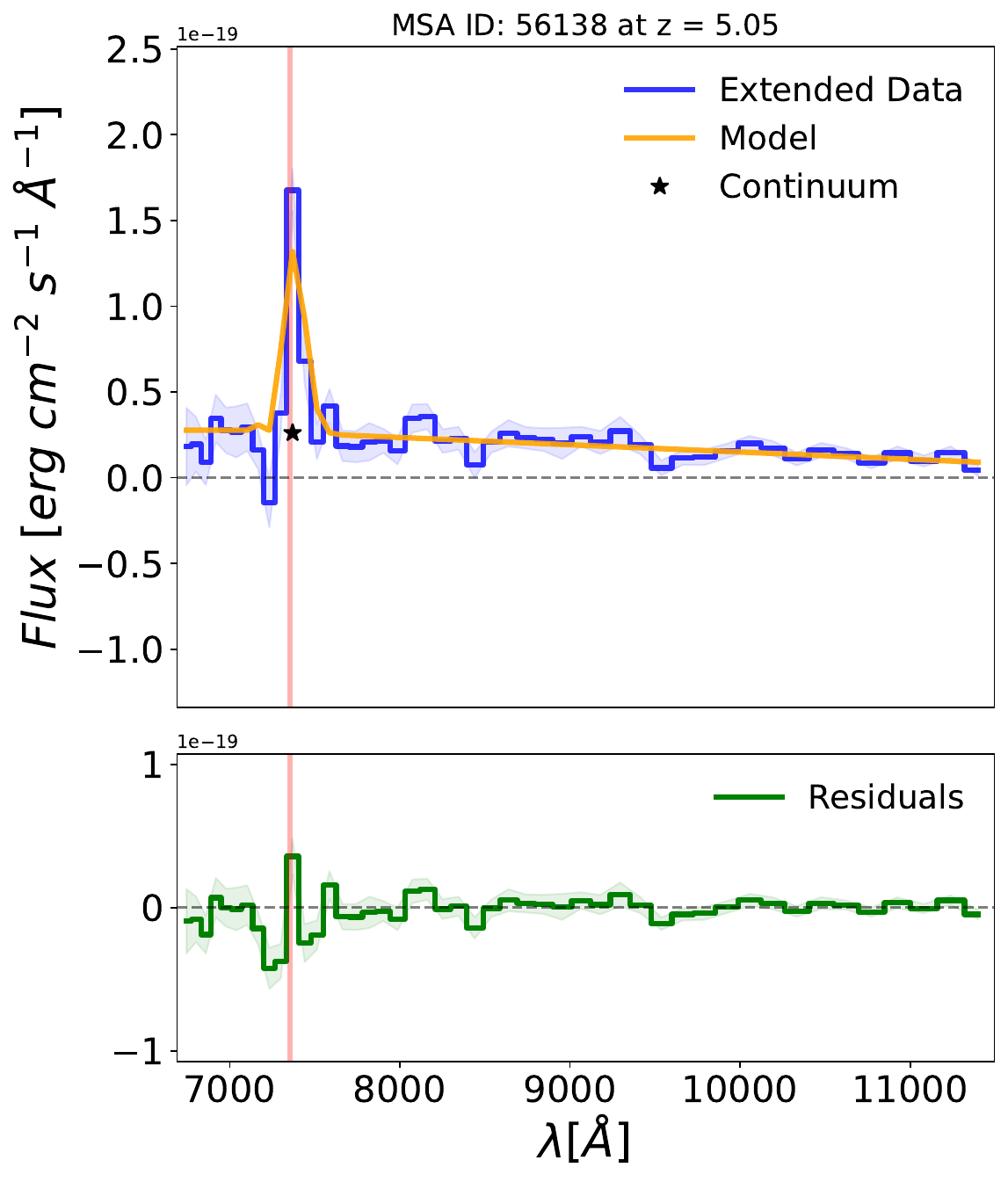}
    \end{subfigure}
    \begin{subfigure}[b]{0.22\textwidth}
        \includegraphics[width=\linewidth]{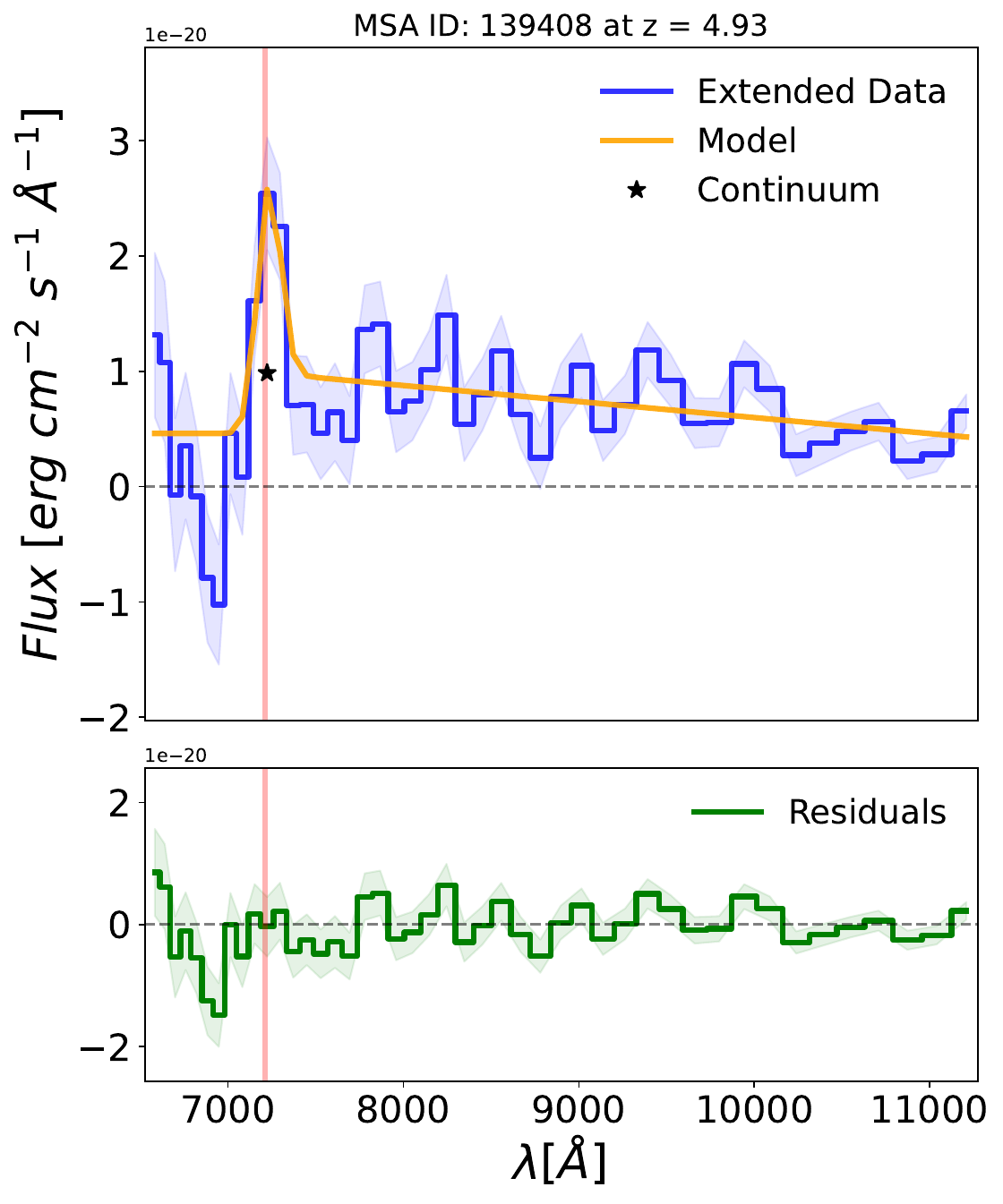}
    \end{subfigure}

    \vspace{0.1cm}

    \begin{subfigure}[b]{0.22\textwidth}
        \includegraphics[width=\linewidth]{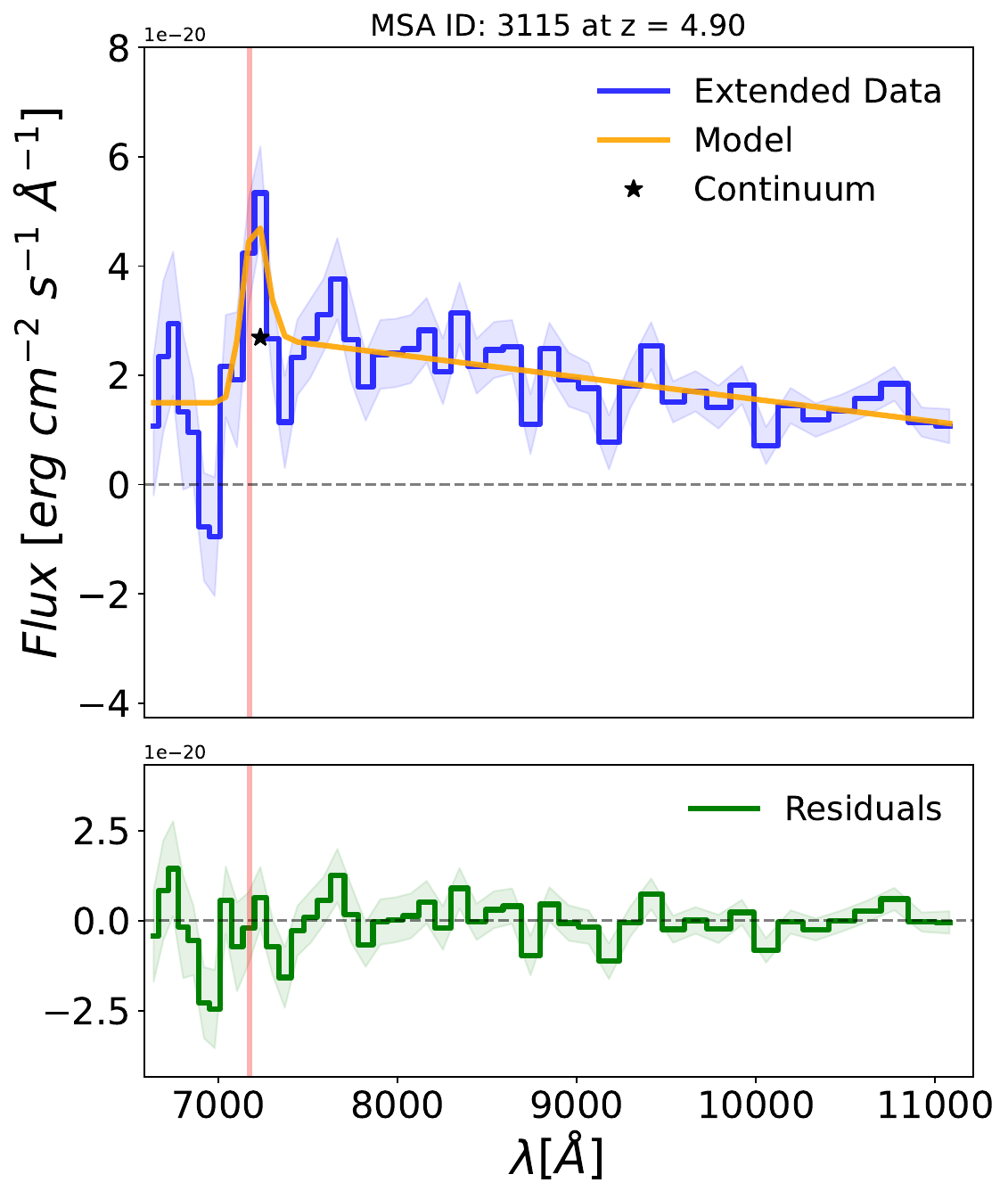}
    \end{subfigure}
    \begin{subfigure}[b]{0.22\textwidth}
        \includegraphics[width=\linewidth]{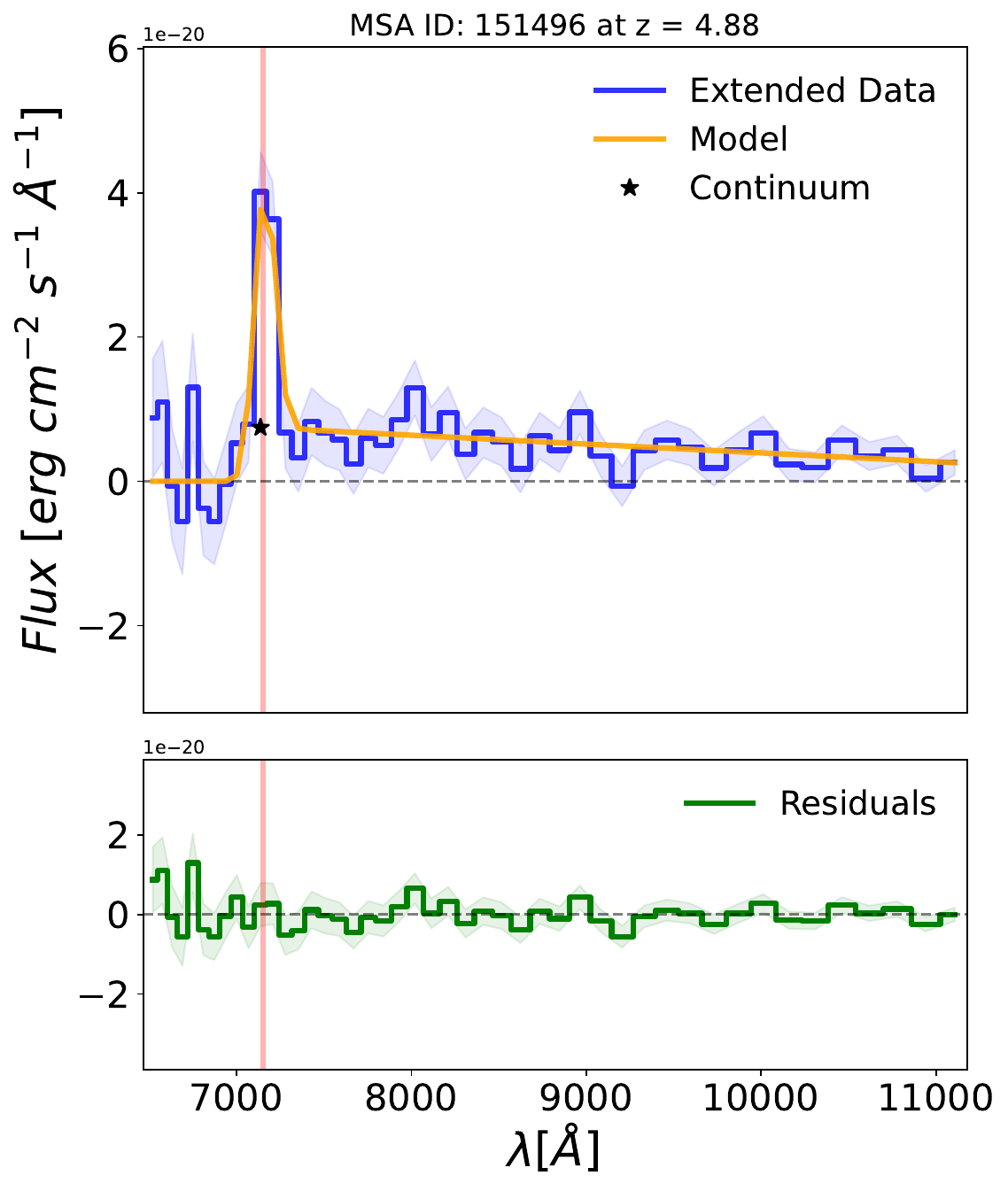}
    \end{subfigure}
    \begin{subfigure}[b]{0.22\textwidth}
        \includegraphics[width=\linewidth]{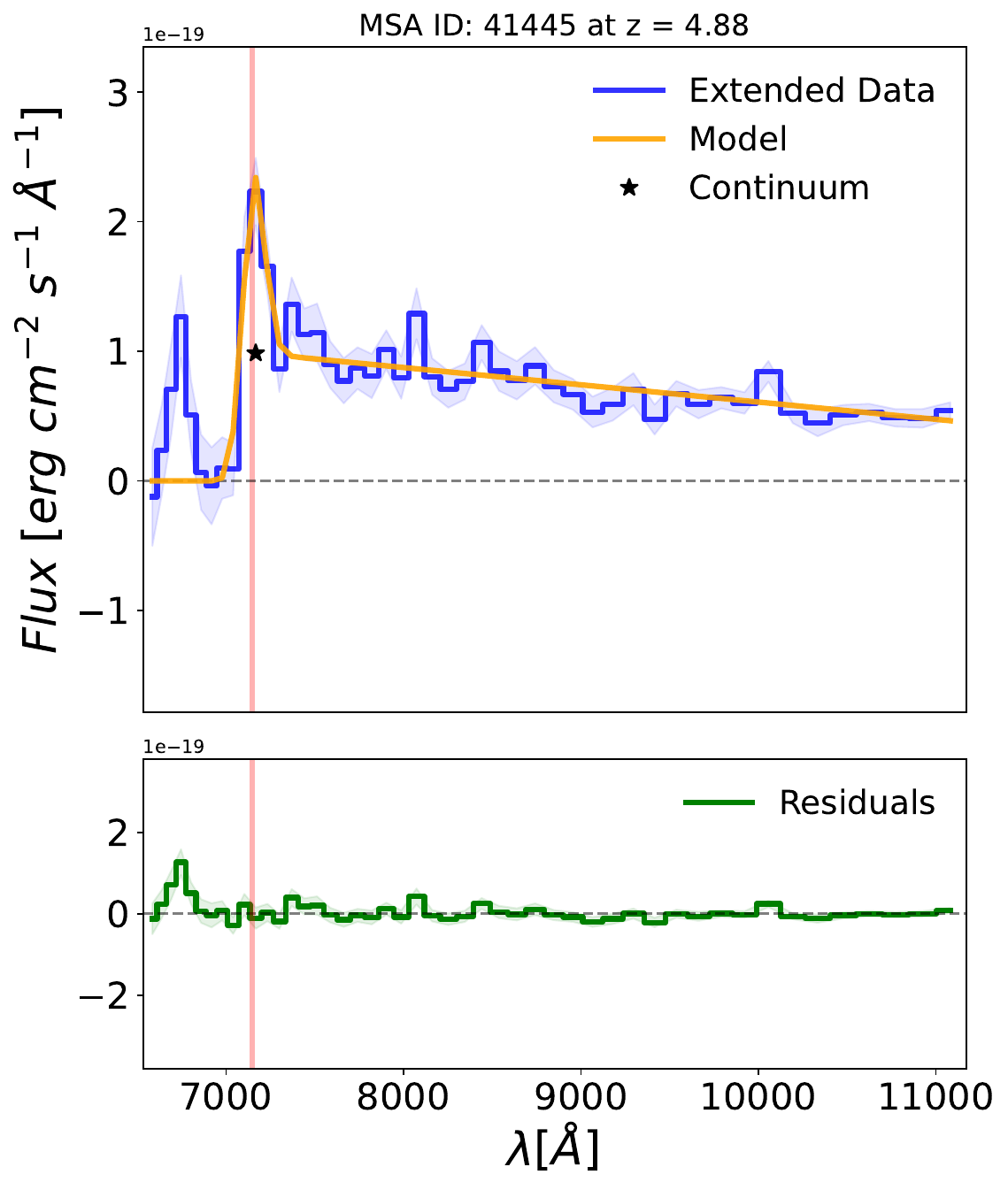}
    \end{subfigure}
    \begin{subfigure}[b]{0.22\textwidth}
        \includegraphics[width=\linewidth]{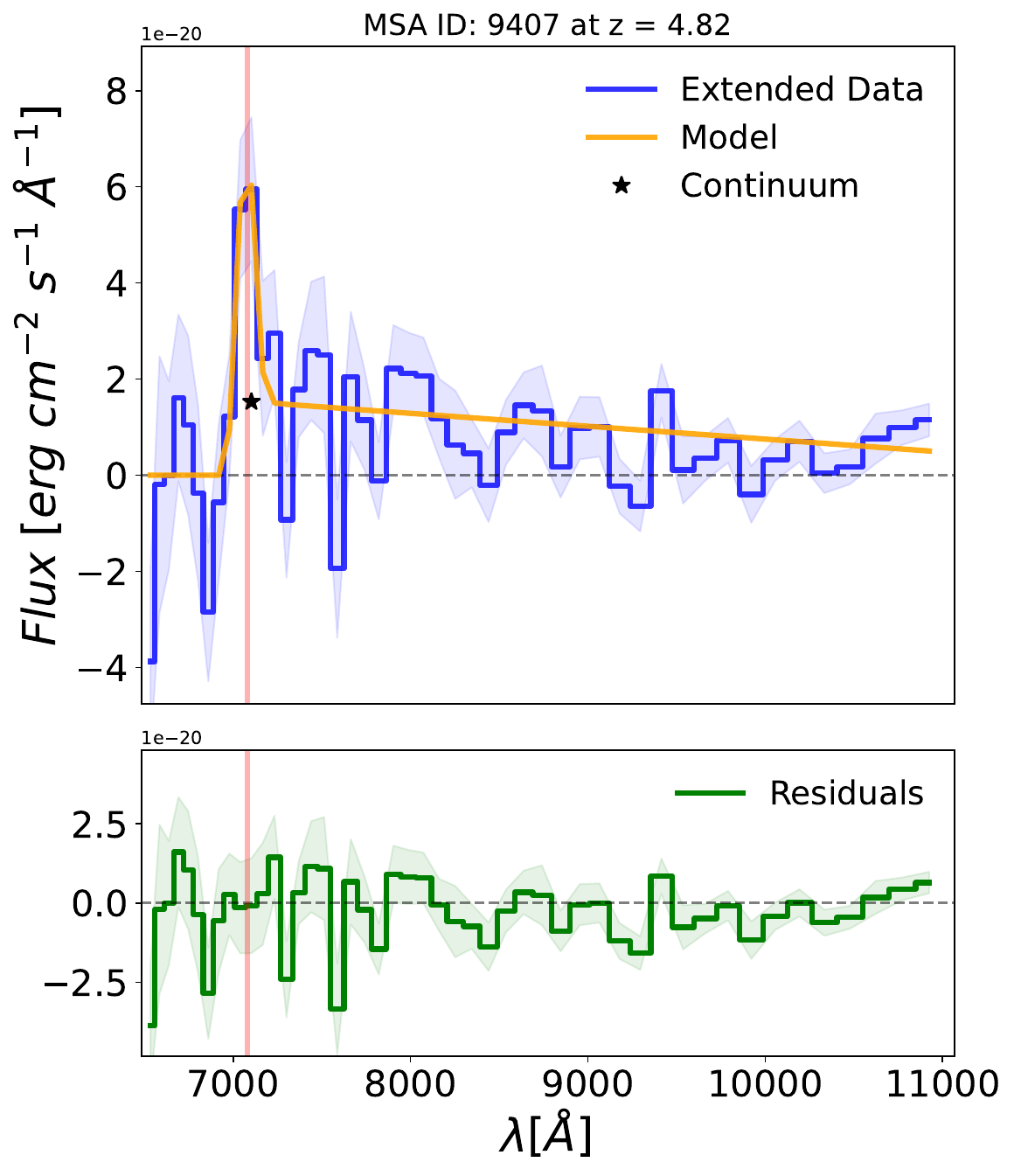}
    \end{subfigure}

    \caption{continued.}
    \label{fig:Lya_emitters_continued3}
\end{figure*}

\begin{figure*}[ht!]
    \ContinuedFloat

    \centering

    \begin{subfigure}[b]{0.22\textwidth}
        \includegraphics[width=\linewidth]{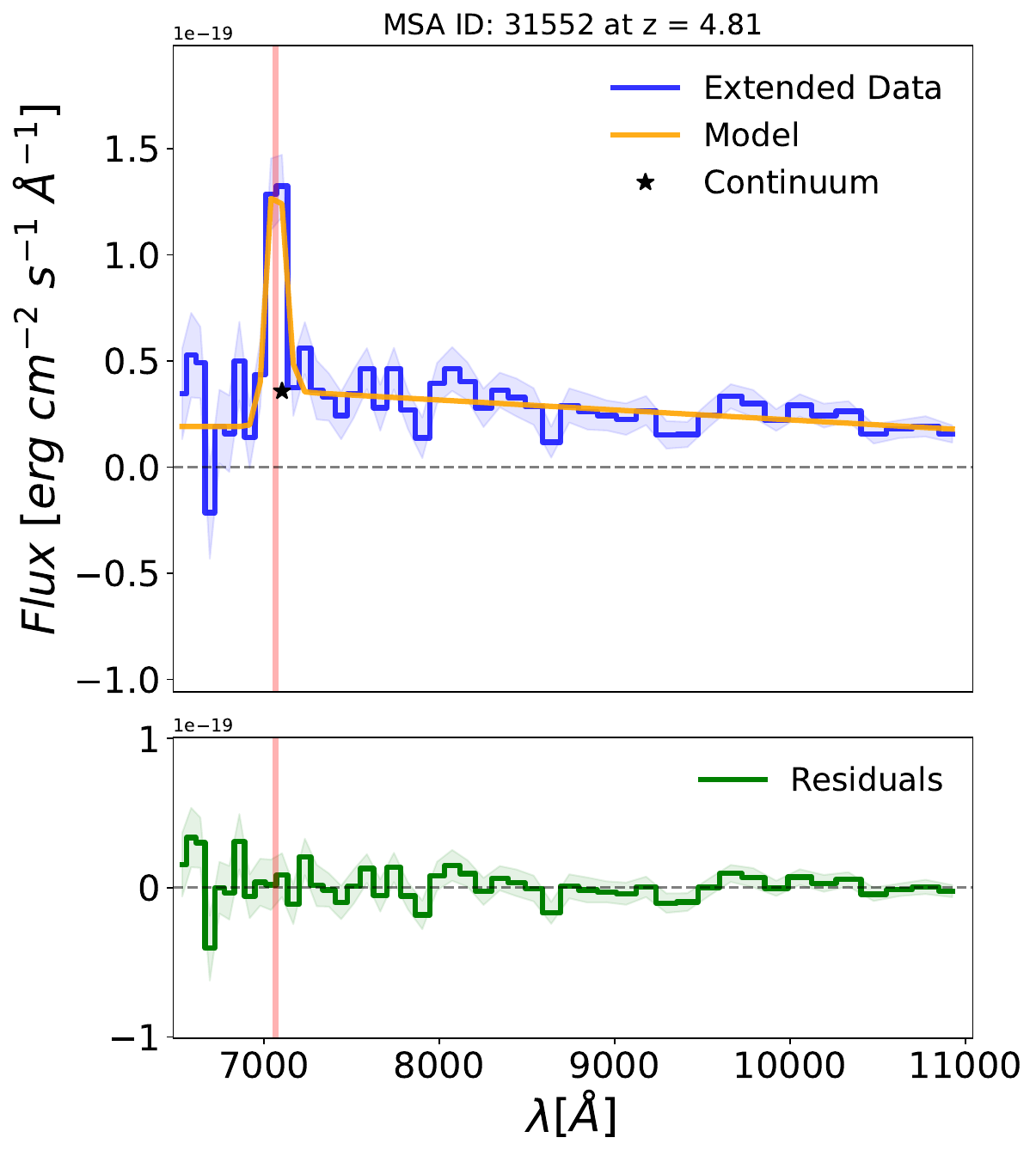}
    \end{subfigure}
    \begin{subfigure}[b]{0.22\textwidth}
        \includegraphics[width=\linewidth]{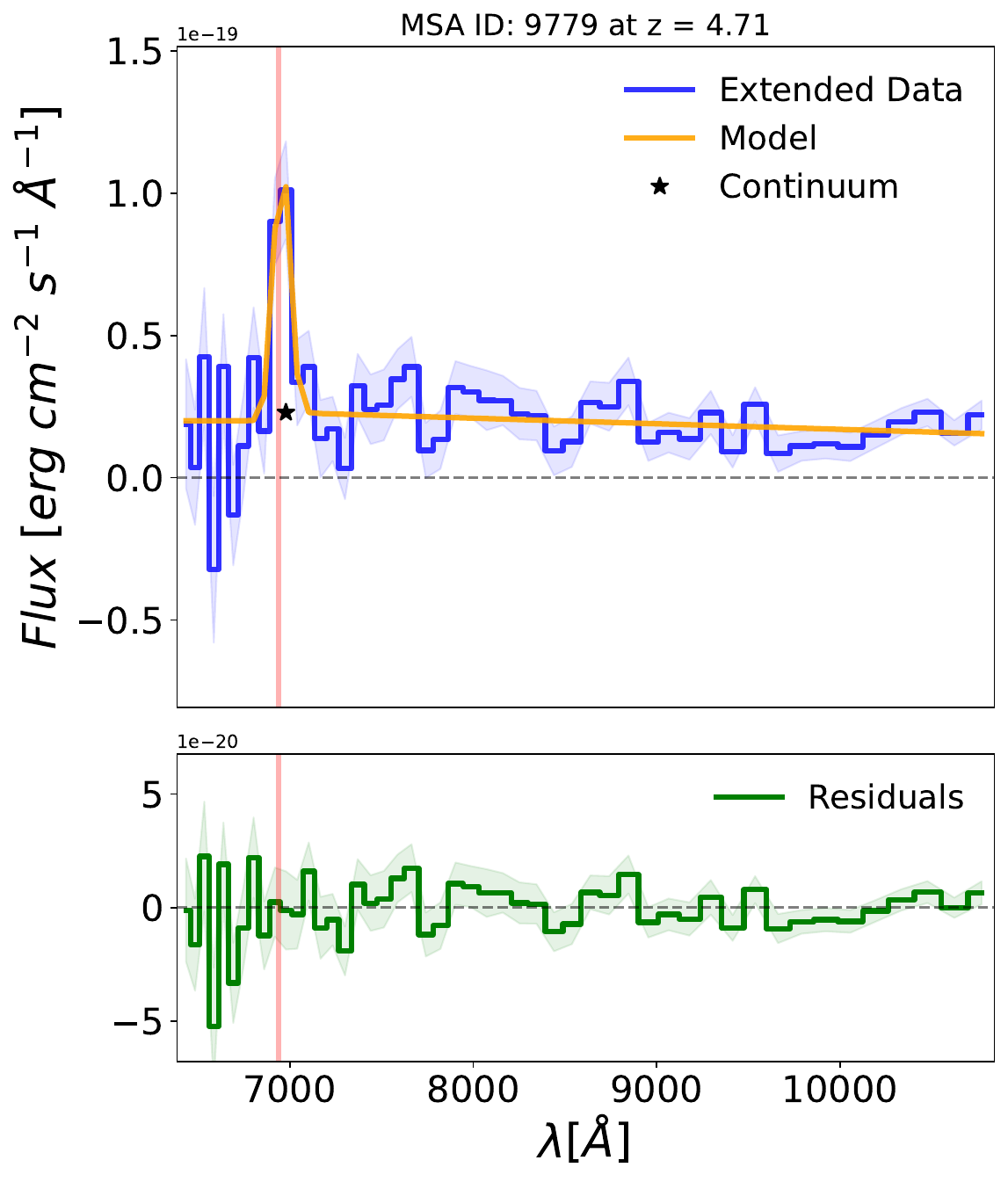}
    \end{subfigure}
    \begin{subfigure}[b]{0.22\textwidth}
        \includegraphics[width=\linewidth]{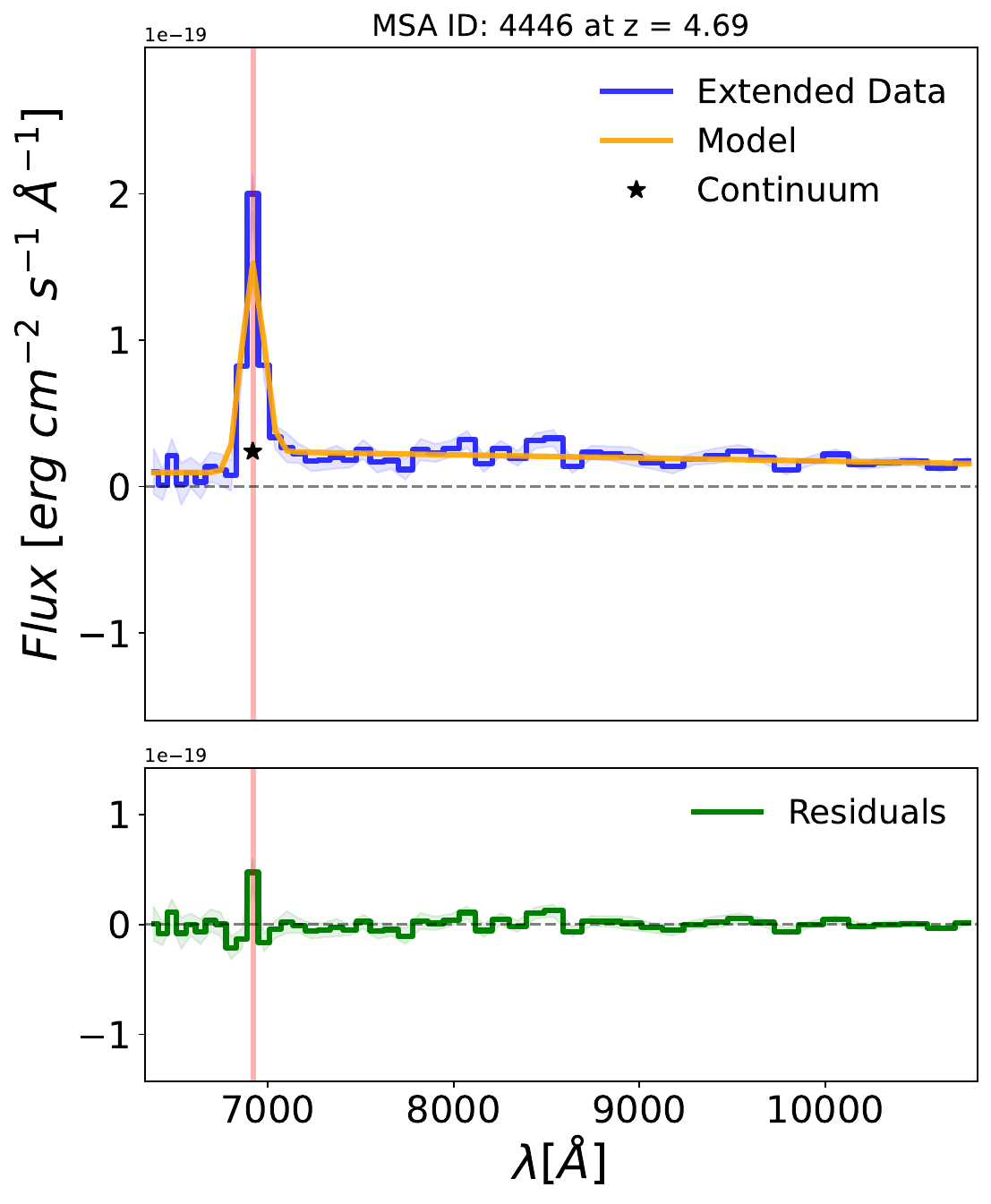}
    \end{subfigure}
    \begin{subfigure}[b]{0.22\textwidth}
        \includegraphics[width=\linewidth]{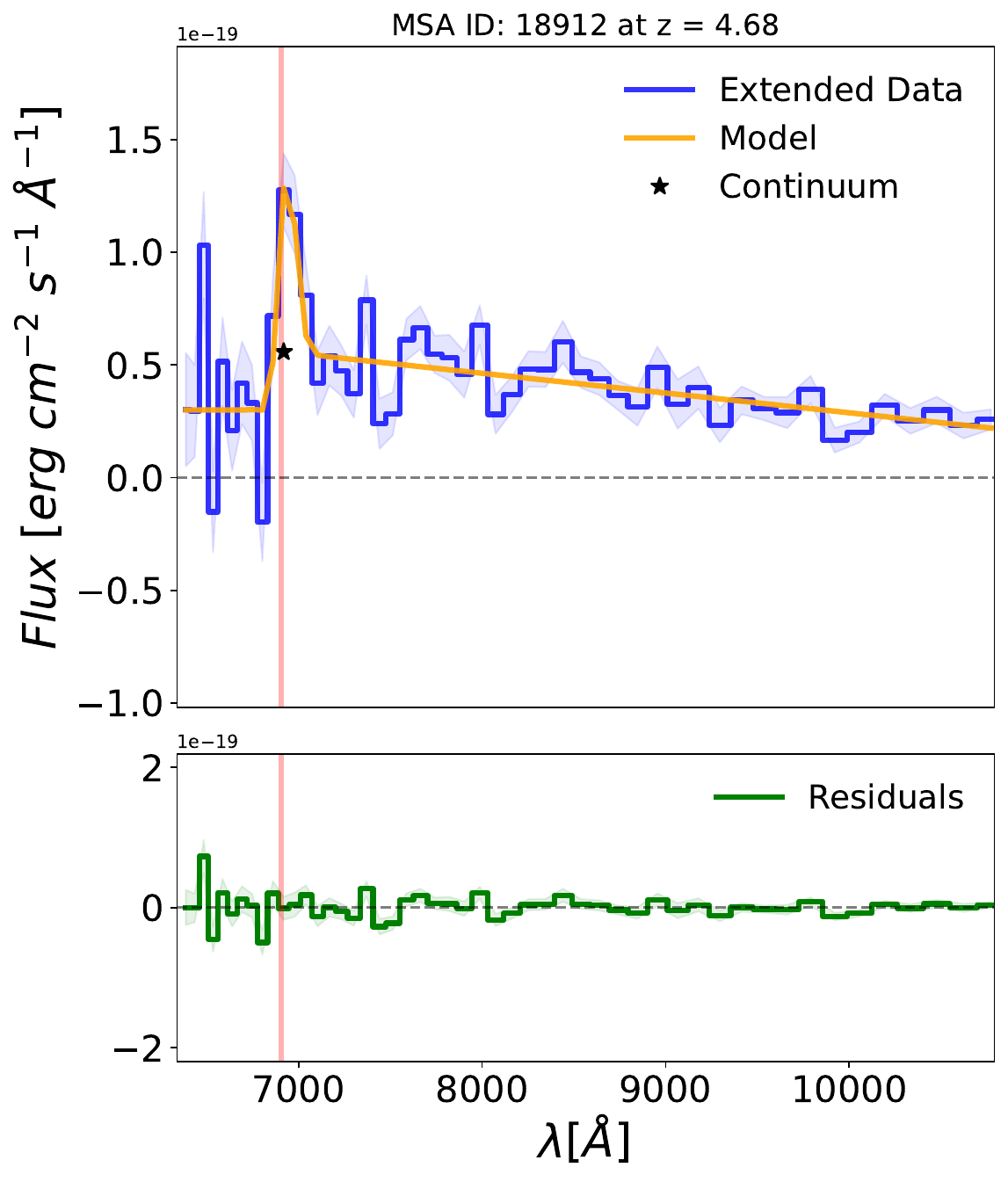}
    \end{subfigure}

    \vspace{0.1cm}

    \begin{subfigure}[b]{0.22\textwidth}
        \includegraphics[width=\linewidth]{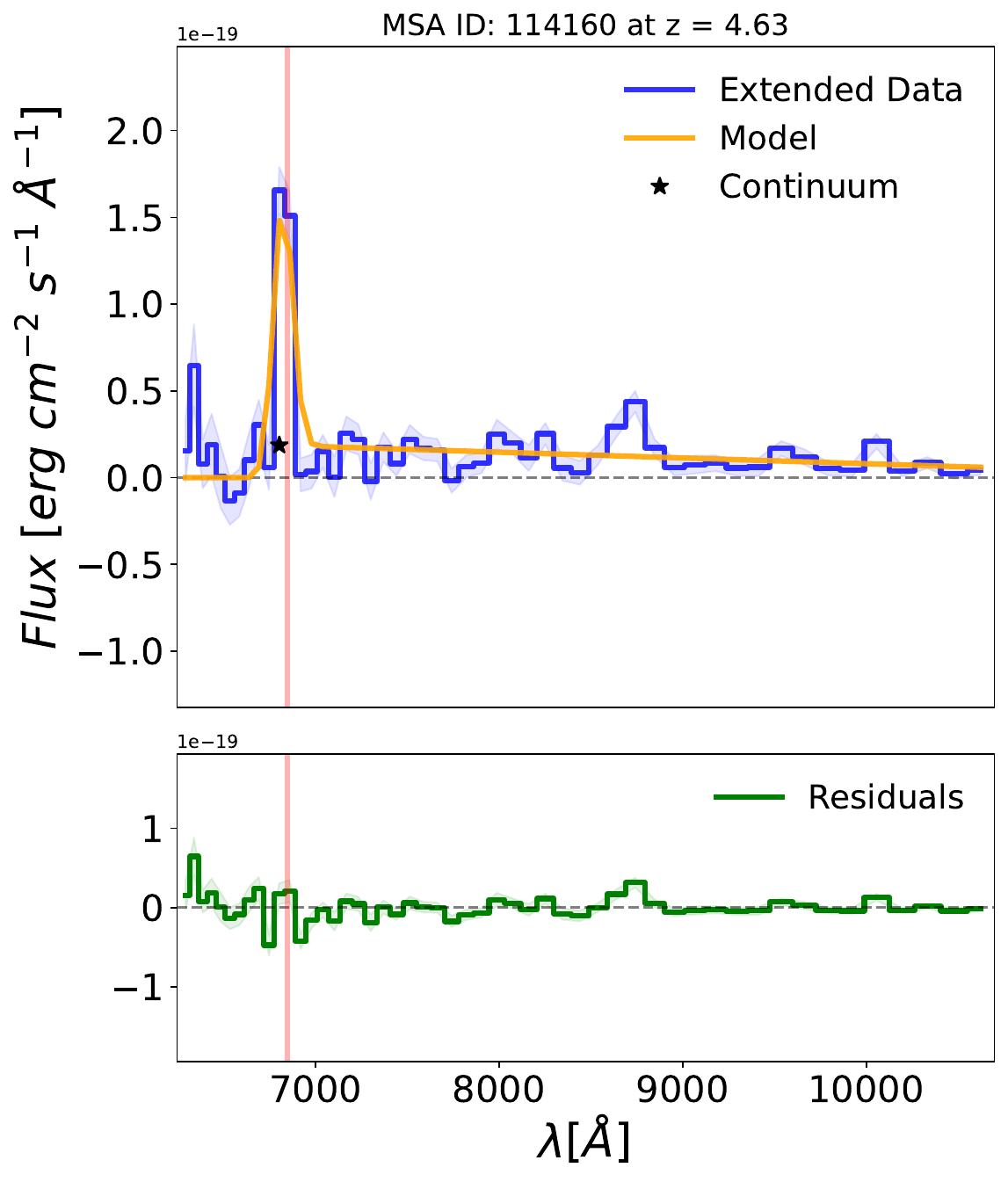}
    \end{subfigure}
    \begin{subfigure}[b]{0.22\textwidth}
        \includegraphics[width=\linewidth]{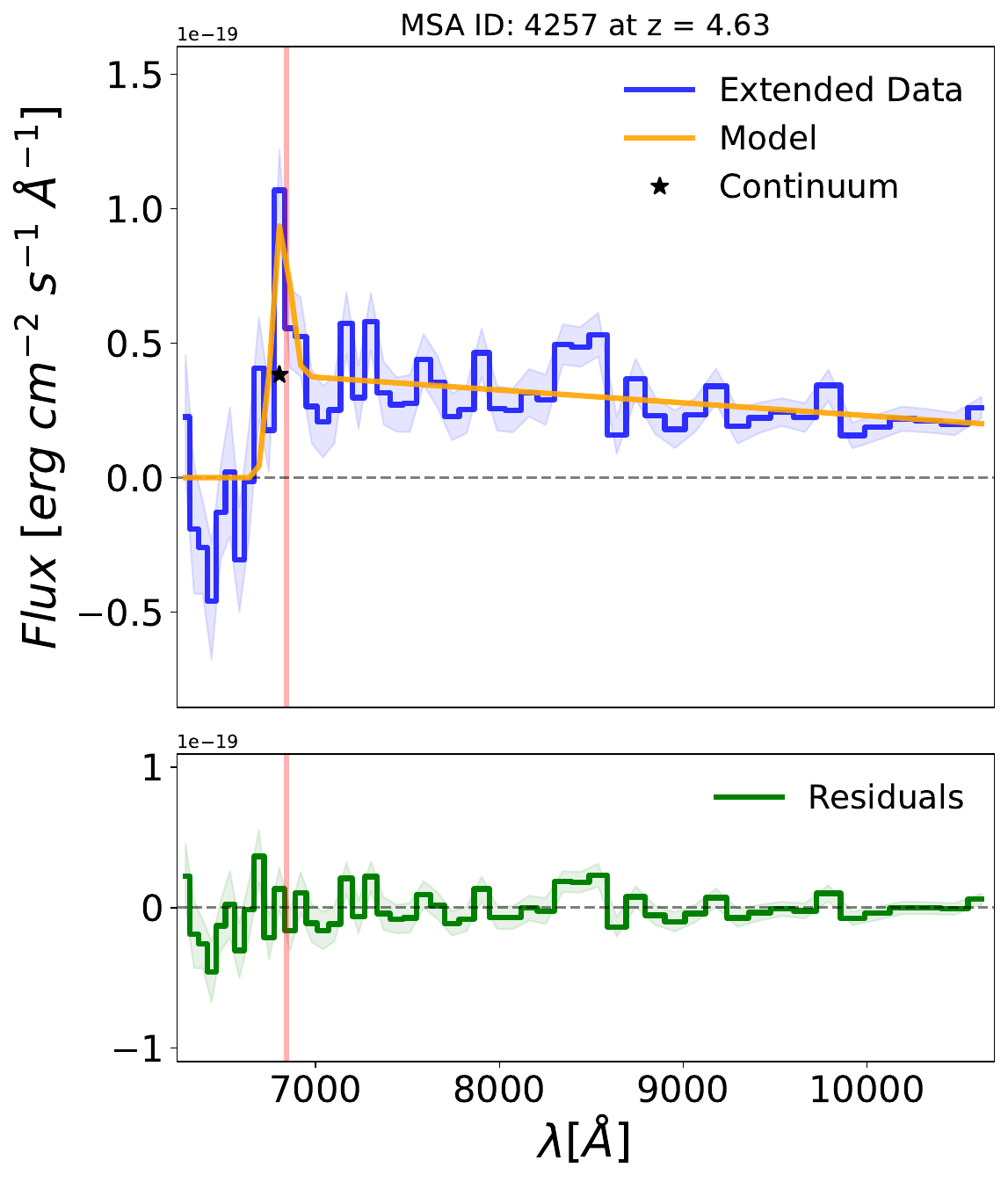}
    \end{subfigure}
    \begin{subfigure}[b]{0.22\textwidth}
        \includegraphics[width=\linewidth]{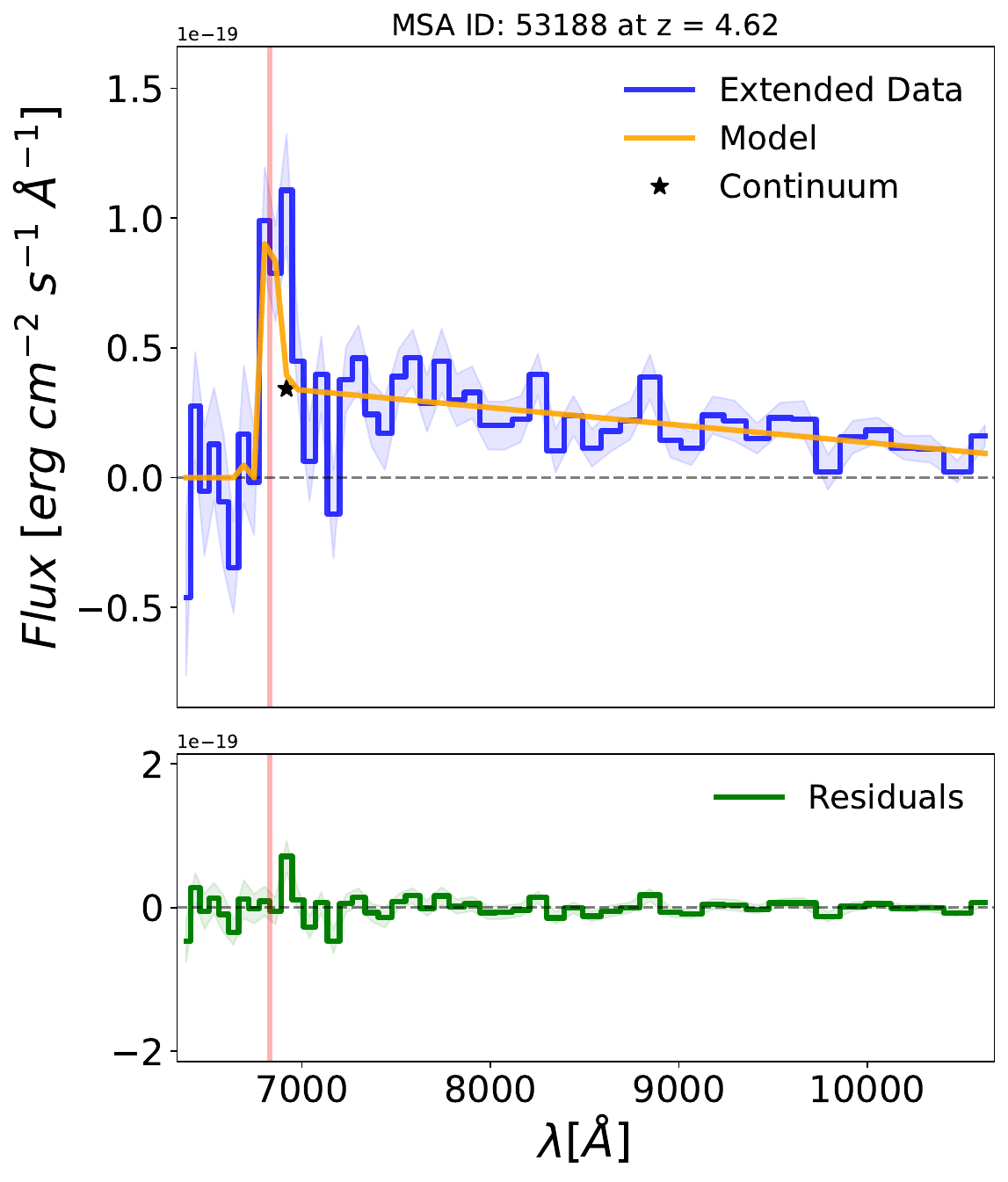}
    \end{subfigure}
    \begin{subfigure}[b]{0.22\textwidth}
        \includegraphics[width=\linewidth]{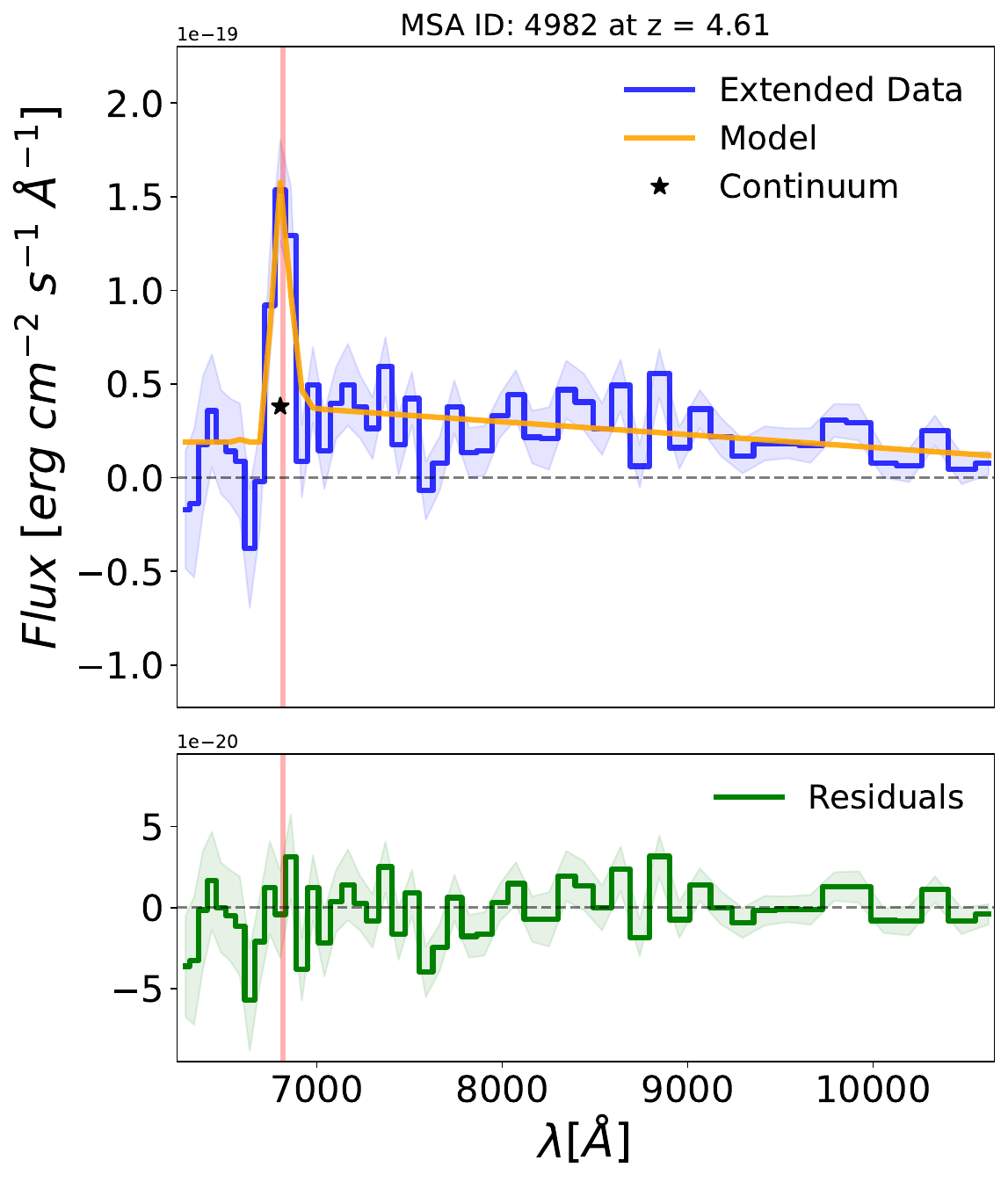}
    \end{subfigure}

    \vspace{0.1cm}

    \begin{subfigure}[b]{0.22\textwidth}
        \includegraphics[width=\linewidth]{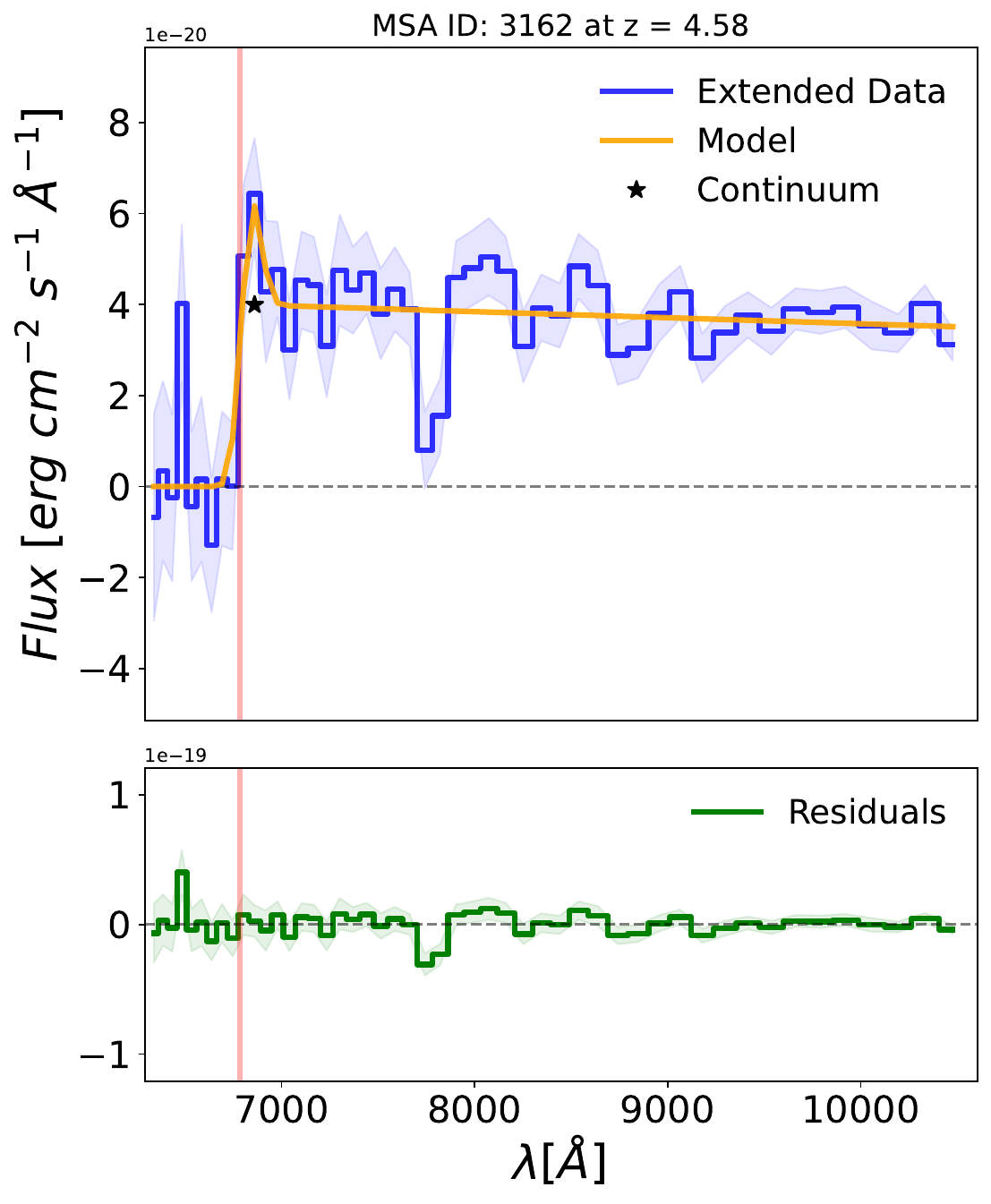}
    \end{subfigure}
    \begin{subfigure}[b]{0.22\textwidth}
        \includegraphics[width=\linewidth]{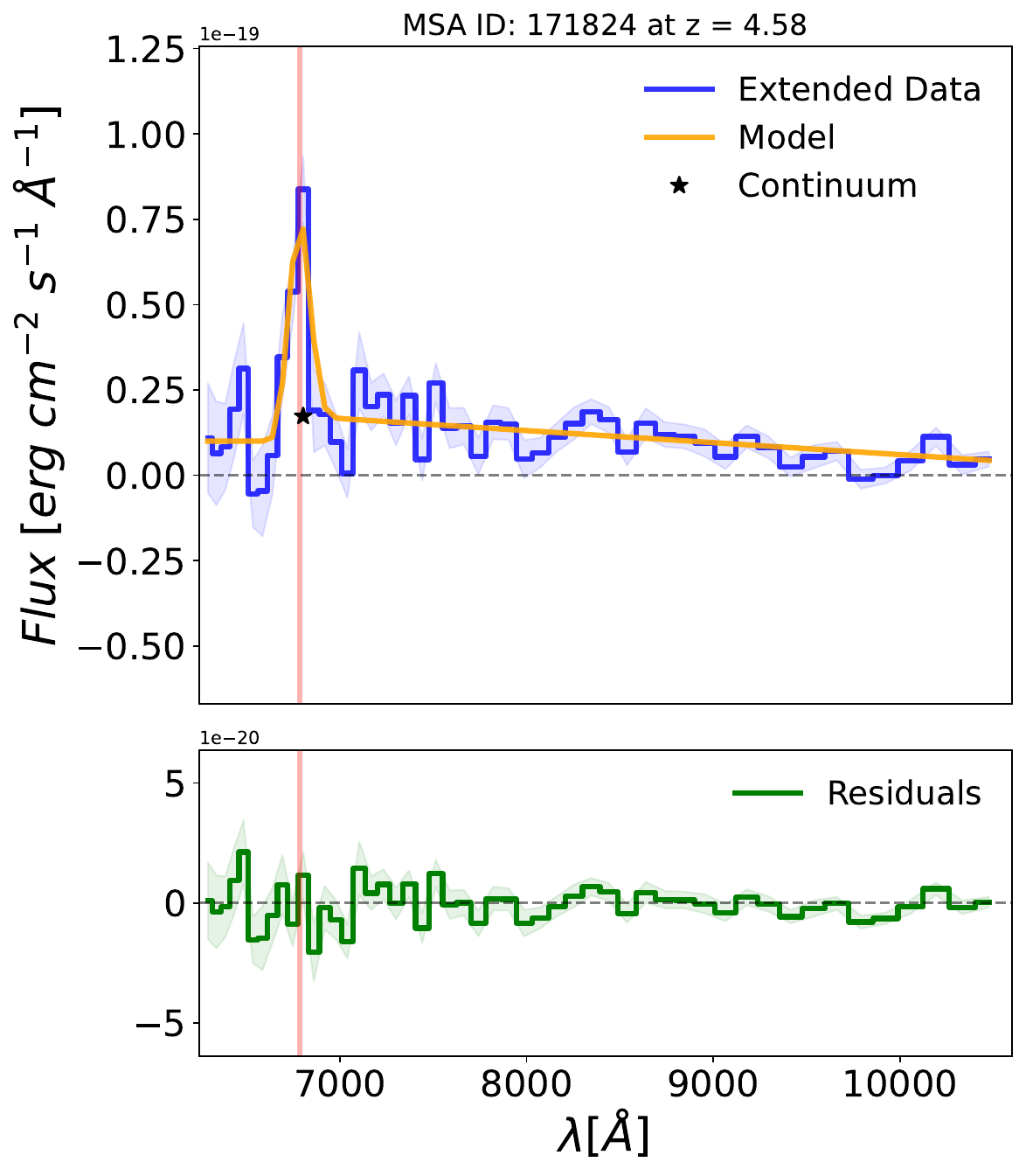}
    \end{subfigure}
    \begin{subfigure}[b]{0.22\textwidth}
        \includegraphics[width=\linewidth]{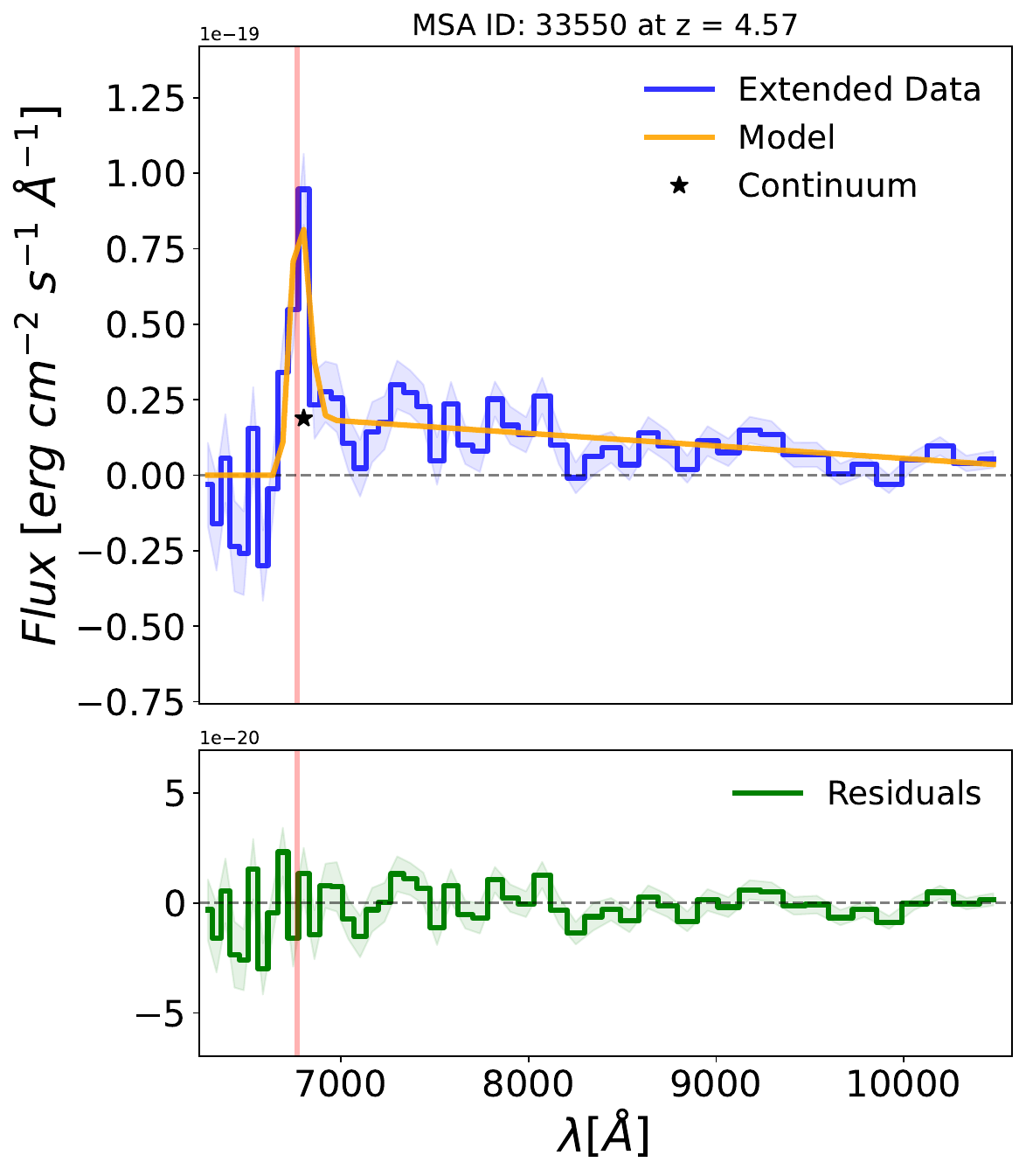}
    \end{subfigure}
    \begin{subfigure}[b]{0.22\textwidth}
        \includegraphics[width=\linewidth]{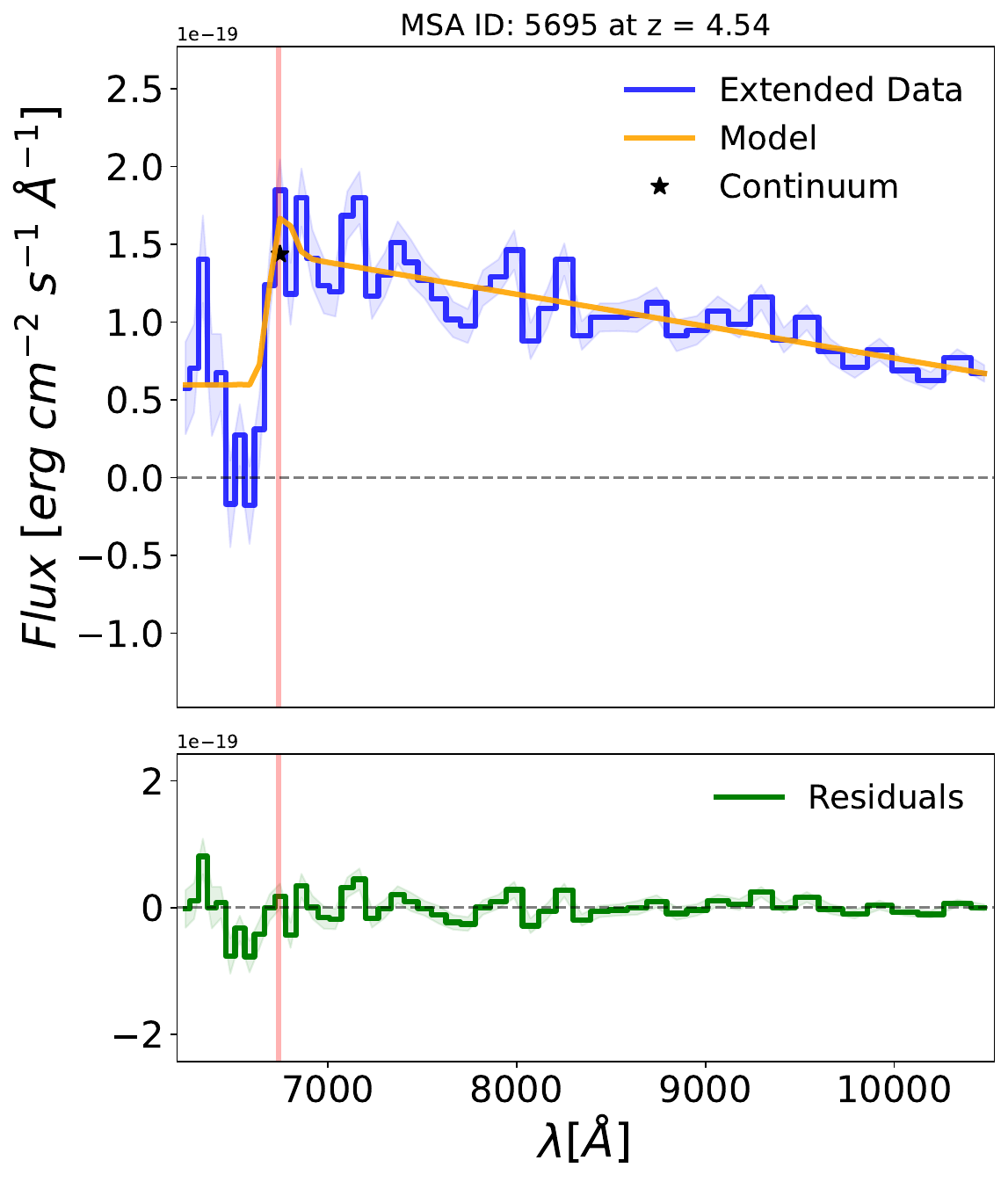}
    \end{subfigure}

    \vspace{0.1cm}

    \begin{subfigure}[b]{0.22\textwidth}
        \includegraphics[width=\linewidth]{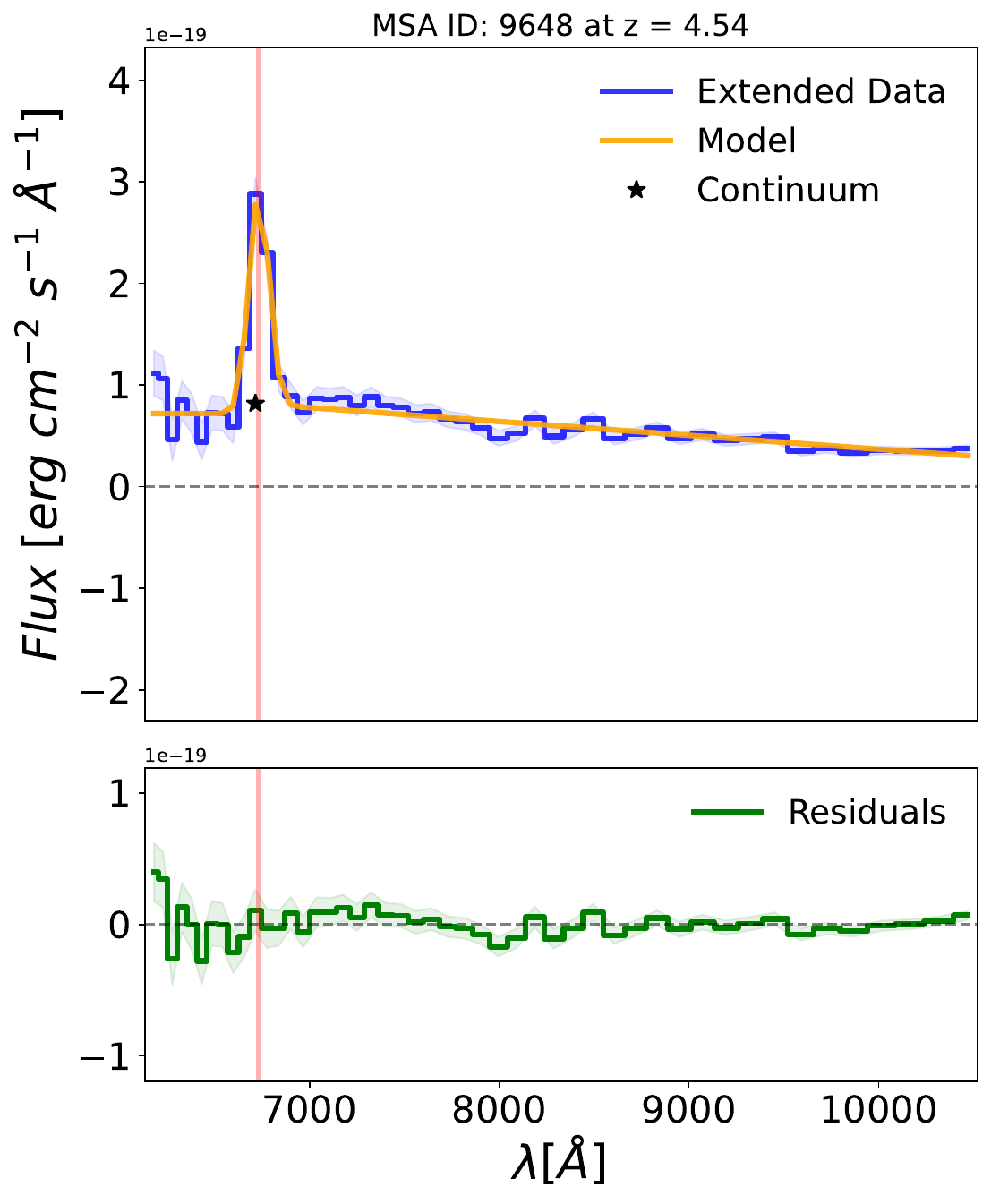}
    \end{subfigure}

    \caption{continued.}
    \label{fig:Lya_emitters_continued4}
\end{figure*}

\end{appendix}

\end{document}